\documentclass[nofootinbib,showkeys,floatfix,times]{revtex4-2}
\usepackage{bm,morefloats}
\usepackage{dcolumn}
\usepackage[T2A,T1]{fontenc} %
\usepackage{fonttable}
\usepackage{graphicx}
\usepackage{balance}
\usepackage[latin1]{inputenc}
        \usepackage{cellspace}
\usepackage{latexsym,color}
\usepackage{centernot}
\usepackage{mathtools}
\usepackage{stmaryrd}
\usepackage{amssymb}
\usepackage[T1]{fontenc}
\usepackage{natbib}
\usepackage{amsmath,amssymb,amsthm,mathrsfs,amsfonts,dsfont}

\usepackage{color}

\usepackage{rotating}
\usepackage[dvipsnames]{xcolor}
\definecolor{LinkBlue}{RGB}{6,69,173}
\definecolor{DarkBlue}{RGB}{11,0,128}
\definecolor{red}{rgb}{1,0.,0.}
\usepackage[colorlinks=true,linkcolor=LinkBlue,urlcolor=magenta,
	citecolor=green,hyperfootnotes=true,urlcolor=red]{hyperref}

\usepackage{enumitem}

\begin{document}

\count\footins = 1000

\title{On  dark matter   effect  on  BH   accretion disks}
\author{D. Pugliese\&Z. Stuchl\'{\i}k}
\email{}
\affiliation{
Research Centre for Theoretical Physics and Astrophysics, Institute of Physics,
  Silesian University in Opava,
 Bezru\v{c}ovo n\'{a}m\v{e}st\'{i} 13, CZ-74601 Opava, Czech Republic
}

\begin{abstract}
Comparing     different dark matter (DM) models,
we explore the  DM  influence on the black hole  ({BH}) accretion disk physics,  considering      co-rotating and counter-rotating     thick  accretion  tori orbiting a central spinning  {BH}.  Our results point out  accretion onto a central BH as a good indicator  of the {DM} presence, signaling  possible DM tracers in the accretion physics.
We analyze accretion around a spinning BH immersed in perfect-fluid dark matter,  cold dark matter  and scalar field dark matter.
 Our investigation   discusses   observational evidence of  distinctive  DM effects on  the toroidal  accretion disks and  proto-jets configurations, proving that
BHs accretion tori, immersed in DM, can present characteristics, such as  inter--disks cusp or double  tori,  which  have been  usually considered as tracers for super-spinars and naked singularity attractors. Therefore in this context    DM influence on the BH geometry could manifest as  super-spinars mimickers.  DM affects  also the central spinning attractor energetics  associated with the accretion physics,    and  its  influence on  the accretion disks  can be searched  in a variation   of the central BH energetics as  an increase of the mass accretion rates.
\end{abstract}
\date{\today}

\maketitle

\def\be{\begin{equation}}
\def\ee{\end{equation}}
\def\bea{\begin{eqnarray}}
\def\eea{\end{eqnarray}}
\newcommand{\tb}[1]{\textbf{\texttt{#1}}}
\newcommand{\actaa}{Acta Astronomica}
\newcommand{\laa}{\mathcal{L}}
\newcommand{\ba}{\mathcal{B}}
\newcommand{\Sie}{\mathcal{S}}
\newcommand{\Mie}{\mathcal{M}}
\newcommand{\La}{\mathcal{L}}
\newcommand{\Em}{\mathcal{E}}

\newcommand{\mso}{\mathrm{mso}}
\newcommand{\mbo}{\mathrm{mbo}}

\newcommand{\rtb}[1]{\textcolor[rgb]{1.00,0.00,0.00}{\tb{#1}}}
\newcommand{\gtb}[1]{\textcolor[rgb]{0.17,0.72,0.40}{\tb{#1}}}
\newcommand{\ptb}[1]{\textcolor[rgb]{0.77,0.04,0.95}{\tb{#1}}}
\newcommand{\btb}[1]{\textcolor[rgb]{0.00,0.00,1.00}{\textbf{#1}}}
\newcommand{\otb}[1]{\textcolor[rgb]{1.00,0.50,0.25}{\textbf{#1}}}
\newcommand{\non}[1]{{\LARGE{\not}}{#1}}

\newcommand{\cc}{\mathrm{C}}

\newcommand{\il}{~}
\newcommand{\la}{\mathcal{A}}
  \newcommand{\Qa}{\mathcal{Q}}
\newcommand{\Sa}{\mathcal{\mathbf{S}}}
\newcommand{\Ta}{{\mbox{\scriptsize  \textbf{\textsf{T}}}}}
\newcommand{\Ca}{\mathcal{\mathbf{C}}}

\section{Introduction}
In this work we study  the dark matter (DM) influence on the black hole (BH) accretion disk physics, investigating
 the accretion disks morphology for   co-rotating and counter-rotating  geometrically   thick accretion  tori orbiting a central spinning  BH.
Comparing   three  different DM models, our  analysis points out  possible observational evidence of  distinctive  DM effects on  the accretion disks,  which can be traces for the DM presence.
Our  investigation  does not cover all the admissible parametric DM values for the  deformed  metrics but, taking into account  constraints  formerly obtained by  the  study of  orbits, the { DM  BH} shadows, or the emission spectra, we perform a  comparative analysis of the DM  models and an  investigation  on the constraints imposed on the accretion discs, aimed at further restricting  the parameters range and  pointing out the possible DM  mark  in some accretion features.
From a methodological  viewpoint we take advantage of
axial symmetry  of DM metrics,   studying fully general relativistic models of stationary  toroidal orbiting configurations.

The physics of accretion disks around BHs and  super massive black hole (SMBH),  hosted in  quasars and active galactic nuclei ({AGN}), powers the  most energetic processes of our Universe,   often accompanied with  ejection of matter  in  jet-like structures with extremely large radiative energy output. We investigate  the DM effects on these aspects, focusing on SMBH at the center of galaxies and   the accretion discs empowering the emissions.
 We analyze particularly the limiting situation of static (and spherically symmetric) background (with spin parameter $a=0$)  and the  DM deformation on the Kerr extreme  BH spacetime (with spin value $a=M$), seeing  significant qualitative detectable   variations with respect to  the  standard vacuum BH case.
 Many astrophysical observations lead to SMBH  hosted at the  galactic center embedded in
a  DM halo, and in  metric models considered here    the central BH  is surrounded by DM  envelope   that   modifies
the geometry around the BH, not interacting directly with the accreting  matter or its  radiation.

 More generally there is a large amount of observational evidence of the presence of  some kind of  DM component in our Universe,  for example   from the galactic rotation curves and  galaxy cluster dynamics. 
However, within the variety of the  different observations pointing at the  DM presence, there is no single metric that encompasses all the DM effects in a single model, and which could also explain the absence of DM  observed at different scales\footnote{An important issue is then  the  mass (or space) scale  when DM effects became significant.
For example  DM  results   missing    in some galaxies (ex. {AGC 114905} --\cite{2022MNRAS.512.3230M,Pavel}), and an explanation for this situation is that the galaxy
  may have been stripped of dark matter from nearby massive galaxies, while  the   DM  presence in the  Solar system is still an open problem\cite{belbruno}.}. 
In this work  we consider  a spinning BH immersed in perfect fluid DM (PFDM) \cite{Ech-Qo,pfdm,Kerr-desditte,Das},  cold dark matter (CDM)  and scalar field dark matter   (SFDM)--\cite{Enta,relative}.

These DM models have been extensively studied  in  recent  literature.
PFDM model was considered in \cite{pfdm} and the effects of PFDM  on particle motion around a static
BH  in an external magnetic field were studied in
\cite{PDUM}. 
The shadow of the spinning BH in PFDM  was studied in \cite{Ech-Qo}, whereas in \cite{Kerr-desditte}
geodesic motion   in PFDM Kerr and  Kerr-anti-de Sitter/de Sitter BH were studied--see also \cite{Das}. For studies of the geodesic in the Kerr-de Sitter spacetimes see \cite{1983BAICz..34..129S,1999PhRvD..60d4006S,2020Univ....6...26S,2000CQGra..17.4541S,2005MPLA...20..561S}.
In \cite{KP}
superradiance and stability of Kerr DM enclosed by anisotropic
fluid matter  were studied.
Spinning  BH solutions with quintessential energy have  been  discussed in \cite{ZS}.
In \cite{Enta}
BH  in  DM halo was considered.
Rotating black holes with an anisotropic matter field  was considered in \cite{Kform}, while
in \cite{relative} there is a discussion of the BH shadow of {Sgr A*} in DM halo.
Superradiance and instabilities in {BHs} surrounded
by anisotropic fluids   were considered in  \cite{aolog}.
Galactic dark matter in the phantom field model was considered in \cite{Hsun}.
The case of rotating (Kerr) naked singularities was treated in \cite{2013CQGra..30g5012S,1980BAICz..31..129S,2016PhRvD..94h6006B,2011CQGra..28o5017S}. Here we focus  on the influence of DM on BHs governing toroidal accretion structures.

{There  is an extensive literature exploring the DM effects    on the BH and BH accretion  physics. Since  DM  influences different aspects of the singularity, from the characteristics of the horizon (for example it can manifest itself in the BH shadows) to the  energetics properties of the surrounding matter, there are various assessments of the DM effects and parameter constraints  on the  models describing  DM presence around BHs.
In  \cite{Traykova} for example
 DM clouds of  axions  around BHs  were studied with  superradiant instabilities and
 accretion which could manifest on  the  gravitational wave signal induced by  a small compact object in the field of the central BH--see also \cite{Clough,Bamber}.
More recently
in \cite{Davoudiasl}
the  formation of SMBHs at high redshifts  was studied in connection with ultralight DM,
see also \cite{Padilla,BAMBER} for growth of accretion driven scalar DM  hair around a Kerr BHs.
The  DM  effect on the quasinormal modes  of massless scalar field and electromagnetic field
perturbations in a BH spacetime surrounded by PFDM   was considered in  \cite{Jusufi}.
An analysis of DM   in the  M87  core in relation to BH shadows effects  was presented in \cite{RKra}, see also \cite{JJS} and
shadows of Sgr A*  BH  surrounded by superfluid DM halo was studied in \cite{SgrAs},  and shadow from a  charged rotating BH in presence
of PFDM is explored in \cite{Atamurotov}.}

The DM candidates are   various, including  string  and brane theory effects, boson clouds, hypothetical new particles,
 primordial BHs and  alternative theories of gravity\footnote{Dark matter was also explained with  diffuse clouds of scalar bosons interacting with gravity  and with gravitational waves. However recent  results of \cite{abbot} sets  constrains to this hypothesis
showing that   there are no young scalar boson clouds in our galaxy.}.
 From the observational view-point, dark matter can be  detectable  from the products of its   decay or annihilation in   cosmic rays, gamma rays, neutrinos or even gravitons--see also \cite{Das-stella}, and gravitational--wave  and neutrino astronomy  can then  open different windows in the  DM analysis\footnote{Concerning possible DM constituents and presence in our Galaxy we mention that
 the DM Milky Way has been recently studied in \cite{2022ApJ...928...30L} and
  \cite{Rohan}. Whereas, DM and primordial BHs are studied in \cite{2022ApJ...926..205C,Basak} constraints on DM rule out BHs as  constituting  only a very small possible fraction of the dark matter.
  Finally hypothesis suggesting antimatter and DM be linked  have been recently studied   in \cite{nature}, posing  limits on the interaction of antiprotons with axion-like DM, see also \cite{Afach}.}. DM
 comprehension, particularly  focused on sub-galactic DM halos,   is also  a goal of the  Webb Telescope
\footnote{\textbf{https://jwst.nasa.gov.}}.

Nevertheless, despite the variety of DM models,    the standard cosmological model  is  in fact  the  $\Lambda$-CDM, which includes  a  cosmological constant  ($\Lambda$) (with negative pressure), encoding    Dark energy ({DE})  in empty space (or vacuum energy)  explaining  the  Universe  accelerating expansion.
(In this scenario  the effects of the  cosmological constant are also  treated as quintessence\footnote{See \cite{Pett} for a recent analysis constraining  the fraction of early dark energy, present   during the early ages of the Universe.}.) Polytropic  models of DM halos in $\Lambda$-CDM cosmology were individuated in \cite{2016PhRvD..94j3513S}.
 In this model,  the  DM  velocity is  less than the speed of light  (in this respect  neutrinos component are  excluded, being non-baryonic but not necessarily  cold) and  it  is dissipationless  as  it is not cooled by  radiating photons.  CDM  may be constituted by an hypothetical  weakly interacting massive particle ({WIMPS}), or  primordial BHs, or  axions.

Although  considered a DM standard model, CDM  is not exempt from various problems, emerging  for example   from the  observations of galaxies  and  galaxy clusters and clusterization    emerging  from the  rotation curves and morphological studies  (as the  cuspy halo problem). There is also a more general problem in describing the effects and presence of DM at large and small scales.
The CDM model   collides therefore   with  small-scale structure
observations. For all these reasons  the search  for alternative DM  models is still an open issue, and in this respect   the  SFDM model seems to adapt to
 both large-scale and small-scale structure observations while  the PFDM seems capable to   explain the asymptotically flat rotation velocity characterizing the
spiral galaxies.

In this analysis we study  geometrically thick  accretion disk  models, Polish Doughnuts (P-D),  {orbiting the central attractor}, whose  center  coincides  with the equatorial plane of the central axisymmetric attractor\citep{Koz-Jar-Abr:1978:ASTRA:,Abr-Jar-Sik:1978:ASTRA:,Jaroszynski(1980),PuMonBe12,PuMon13}. These  thick  accretion  tori are characterized by   very high (super-Eddington) accretion rates and  high optical depth.  Tori  morphology and stability are  essentially governed by  the pressure gradients on the equatorial  plane \cite{Abr-Jar-Sik:1978:ASTRA:}. The thin (Keplerian)  disks can be considered as  P-D limiting configurations regulated  by the background geodesic structure.
The DM background metric has a characteristic geodetic structure constituting a first major constraint on accretion physics.
The tori, described by purely hydrodynamic (barotropic) models, are governed by the equipressure surfaces that can be closed, giving stable equilibrium configurations, and open, giving unstable, jet-like  (proto-jets) structures caused by the relativistic instability due to the Paczynski mechanism where the effects of strong gravitational fields are dominant\footnote{The time scale of the dynamical processes (regulated by the gravitational and inertial forces) is much lower than the time scale of the thermal ones (heating and cooling processes, radiation) that is lower than the time scale of the viscous processes. The entropy is constant along the flow and, according to the von Zeipel condition, the surfaces of constant angular velocity $\Omega$ and of constant specific angular momentum $\ell$ coincide. This implies that the rotation law $\ell=\ell(\Omega)$ is independent of the equation of state \cite{Lei:2008ui,Abramowicz:2008bk}.}
 with respect to the  dissipative ones
 and predominant to determine  the unstable phases of the systems \cite{F-D-02,Igumenshchev,abrafra,pugtot,Pac-Wii,Fragile:2007dk,DeVilliers,Fon03}.

Many features of the tori dynamics and morphology like their thickness, their stretching in the equatorial plane, and the location of the tori are predominantly determined by the geometric properties of spacetime via a fluid  effective potential function.
Consequently, in models  where  DM is  geometrized as a metric deformation, DM has a clear impact  in the tori structure, modifying  the fluid effective potential.
The gradients of the effective potential on the tori equatorial and symmetry plane regulate the pressure gradient of the fluid in the Euler law governing dynamics of the perfect fluid \citep{Koz-Jar-Abr:1978:ASTRA:}.  The special case of cusped equipotential surfaces is related to the  accretion phase  onto the central attractor  \citep{Koz-Jar-Abr:1978:ASTRA:,Abr-Jar-Sik:1978:ASTRA:,
Jaroszynski(1980),Pac-Wii,Abr-Cal-Nob:1980:ASTRJ2:}. The outflow of matter through the cusp occurs due to an instability in the balance of the gravitational and inertial forces and the pressure gradients in the fluid, i.e., by the so called Paczynski mechanism of violation of mechanical equilibrium of the tori \citep{Jaroszynski(1980)}.

DM affects  the cusp formation and cusp location with respect to the central singularity, modifying  the  disk accretion   throat, constraining  the thickness of the accretionary flow  and the maximum amount of matter swallowed by the central BH. Consequently DM  will influence  the  energetic  characteristics of the  BH in accretion and  the disk characteristics, as accretion rates or cusp luminosity \cite{Japan,letter,Multy}.

\medskip

More in details  plan of the article is as follows:
Thick disks in axially symmetric spacetimes are  discussed in Sec.\il(\ref{Sec:intr-thickdissk}).
The Kerr metric is introduced in Sec.\il(\ref{Sec:kerr-metric}).
The Polish doughnut tori  models are detailed in Sec.\il(\ref{Sec:PD}).
The fluid effective potential is the subject of Sec.\il(\ref{Sec:fep}).
Extended geodesic structure, constrained the tori modes is explored in Sec.\il(\ref{Sec:extended-geode}).
Dark matter models are discussed in Sec.\il(\ref{Sec:DMM}).
In Sec.\il(\ref{Sec:PFDM})  the perfect fluid dark matter  is considered.
Cold   and scalar field dark matter models  are studied  Sec.\il(\ref{Sec:SFDMCDM}).
Discussion and conclusions follow in Sec.\il(\ref{Sec:discussion}).
\section{Thick disks in axially symmetric spacetimes}\label{Sec:intr-thickdissk}
We study  geometrically thick tori in axially symmetric DM--BH spacetimes,  considered as a  DM-induced deformation of  the Kerr geometry. Therefore it is useful here to review the properties of the Kerr metric and the construction of tori in this geometry.
More specifically, in Sec.\il(\ref{Sec:kerr-metric}) the Kerr metric is introduced, while  the Polish doughnut  tori models are discussed in Sec.\il(\ref{Sec:PD}).
\subsection{The Kerr metric}\label{Sec:kerr-metric}
The Kerr metric is an axially symmetric, asymptotically flat, vacuum exact solution of the Einstein equation describing  the spacetime  of  central spinning compact object.  According to the metric parameter values (dimensionless spin $a/M$), the Kerr metric  describes  naked singularities (NSs) for $a>M$ and black holes  (BHs) for $a\in[0,M]$. The Kerr BH geometry  has  the   limiting static solution of Schwarzschild  for  $a=0$ and the extreme Kerr BH spacetime  for $a=M$.

In the
 Boyer-Lindquist (BL)  coordinates
\( \{t,r,\theta ,\phi \}\), the  metric tensor  reads\footnote{We adopt the
geometrical  units $c=1=G$ and  the $(-,+,+,+)$ signature, Latin indices run in $\{0,1,2,3\}$.  The radius $r$ has unit of
mass $[M]$, and the angular momentum  units of $[M]^2$, the velocities  $[u^t]=[u^r]=1$
and $[u^{\phi}]=[u^{\theta}]=[M]^{-1}$ with $[u^{\phi}/u^{t}]=[M]^{-1}$ and
$[u_{\phi}/u_{t}]=[M]$. For the seek of convenience, we always consider the
dimensionless  energy and effective potential $[V_{eff}]=1$ and an angular momentum per
unit of mass $[L]/[M]=[M]$.}:
%
\bea \label{alai}&& ds^2=-\left(1-\frac{2Mr}{\Sigma}\right)dt^2+\frac{\Sigma}{\Delta}dr^2+\Sigma
d\theta^2+\left[(r^2+a^2)+\frac{2M r a^2}{\Sigma}\sin^2\theta\right]\sin^2\theta
d\phi^2-\frac{4rMa}{\Sigma} \sin^2\theta  dt d\phi,
\eea
where
\bea
\Delta\equiv a^2+r^2-2 rM;\quad \Sigma\equiv a^2(1-\sigma)+r^2,\quad \sigma\equiv\sin^2\theta,
\eea
with  $G=c=1$.
  The horizons $r_-<r_+$ are respectively given by
\bea&&
r_{\pm}\equiv M\pm\sqrt{M^2-a^2},
\eea
the horizons can be
found\footnote{Quantities  $a_\pm$ and $a_\epsilon^\pm$  turn to be  very useful  in the comparison of the dark matter (DM) solutions of Sec.\il(\ref{Sec:DMM})  with respect to Kerr solutions in absence of DM.} by solving  the equation $ a=a_{\pm}\equiv \sqrt{r(2M-r)}$ for $r\in[0,2M]$ . The outer  and inner stationary  limits $r_{\epsilon}^\pm$ (ergosurfaces)  (solutions of $g_{tt}=0$)  are respectively
\bea
&& r_{\epsilon}^{\pm}\equiv M\pm\sqrt{M^2- a^2 (1-\sigma)},
\eea
the ergosurfaces  can be
found  by solving  the equation $a= a_{\epsilon}^\pm\equiv {a_{\pm}}/{\sqrt{1-\sigma}}$ (for $\sigma\neq1$),
where $r_+<r_{\epsilon}^+$ on   $\theta\neq0$  and  $r_{\epsilon}^+=2M$  in the equatorial plane $\theta=\pi/2$ ($\sigma=1$). %
 Static  observers, with  four-velocity   $\dot{\theta}=\dot{r}=\dot{\phi}=0$(where
$\dot{q}$ indicates the derivative of any quantity $q$  with respect  the proper time (for time-like particles) or  a properly defined  affine parameter for the light-like orbits)
cannot exist inside the (outer)  ergoregion\footnote{The ergoregion is  the  range   $[r_{\epsilon}^-,r_{\epsilon}^+]$  (where $r_{\epsilon}^\pm$ are functions of the plane $\sigma\in[0,1]$). Here we often intend the outer   ergoregion (or  simply ergoregion) in the BH spacetimes as  the region  $]r_+,r_\epsilon^+]$.
Then on the equatorial planes, in the Kerr spacetime there is  $r_\epsilon^-=0$  and  the outer ergosurface is $r_{\epsilon}^+=2M$.  
}, but  trajectories   $\dot{r}\geq0$, including particles  crossing the stationary  limit and escaping outside
in the region $r\geq r_{\epsilon}^+$ are possible.

The constants of the geodesic motions are
\bea&&\label{Eq:EmLdef}
\Em=-(g_{t\phi} \dot{\phi}+g_{tt} \dot{t}),\quad \La=g_{\phi\phi} \dot{\phi}+g_{t\phi} \dot{t},\quad  g_{ab}u^a u^b=-\mu^2,
\eea
with\footnote{The other constant of geodesic  motion of the Kerr metric is the Carter constant $
\Qa=(\cos\theta)^2 \left[a^2 \left(\mu^2-\Em^2\right)+\left(\frac{\La}{\sin\theta}\right)^2\right]+(g_{\theta\theta} \dot{\theta})^2$. In this work, where  tori share symmetry plane with the equatorial plane of the central BH, this constant is irrelevant. }   $u^a\equiv\{ \dot{t},\dot{r},\dot{\theta},\dot{\phi}\}$.
In Eqs\il(\ref{Eq:EmLdef}) quantities  $\Em$ and $\La$  represent the total energy  and momentum of the test particle
 coming from radial infinity, as measured  by  a static observer at infinity.

The relativistic angular velocity and the specific  angular momentum are
 \bea&&\label{Eq:flo-adding}
\Omega \equiv\frac{u^\phi}{u^{t}}=-\frac{\Em g_{\phi t}+g_{tt} \La}{\Em g_{\phi \phi}+g_{\phi t} \La}= -\frac{g_{t\phi}+g_{tt} \ell}{g_{\phi\phi}+g_{t\phi} \ell},\quad
\ell\equiv\frac{\La}{\Em}=-\frac{u_\phi}{u_{t}}=-\frac{g_{\phi\phi}u^\phi  +g_{\phi t} u^t }{g_{tt} u^t +g_{\phi t} u^\phi } =-\frac{g_{t\phi}+g_{\phi\phi} \Omega }{g_{tt}+g_{t\phi} \Omega},
\eea
 respectively. The  sign of $\La(\ell)$ defines the co-rotation/counter-rotation
  of the particles (fluid).
The DM models are axis-symmetric and stationary and we define similarly   notion of co-rotating and counter-rotating motions.

\subsection{Geometrically thick tori: the Polish doughnut models}\label{Sec:PD}
We specialize our analysis to   the  Polish doughnut (P-D)  tori,  general relativistic hydrodynamic (GRHD) toroidal  configurations centered on the central  BH  equatorial plane, which is  coincident with  the tori equatorial symmetry  plane.

These toroidal models are well known and  used  in different contexts. They are analytic and general relativistic  models  defined and integrable in  axis symmetric spacetimes, where the  results known as  the  "von Zeipel theorem"  hold, ensuring  the integration condition on the equations for the fluid. For this reason we here apply these results to the  stationary DM metric  models\footnote{The toroids are
constant pressure  surfaces, whose construction in the axis--symmetric spacetimes  is   based  on the application  of the von Zeipel theorem, for which
 the surfaces of constant angular velocity $\Omega$ and of constant specific angular momentum $\ell$ coincide and  the toroids  rotation law $\ell=\ell(\Omega)$ is independent of  the  details of the equation of state.
More precisely,  the von Zeipel theorem   reduces to an   integrability condition on  the Euler equation, in the case of  barotropic fluids, where
$\ell=\ell(\Omega)$ and, consequently, in the geometrically thick disks  the functional form of the angular
momentum and entropy distribution, during the evolution of dynamical processes, depends on the initial conditions of the system and not on
the details of the dissipative processes\cite{abrafra}.}.

Tori are  composed by a  one particle-species
 perfect fluid,   where
\be\label{E:Tm}
T_{a b}=(\varrho +p) u_{a} u_{b}+p g_{a b},
\ee
is the fluid energy momentum tensor,  $\varrho$ and $p$ are  the total energy density and
pressure, respectively, as measured by an observer moving with the fluid.
The timelike flow vector field  $u^a$  denotes  the fluid
four-velocity.
The  fluid dynamics  is described by the continuity  equation and the Euler equation respectively:
\bea
&&
u^a\nabla_a\varrho+(p+\varrho)\nabla^a u_a=0,
\quad
(p+\varrho)u^a \nabla_a u^c+ \ h^{bc}\nabla_b p=0,
\eea
where the projection tensor $h_{ab}=g_{ab}+ u_a u_b$ and $\nabla_a g_{bc}=0$.

We assume  a barotropic equation of state (\textbf{EoS})
$p=p(\varrho)$ and the   stationary and axially symmetric matter distribution moves on  circular trajectories.
We investigate   the case of a fluid toroidal configuration  defined by the constraint
$u^r=0$,  as for the circular test particle motion no
motion it is assumed in the $\theta$ angular direction, which means $u^{\theta}=0$.
Because of these symmetries, the  continuity equation
is  identically satisfied and  the orbiting configurations are  regulated  by the Euler equation for the pressure $p$ only, which can be  written as
\bea\label{Eq:scond-d}
\frac{\partial_{a}p}{\varrho+p}=-{\partial_a}W+\frac{\Omega \partial_{a}\ell}{1-\Omega \ell},\quad\mbox{with}\quad W\equiv\ln V_{eff},\quad\mbox{and}\quad V_{eff}=u_t,
\eea
where $V_{eff}$ is the torus effective potential.
Tori are regulated by  the balance of the    hydrostatic  and   centrifugal  factors due to the fluid  rotation and by the curvature  effects  of the   background, encoded in the effective potential function $V_{eff}$.

 Assuming  the fluid  is   characterized by the  specific  angular momentum  $\ell$  constant (see also discussion  \cite{Lei:2008ui}),  we consider the equation for  $W:\;  \ln(V_{eff})=\rm{c}=\rm{constant}$ or $V_{eff}=K=$constant.
By setting  $\ell=$constant  as a torus parameter, the maximum density points in the disk,
the pressure gradients (from the Euler equation) are  determined  by  the gradients of the tori effective potential function\footnote{
The procedure adopted   here
borrows from the  Boyer theory on the equipressure surfaces applied to a  thick  torus \citep{Boyer69,abrafra}.
  The Boyer surfaces  tori are given by the surfaces of constant pressure.}.
  The maximum points of the tori effective potential as function of the radial coordinate provide  the minimum points of pressure, where fluid particles are free on unstable circular geodetic orbits.
\subsubsection{The fluid effective potential}\label{Sec:fep}
The fluid  effective potential (\ref{Eq:scond-d}) is  explicitly  \cite{abrafra,ringed}
\bea\label{Eq:Veff}
V_{eff}^2=\left(\frac{\Em}{\mu}\right)^2={\frac{g_{t\phi}^2-g_{\phi\phi} g_{tt}}{g_{\phi\phi}+2 g_{t\phi} \ell +g_{tt} \ell ^2}}.
\eea
The extremes of  the pressure  are  regulated by  the angular momentum distributions $\ell^\pm:\partial_r V_{eff}=0$, on the equatorial plane $\theta=\pi/2$
for co-rotating $(-)$  and counter-rotating $(+)$  fluids respectively\footnote{Note, in the  test particles analysis and  accretion tori models, for slowly spinning NSs ($a\in ]M,1.29M[$),  there are  circular geodesic orbits with $(\Em<0,\La<0)$ and $(\Em>0,\La<0)$  on the equatorial plane of the ergoregion--see Figs\il(\ref{Fig:PlotconflKerr}). These solutions correspond to  the relativistic angular velocity
(the Keplerian velocity with respect to static observers at infinity  $\Omega=d\phi/dt$)    $\Omega>0$; therefore, in this sense, they are all co-rotating with respect to the static observers at infinity but they can be  counter-rotating according to $\La<0$ and $\ell<0$ or   counter-rotating according to $\La<0$ but co-rotating according to $\ell>0$,  there can also be orbits with $\ell=\Omega=0$ --see, for example, \cite{1980BAICz..31..129S,1981BAICz..32...68S,Pu:Kerr,Adamek,slany,retro-inversion,submitted}. This possibility has not been discussed in  the analysis of DM models.
}.

 Torus cusp $r_\times$ is the  minimum point of pressure and density in the torus  corresponding  to the maximum point of the  fluid effective potential. The torus center $r_{center}$ is the maximum point of pressure and density in the torus,  corresponding  to the minimum point of the fluid  effective potential.
At the cusp ($r\leq r_\times$) the fluid may be considered pressure-free.
Fluid effective potential defines  the function 
  $K(r)=V_{eff}(\ell(r))$. Cusped tori have parameter  in the open ranges\footnote{The notation is as follows:
the  (closed) interval  between quantities  $q_l$ and $q_r$, including $q_l$ and $q_r$,
 is  denoted as$[q_l,q_r]=\{ q\in
\mathbb{R}: q_l\leq q\leq q_r\}$,
the notation $]q_l,q_r[=\{ q\in
\mathbb{R}: q_l<q<q_r\}$  denotes the open interval. Similarly there is $]q_l,q_r]=\{ q\in
\mathbb{R}: q_l< q\leq q_r\}$ and $[q_l,q_r[=\{ q\in
\mathbb{R}: q_l\leq q< q_r\}$.}$K=K_\times\equiv K(r_{\times})\in]K_{center}, 1[\subset ]K_{mso}, 1[$, where $K_{center}\equiv K(r_{center})$. (We adopt the notation $q_{\bullet}\equiv q(r_{\bullet})$ for any quantity $q$ evaluated on a radius $r_{\bullet}$.)
\subsubsection{Extended geodesic structure and notable radii}\label{Sec:extended-geode}
\begin{figure}
\centering
     \includegraphics[width=8.5cm]{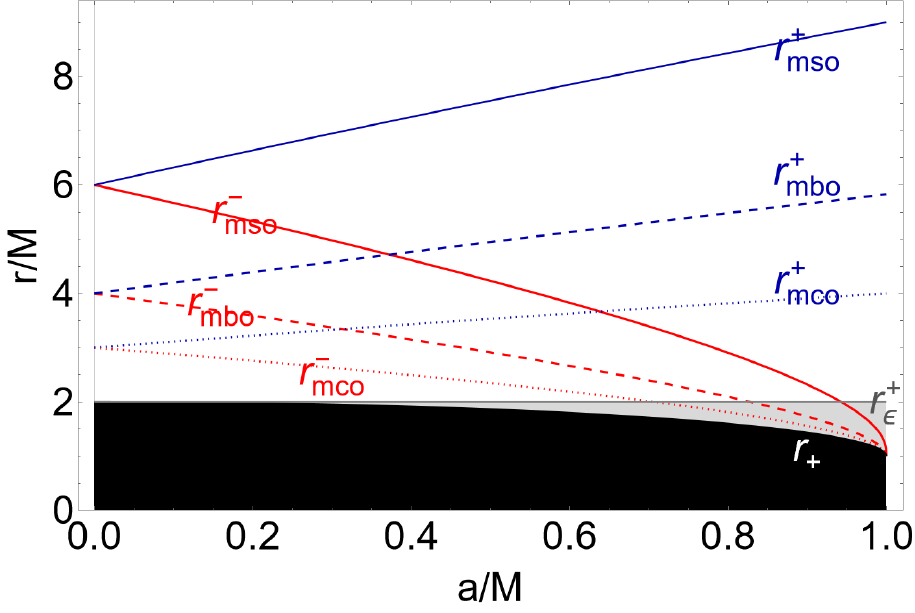}
       \includegraphics[width=8.5cm]{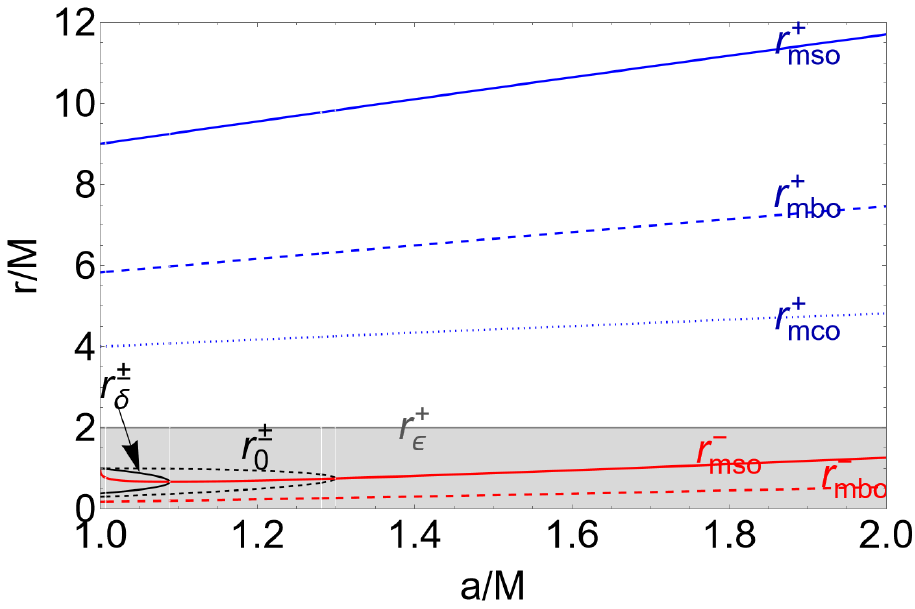}
           \caption{Geodesic equatorial   circular structure  of  the Kerr  geometry for spin  $a\in[0,M]$ (left panel) and $a>M$ (right panel), for co-rotating ($(-)$-red-curves) and counter-rotating ($(+)$- blue curves) orbits.  The marginally stable  orbits ($mso$) are the solid  curves, the  marginally bounded orbits ($mbo$) are dashed curves,  and marginally circular   orbits $r_{mco}^\pm$ (which are also photon circular orbits) are the dotted curves. (Note for $a>M$ there is no last  co-rotating circular  orbit (i.e. $r_{mco}^-=0$.) Radius $r_+$ is the outer horizon, $r_{\epsilon}^+$ is the outer ergosurface on the equatorial plane. Black region is $r<r_+$ and gray region is  $r\in[r_+,r_{\epsilon}^+]$ for Kerr  BHs and $r\in[0,r_{\epsilon}^+]$ for Kerr  NSs. On radii $r_{0}^\pm$ (black dashed curve-right panel)  there is $\La=0$ (where $\La$ is the test particle angular momentum) and on   $r_{\delta}^\pm$ (black  curve-right panel)  there is $\Em=0$ (where $\Em$ is the test particle energy).}\label{Fig:PlotconflKerr}
\end{figure}

The   geometry equatorial   circular geodesic structure   constrains the accretion disk physics governing, in the P-D model,  the tori cusps  and centers locations. In the Kerr geometry  the  geodesic structure is constituted by   the marginally  circular orbit for timelike particles  $r_{mso}^\pm$,  which is also a photon circular  orbit, $r_{mco}^\pm\equiv r_{\gamma}^{\pm}$, the marginally  bounded orbit, $r_{mbo}^{\pm}$, and the marginally stable circular orbit, $r_{mso}^{\pm}$ (see Figs\il(\ref{Fig:PlotconflKerr}))\footnote{In the Kerr  spacetime $r_{mco}^{\pm}$ is  the marginal circular orbit   (a photon circular orbit) where  timelike  circular orbits  can fill  the spacetime region $r>r_{mco}^{\pm}$. Stable circular  orbits are in $r>r_{\mso}^{\pm}$ for counter-rotating and co-rotating test particles respectively.  The marginal  bounded circular  orbit  is $r_{\mbo}^{\pm}$, where
 $\Em^{\pm}(r_{\mbo}^{\pm})=1$. More details and the exact forms of these radii can be found for example in  \cite{Pu:Kerr}.}. Radii $\{r_{mso}^\pm,r_{mbo}^\pm,r_{mco}^\pm\}$ constrain the  location of the  tori cusps (inner edges) with fluid specific angular momentum  $\ell=\ell^\pm$ respectively
\bea
&&\label{Eq:def-nota-ell}
\mbox{where}\quad r_{mco}^{\pm}<r_{\mathrm{mbo}}^{\pm}<r_{\mathrm{mso}}^{\pm}<
 r_{\mathrm{(mbo)}}^{\pm}<
  r_{(mco)}^{\pm},
  \eea
  where we introduced also the radii  $(r_{\mathrm{(mbo)}}^{\pm},
  r_{(mco)}^{\pm})$ defined as
\bea&&\nonumber
r_{\mathrm{(mbo)}}^{\pm}:\;\ell^{\pm}(r_{\mathrm{mbo}}^{\pm})=
 \ell^{\pm}(r_{\mathrm{(mbo)}}^{\pm})\equiv \mathbf{\ell_{\mathrm{mbo}}^{\pm}},\quad
  r_{(mco)}^{\pm}: \ell^{\pm}(r_{mco}^{\pm})=
  \ell^{\pm}(r_{(mco)}^{\pm})\equiv \mathbf{\ell_{mco}^{\pm}},
  \eea
   significant as governing  the location  of the tori centers, More precisely
ranges  $(\mathbf{L_1,L_2,L_3})$  of fluids specific angular momentum  $\ell$ govern the  tori    topology, according to the geodesic structure of Eqs\il(\ref{Eq:def-nota-ell}),  as follows:
\begin{description}
\item[\textbf{$ \mathbf{L_1} $}:] for $\ell\in \mathbf{L_1} $ there are  quiescent (i.e. not cusped)  and cusped tori--where there is $
\mp \mathbf{L_1}^{\pm}\equiv[\mp \ell_{mso}^{\pm},\mp\ell_{mbo}^{\pm}[$.
  The cusp is   $r^{\pm}_{\times}\in]r^{\pm}_{mbo},r^{\pm}_{mso}]$ (with $K_{\times}^{\pm}<1$)) and  the center with maximum pressure in $r^{\pm}_{center}\in]r^{\pm}_{mso},r^{\pm}_{(mbo)}]$.
\item[\textbf{$\mathbf{L_2}$:}] for $\ell\in \mathbf{L_2}$ there are  quiescent  tori and proto-jets (open-configurations) --where there is $\mp \mathbf{L_2}^{\pm}\equiv[\mp \ell_{mbo}^{\pm},\mp\ell_{mco}^{\pm}[ $.
The    cusp  $r_{\times}^{\pm}\in]r_{mco}^{\pm},r_{mbo}^{\pm}]$  is associated to the proto-jets,  with $K_{\times}>1$,  and the  center with maximum pressure is in $r_{center}^{\pm}\in]r_{(mbo)}^{\pm},r_{(mco)}^{\pm}]$. Proto-jets  are   associated to (not-collimated) open  structures,    with matter funnels along the BH rotational axis--see \cite{pugtot,open,proto-jets,ella-jet};
\item[\textbf{ $\mathbf{L_3}$:}] for $\ell\in \mathbf{L_3}$ there are only quiescent  tori where there is   $\mp \mathbf{L_3}^{\pm}\equiv\mp \ell \geq\mp\ell_{mco}^{\pm}$
and the torus center is at  $r^{\pm}_{center}>r_{(mco)}^{\pm}$.
\end{description}
 In the metric models we consider, the   DM affects  the orbiting  fluids   modifying   the  Kerr axially symmetric geometry and the  fluid effective potential--see also \cite{LH}.
Therefore  we study  the    radii  limiting the tori construction, defined through  the  fluid effective potential   for the geometries modified by the DM.
More precisely  we  identify  the    marginally circular orbit, $r_{mco}^\pm$, as the radius  $r_{mco}^\pm:K^\pm(r)=\infty$, the  marginally bounded orbit  defined by $r_{mbo}^\pm:K^\pm(r)=1$  (asymptotically flat spacetimes) and  the marginally stable orbits $r_{mso}^\pm: \partial_r\ell^\pm=0$.

The orbiting fluid  is governed by the geodesic structure of the considered spacetime. The tori are specified by the profile of the distribution of the specific angular momentum of the orbiting matter, (in the equator) and its relation to the radial profile of the specific angular momentum of equatorial circular geodesics. The so called Keplerian distribution of the circular geodesic  angular momentum related to a given spacetime is generally given by the relation
\bea&&
\ell=\ell_K\equiv \frac{\Phi_o\mp\sqrt{ \Phi\Phi_{(-)}^2}}{\Phi_\star},\quad\mbox{where}\quad
\Phi_{(\mp)}\equiv g_{t\phi}^2\mp g_{tt} g_{\phi \phi },\quad {\Phi}\equiv \left(g_{t\phi}'\right)^2-g_{tt}' g_{\phi \phi }',\\\nonumber
&& \Phi_o\equiv g_{t\phi}'\Phi_{(+)}-g_{t\phi}\left(g_{\phi \phi } g_{tt}\right)',\quad \Phi_\star\equiv g_{t\phi}^2 g_{tt}'+g_{tt}^2 g_{\phi \phi }'- g_{tt} (g_{t\phi}^2)',
\eea
for $\theta=\pi/2$,
where $(')$ is for the derivative with respect to $r$. The Keplerian profile intersection with the tori profile determines centers and cusps of the tori. In the standard Kerr spacetime it takes the well known form:
\bea
\ell_K(r,a)=\ell^{\mp}\equiv\frac{a^3\mp r^{3/2} \Delta-a (4M-3 r) r}{a^2-(r-2M)^2 r}.
\eea
\section{Dark matter models}\label{Sec:DMM}
We analyze  accretion tori  orbiting   spinning BHs  with spacetimes influenced by  different   DM
models. The metrics reduce, for some limiting  values of the DM parameters reduces to  the  Kerr BH geometry. Thus, using  Eqs.\il(\ref{Eq:Veff})   we  consider the  {BH DM} metric components $\{g_{tt},g_{t\phi},g_{\phi\phi}\}$ in BL coordinates,
and we  refer to the literature for details on the metric tensor and  the geometry properties.
We  investigate  the equatorial circular  geodesic structures for the  fluid effective potential, the effective potential function and the tori structure for  co-rotating and counter-rotating tori, in  three  DM models: in Sec.\il(\ref{Sec:PFDM}) we  address  the perfect fluid DM (PFDM) model of  \cite{Ech-Qo}.
Cold   and scalar field  DM models  of \cite{relative,Enta} are discussed Sec.\il(\ref{Sec:SFDMCDM}).
\subsection{Perfect fluid dark matter  (PFDM)}\label{Sec:PFDM}
 A rotating BH  solution in perfect fluid dark matter (PFDM)  has been discussed in \cite{Ech-Qo},
with
\bea\label{Eq:metricDM}&& g_{tt}=-\left[1-\frac{2M r-f_D(r)}{\Sigma}\right], \quad g_{t\phi}=-\frac{\sigma  a [2 Mr-f_D(r)]}{\Sigma},\quad g_{\phi\phi}=\sigma  \left[\frac{a^2 \sigma  [2 Mr-f_D(r)]}{\Sigma}+\left(a^2+r^2\right)\right],\\
&&g_{rr}\equiv \frac{\Sigma}{\Delta_{{D}}},\quad g_{\theta\theta}=\Sigma
\\&&\nonumber
\mbox{where}\quad \Delta_D\equiv\Delta+f_D(r)\quad\mbox{and}\quad  f_D(r)\equiv k r \log\frac{r}{| k| },
\eea
see  also \cite{pfdm,Kerr-desditte,Das},
where $k$ is the  parameter describing the {intensity} of the PFDM, set in the ranges $ k\in]-7.18M, 2M[$. For $k=0$
the line element
reduces to  the Kerr metric\footnote{Constrains of $k/M$  positive were  obtained by fitting the
rotation curves in spiral galaxies,  with values  $10^{-6}-10^{-7}$ \cite{Ech-Qo}.}.
The metric singularities, $r_\pm$, defining  the DM deformations  on the Kerr horizons, can be  found  by solving  the equation $a=a_{\pm}\equiv \sqrt{ r(2M-r)-f_D(r)},
$
or  from the equation
  $k= k_\pm\equiv  \Delta/[r W\left({\Delta}/{r^2}\right)]$,
where  $W$ is the  Lambert function,  such  that  $W(z)$
gives the principal solution for $w$ in $z=we^w$.

\begin{figure}
\centering
       \includegraphics[width=8cm]{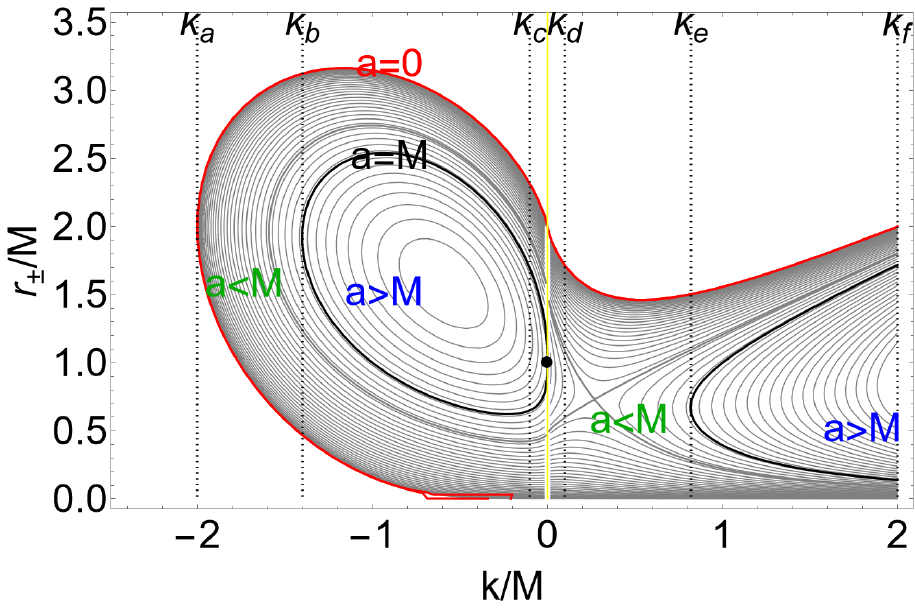}
         \includegraphics[width=8.5cm]{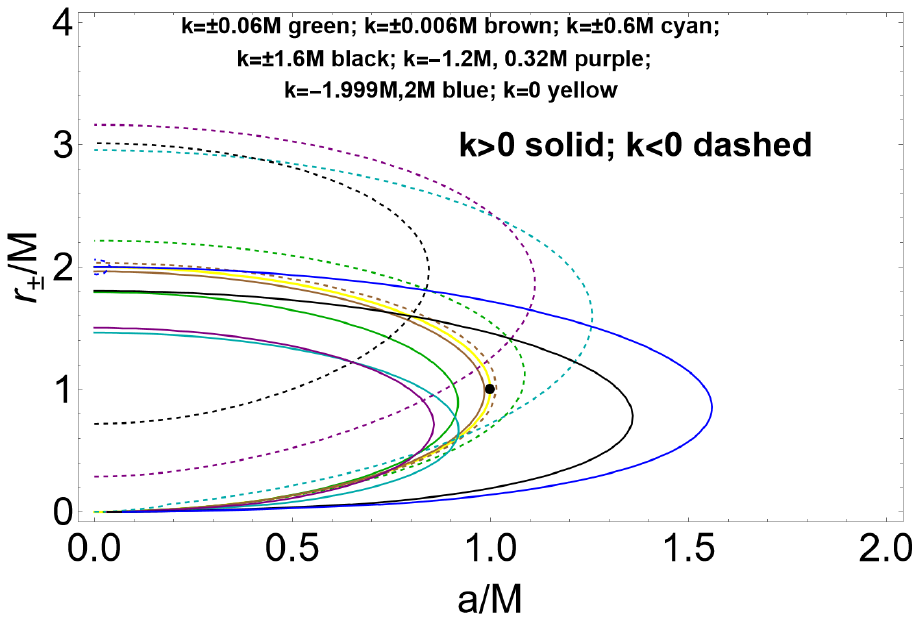}
          \includegraphics[width=8.69cm]{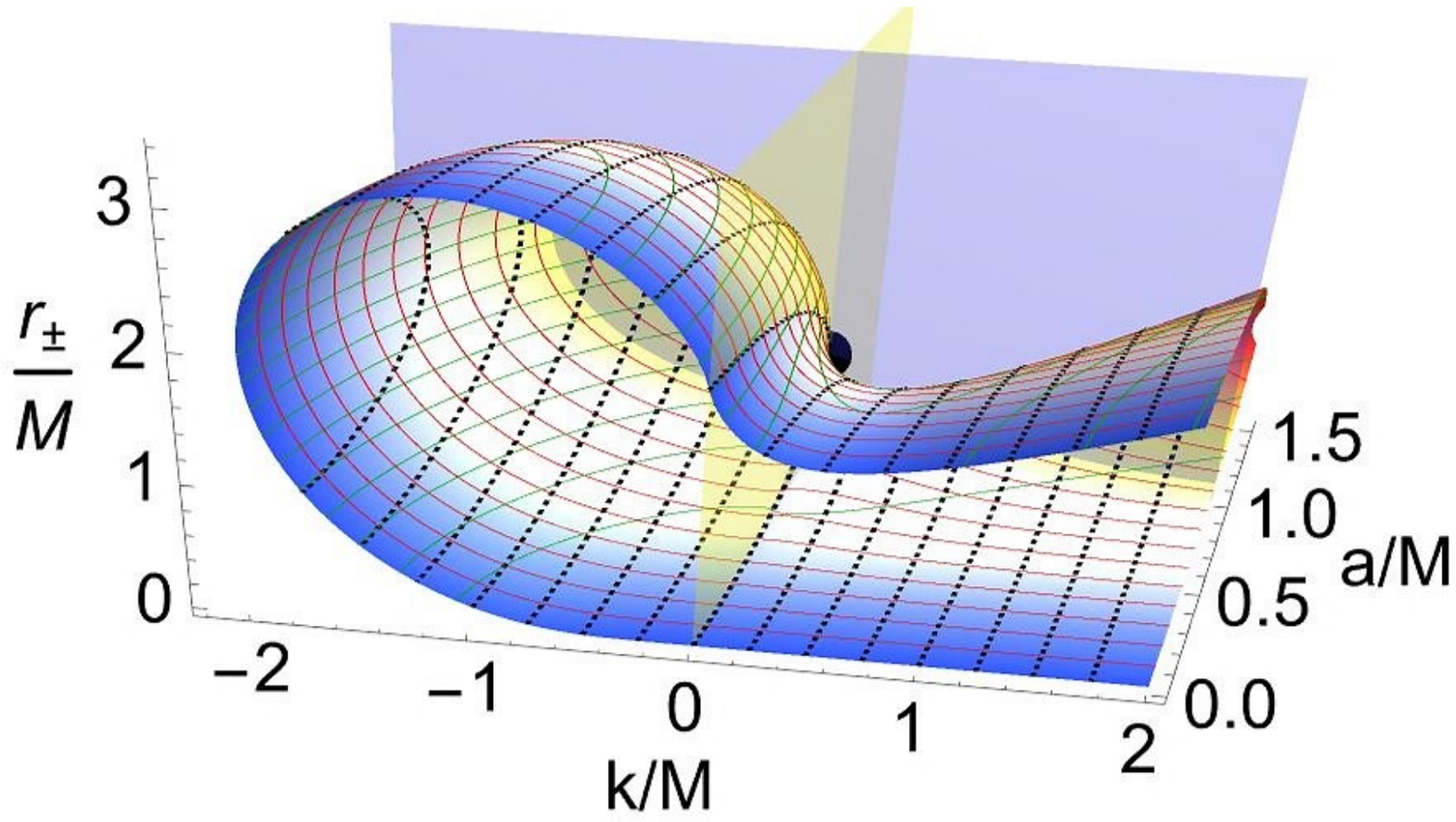}
           \includegraphics[width=8.69cm]{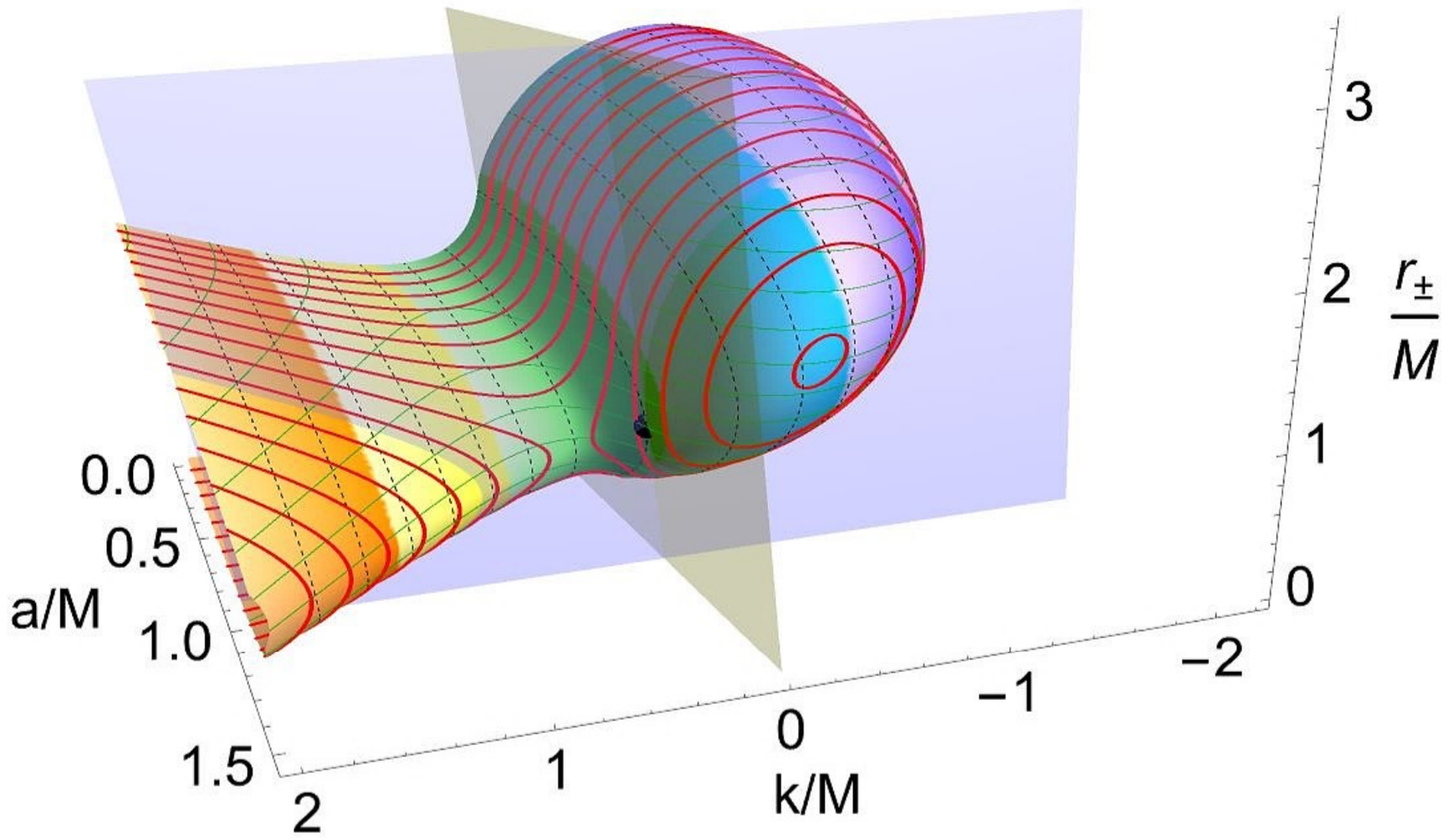}
           \caption{Horizons and ergosurfaces of the BHs in perfect fluid dark matter  (PFDM) of  Eq.\il(\ref{Eq:metricDM}).  Different values of the $ k\in ]-7.18M, 2M[ $  parameter,  describing the PFDM  intensity, are considered. (For $k=0$
the line element describes the  Schwarzschild and Kerr geometries in absence of DM.) Upper left  panel: horizons $r_\pm$ as functions of the $k$ parameter, for different values of the spin $a\geq 0$. Curves at $a \lesseqgtr M$ are shown. Red curves are the horizons  of the static case ($a=0$) (and the  ergosurfaces  $r_{\epsilon}^\pm$ Eqs\il(\ref{Eq:2bmetric2}) on the equatorial plane for $a>0$).  Values $\mathbf{k_6}$ of Eq.\il(\ref{Eq:stra-t-fut}) are signed as dotted vertical lines.  Yellow vertical line labels  the Schwarzschild and Kerr geometries in absence of DM. Upper right panel: the horizons $r_{\pm}/M$ as functions of the spin $a/M$ for different values of $k>0$ (solid curves) and $k<0$ (dashed curves). There is $k=\pm0.06M$ (green), $k=\pm 0.006M$ (brown), $k=\pm 0.6M$ (cyan),
$k=\pm1.6M$ (black), $k=-1.2M, 0.32M$ (purple),
$k=-1.999M,2M$ (blue). Curve $k=0$ (yellow) represents the Schwarzschild and Kerr geometries in absence of DM.  Bottom right and left  panels: different views of  horizons $r_\pm$ as functions of $k/M$ and $a/M$, mesh-functions are curves with constant
$r/M$ (solid green),  $k/M$ (dotted gray),  and $a/M$ (solid red).  Yellow vertical plane labels  the case $k=0$. Function contour colors are according to the $a/M$ values (left panel) and $k/M$ values (right panel).  We have  marked with a black spot the extreme Kerr BH  horizon $(a=M,r=M)$ on \emph{all} panels. To highlight better this point plane $a=M$ (blue) is included in the bottom panels (the crossing of $a=M$ and $k=0$ planes occurs at the point $a=M,k=0,r=M$).}\label{Fig:PlotDardie}
\end{figure}

The  deformed  ergosurfaces  $r_\epsilon^\pm$   can be found, for $\sigma\neq 1$, as solution of the equation $a=a_\epsilon^\pm \equiv  {a_{\pm}}/{\sqrt{1-\sigma }}$ while, on the equatorial plane $(\sigma=1)$ there is
\bea\label{Eq:2bmetric2}
 r_{\epsilon}^{\pm}=
\begin{cases}
k W\left[\frac{e^{2M/k}}{\text{sgn}(k)}\right]<2,\quad \mbox{for}\quad  k\lessgtr0,\\
 k W\left[-1,-e^{2M/k}\right]>2, \quad \mbox{for}\quad k<0,
 \\
2, \quad \mbox{for}\quad k=0
\end{cases}
\eea
(see  red curve in Figs\il(\ref{Fig:PlotDardie})--bottom left panel)
where
  the  Lambert function
$W(s,z)$  gives the $
s^{\text{th}}$ solution  for  $w$ in $z=we^w$,
and $\text{sgn}(k)$ gives the sign of $k$, therefore
it is  $-1, 0, 1$   for  $k$  negative, zero, or positive.

The PFDM horizons are independent of $\sigma$ (like in the Kerr case).
 The PFDM ergosurfaces are independent of spin in the equatorial plane (like
in the Kerr cases).
The horizons $r_\pm$ (and the ergosurfaces $r_{\epsilon}^\pm$ on the equatorial plane of Eq.\il(\ref{Eq:2bmetric2}))  are shown in  Figs\il(\ref{Fig:PlotDardie}).

The red curve in  Figs\il(\ref{Fig:PlotDardie}), which represents the   BH horizon for $a=0$ (\emph{and} the ergosurfaces Eq.\il(\ref{Eq:2bmetric2})  on the equatorial plane for $a\neq0$), bounds  the collections of   horizons  at different $a=$constant  in the plane $(r/M,a/M)$.
The PFDM metric describes solutions with $0$,$1$, and $2$ horizons.
For BH solutions, horizons can be shifted outwardly or inwardly with respect to
the Kerr BH spacetime depending on the value of $k$.
There are also spacetime solutions with horizons for  $a>M$.

Accordingly we select the PFDM metric parameter in the  following  six cases
\bea
\label{Eq:stra-t-fut}\mathbf{k_6}\equiv
\{k_a= -2M,k_b\approx -1.399M,k_c= -0.1M,k_d= 0.1M,k_e\approx 0.82M,k_f\equiv 2M\}.
\eea
 \begin{figure}
\centering
     \includegraphics[width=8.5cm]{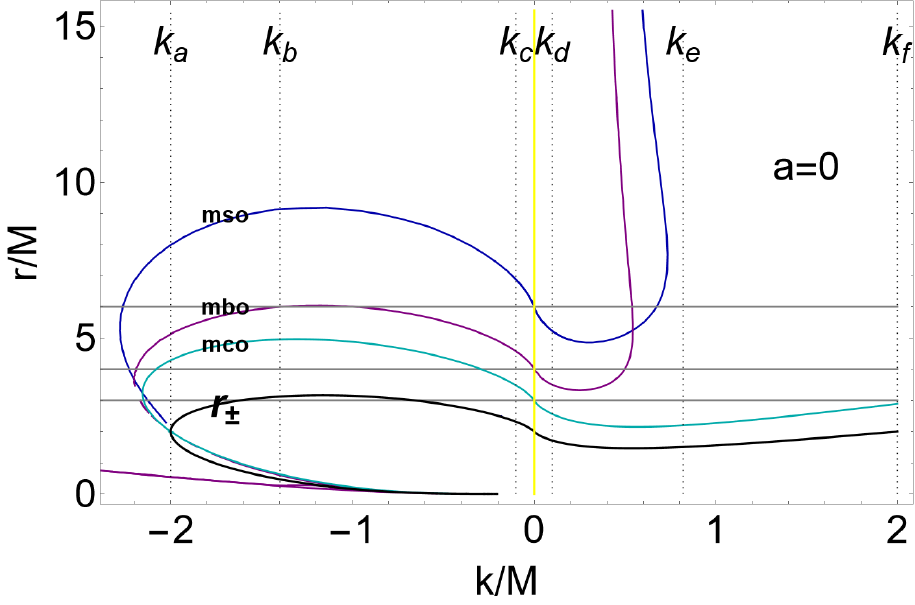}
       \includegraphics[width=8.5cm]{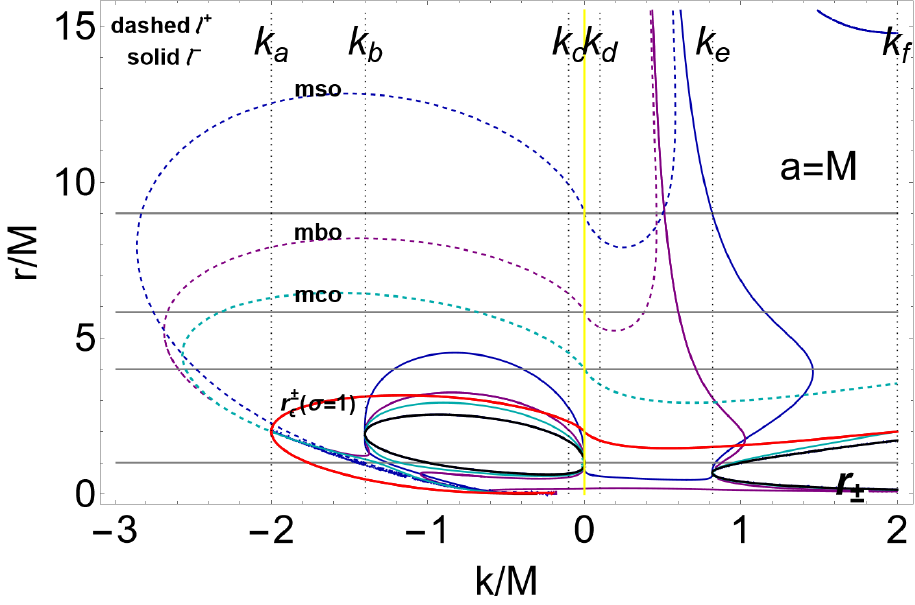}
           \caption{Equatorial circular geodesic structures  of the   Kerr attractor  in perfect fluid dark matter  (PFDM) of  Eq.\il(\ref{Eq:metricDM}).  Different values of the $ k\in ]-7.18M, 2M[ $,  describing the PFDM intensity are considered, where for  $k=0$  (vertical yellow line)
the line element describes the Schwarzchild  (for $a=0$) and extreme  Kerr BH geometry (for $a=M$).  Values $\mathbf{k_6}$ of Eq.\il(\ref{Eq:stra-t-fut}) are signed with  dotted vertical lines. Right panel shows the situation for $a=M$, left panel is for $a=0$. Radii $r_\pm$ are the  horizons, $r_{\epsilon}^\pm$  are   the ergosurfaces  on the equatorial plane (coincident with the horizons of the static case). $mso$ is for marginally stable orbit, $mbo$ is for the marginally bounded orbit, $mco$ is for marginally circular orbit, for co-rotating fluids ($\ell=\ell^-$, solid curves) and counter-rotating fluids ($\ell=\ell^+$, dashed curves).  Horizontal gray lines  show the radii of the geodesic structure for the Kerr  and Schwarzschild BH geometry  in absence of dark matter ($k=0$) for $a=0$ and $a=M$ respectively. See also Figs\il(\ref{Fig:PlotconflKerr}).}\label{Fig:PlotDardieM}
\end{figure}
\begin{figure}
\centering
                \includegraphics[width=8cm]{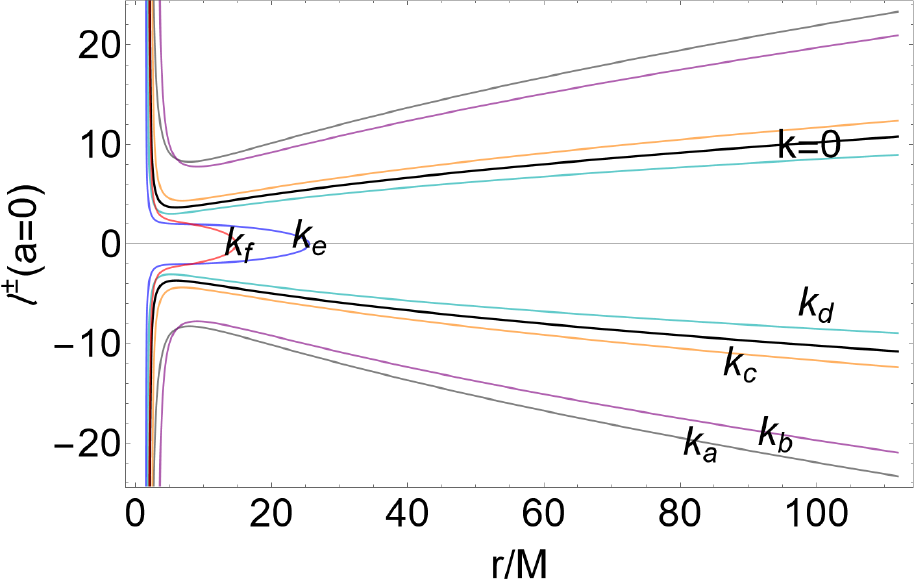}  \includegraphics[width=8cm]{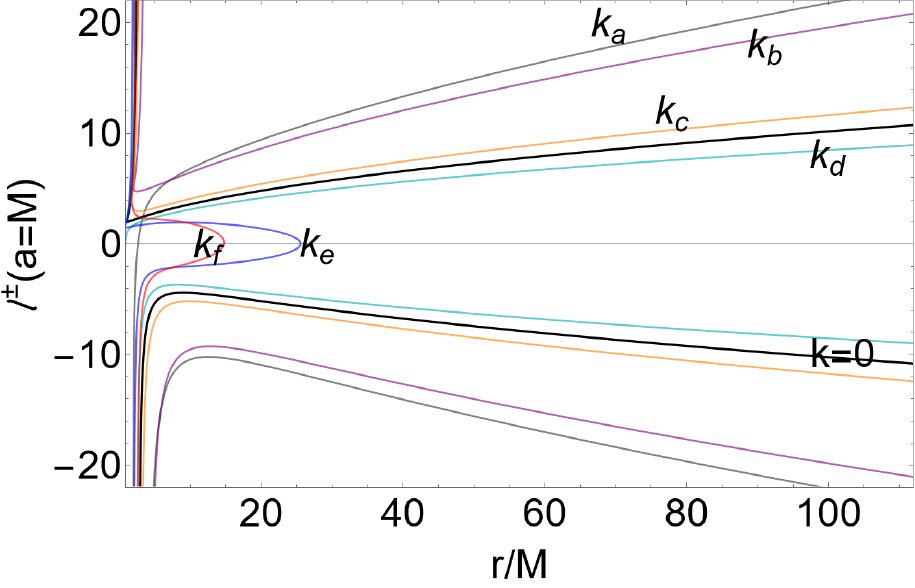}\\
                       \includegraphics[width=8cm]{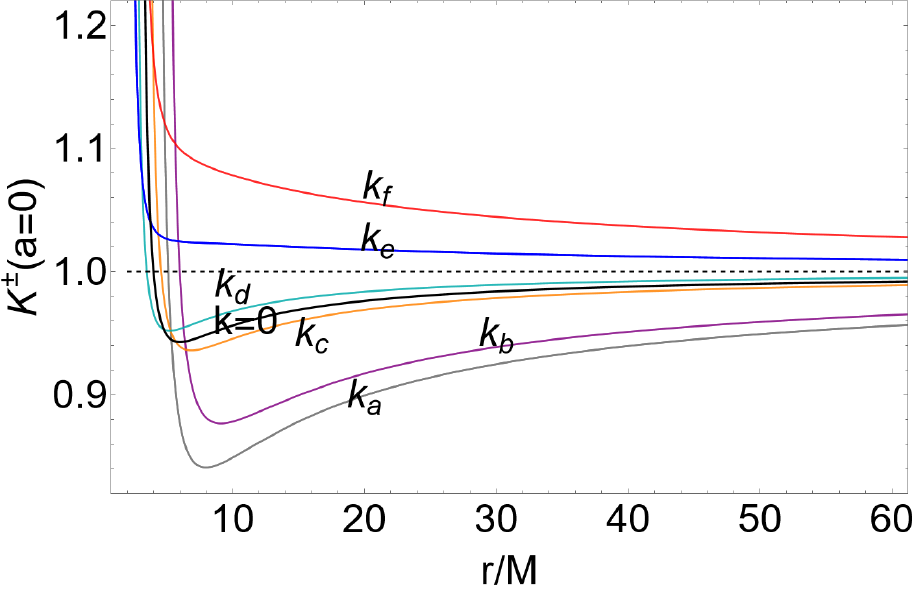}
                         \includegraphics[width=8cm]{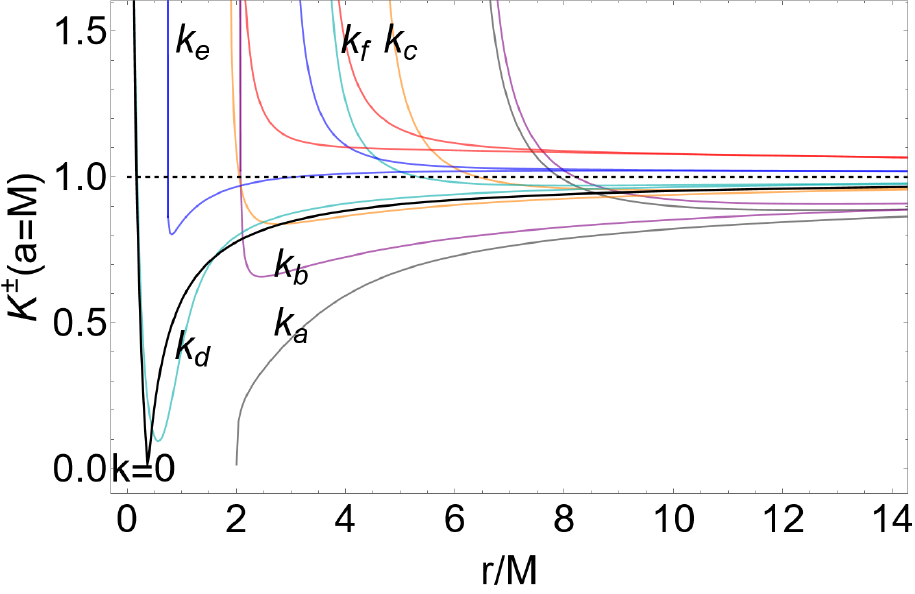}
                      \\
      \includegraphics[width=8cm]{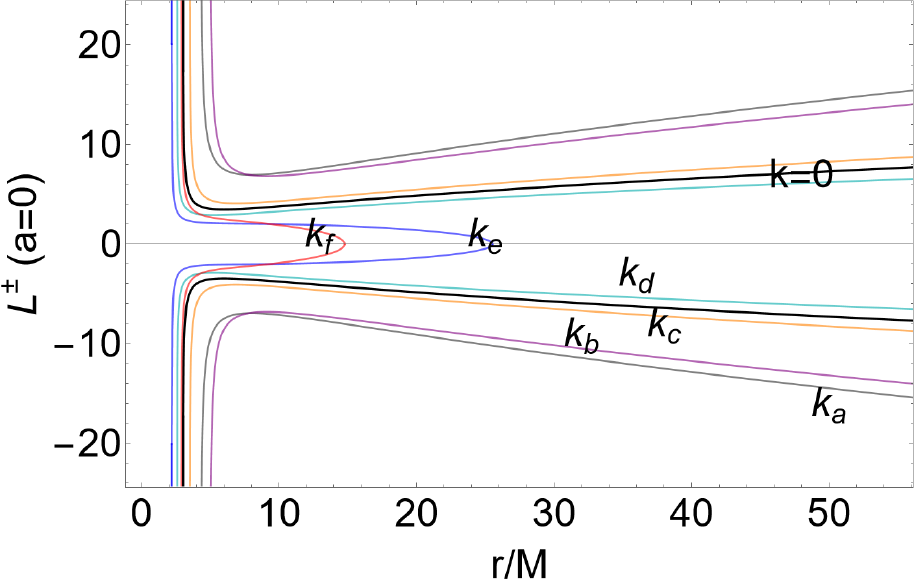}
        \includegraphics[width=8cm]{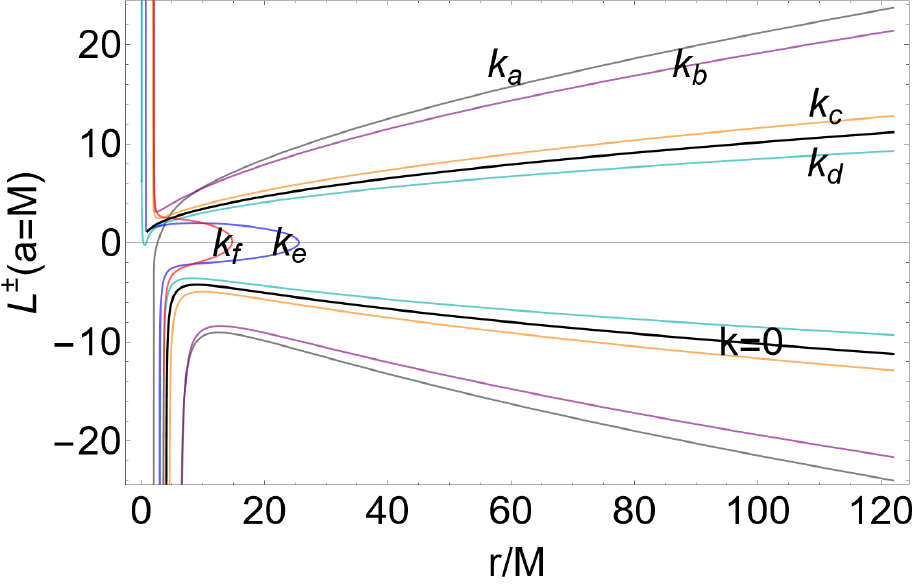}
            \caption{ Fluid specific angular momentum $\ell^\pm$, energy parameter $K^\pm$  and (test particles) Keplerian angular momentum $\mathcal{L}^\pm$ as function of $r/M$  for co-rotating and counter-rotating fluids, different   perfect fluid dark matter  (PFDM) parameters of  Eq.\il(\ref{Eq:metricDM})  signed on the panels.  Different values of the $ k\in ]-7.18M, 2M[ $  parameter,  describing the PFDM  intensity, are considered. Columns are $a=0$ (left) and $a=M$ (right), and
rows are $\ell^\pm$ (top), $K^\pm$ (middle), $\La^\pm$ (bottom).   (For $k=0$
the line element describes the  Schwarzschild  or the Kerr geometry.)   Values $\mathbf{k_6}$ are in Eq.\il(\ref{Eq:stra-t-fut}).}\label{Fig:PlotDardiesenti0}
\end{figure}
The cases
$(k_a,k_f)$  have  as horizon  $r=2M$  for  $a=0$ (the static case) therefore, in this sense, these solutions can be compared to the  Schwarzschild case.  Similarly, for $k=k_b$, the geometry with   $a=M$  has one horizon at  $r=2M$.  There is one horizon when $k=k_e$ and $a=M$\footnote{From Figs\il(\ref{Fig:PlotDardieM}) we note that  there is one horizon at $r\approx1.583M$ for  the special parameters
$(a\gtrapprox 1.25776M, k\approx-0.595M)$ and at $ r\approx 0.73M$ for  $(a\approx0.855M,k\approx 0.27M)$.}.

In Figs\il(\ref{Fig:PlotDardiesenti0}) we show also the  fluid specific angular momentum distribution for  the $\mathbf{k_6}$ parameters,  for the cases   $a=0$ and $a=M$, compared to the distribution on the geometry in absence of DM (see also Figs\il(\ref{Fig:PlotconflKerr})), the  associated $K$ parameter and the Keplerian (test particle) angular momentum $\mathcal{L}^\pm\equiv \ell^\pm K^\pm$ respectively, which we have defined as related to the thick tori counterparts from definition of $K(r)\equiv V_{eff}(\ell(r))$,   showing influence of the DM on the limiting thin Keplerian (geodesic) disk. The thin Keplerian disk is constrained by the geodesic structure of the gravitational background.

\medskip

In general the geodesic structure is qualitatively similar for any spin--Figs\il(\ref{Fig:PlotDardieM}).
 We focus on the two limiting cases  $a=0$ and $a=M$, whose
  equatorial circular geodesic structures for the orbiting configurations are represented   in Figs\il(\ref{Fig:PlotDardieM}).
In this model DM couples   with the BH rotation,  entangling with  the frame dragging,  evidenced in   the  deformed rotational law $\ell^\pm$ of the orbiting matter.
According to the DM parameter $k$ and spin  $a$,  relation (\ref{Eq:def-nota-ell}) i.e.
$r_{mso}^\pm>r_{mbo}^\pm>r_{mco}^\pm$  for  $\ell=\ell^\pm$ cases holds,  similarly to the non deformed BH case.
According to the discussion of  Sec.\il(\ref{Sec:extended-geode}) for the Kerr background, this relation, for  small magnitude of $k$, reflects  in the relative location of the  maximum-minimum points of pressure  of the orbiting disks. For  values of $ k $ where this relation is not verified, for example for  $k>0$ of Figs\il(\ref{Fig:PlotDardieM}), the orbiting toroidal  structures  show large  qualitative divergences with respect to the accretion  tori formation and dynamics in the Kerr BH spacetime.

\begin{description}
\item{\textbf{--The static attractor ($a=0$)}}
The geodesic structure is represented in  Figs\il(\ref{Fig:PlotDardieM})-left panel, compared with the Schwarzschild  case. We note that for $k>0$ the situation changes qualitatively with respect to the case in absence of DM.
In Figs\il(\ref{Fig:PlotDardiesenti0}) is the fluid specific angular momentum $\ell^\pm$, the tori energy parameter $K^\pm$  and (test particles) Keplerian angular momentum $\mathcal{L}^\pm$ as function of $r/M$, for different PFDM parameters of  Eq.\il(\ref{Eq:metricDM}), compared with the case  $k=0$
 describing  the  Schwarzschild  geometry. Notably, for $(k_f,k_e)$, there is  $K(r)>1$.
 From the   $(\ell^\pm, K^\pm)$  analysis it is noted how, for some values of $k$, curves are lower with respect to the corresponding curves in absence of DM--Figs\il(\ref{Fig:PlotDardieM}).
 From the analysis of the curves $L^\pm$, we can note how in some case the test particle angular momentum  is qualitatively different from the corresponding in the Schwarzschild case. Tori and effective potentials in this case are represented in  Figs\il(\ref{Fig:Plotselez}), constrained by the geodesic structure of  Figs\il(\ref{Fig:PlotDardieM}).
\\
\item{\textbf{--The spinning attractor ($a>0$)}}
The geodesic structure  for the spinning attractor geometry in PFDM is in  Figs\il(\ref{Fig:PlotDardieM})--right panel.
The fluid specific angular momentum $\ell^\pm$, energy parameter $K^\pm$  and (test particles) Keplerian angular momentum $\mathcal{L}^\pm$  are shown in Figs\il(\ref{Fig:PlotDardiesenti0}) as functions of $r/M$  for co-rotating and counter-rotating fluids and  PFDM parameters of  Eq.\il(\ref{Eq:metricDM}), compared to the case  $k=0$
describing  the  Kerr geometry in absence of DM.
Tori effective potentials for the counter-rotating (co-rotating) fluids are in Figs\il(\ref{Fig:PlotDardieMVm}). Tori are in Figs\il(\ref{Fig:PlotDardieMVmw}) and Figs\il(\ref{Fig:PlotDardieMVmwmariu}).

As clear from Figs\il(\ref{Fig:PlotDardieM})  the geodesic structure in the DM geometry with  $a=0$ is  similar to the counter-rotating  geodesic structure  in  the axially symmetric spacetime $(a\neq0)$. The co-rotating case (for $a=M$), $\ell=\ell^-$, especially for    $k>0$, is  remarkably different from the geodesic structure in the geometries with $a=0$  and it is further complicated by the presence of the ergoregion deformed by the PFDM with, however,  quantitative discrepancies with respect to the Kerr spacetime for large part of the DM parameter values   $k<0$.
 For larger $k>0 $ there are also NS solutions (also for $a\in]0,M]$) and, for even large values of $k>0$,  there are BH solutions (also for $a>M$).
 \begin{figure}
\centering
   \includegraphics[width=8cm]{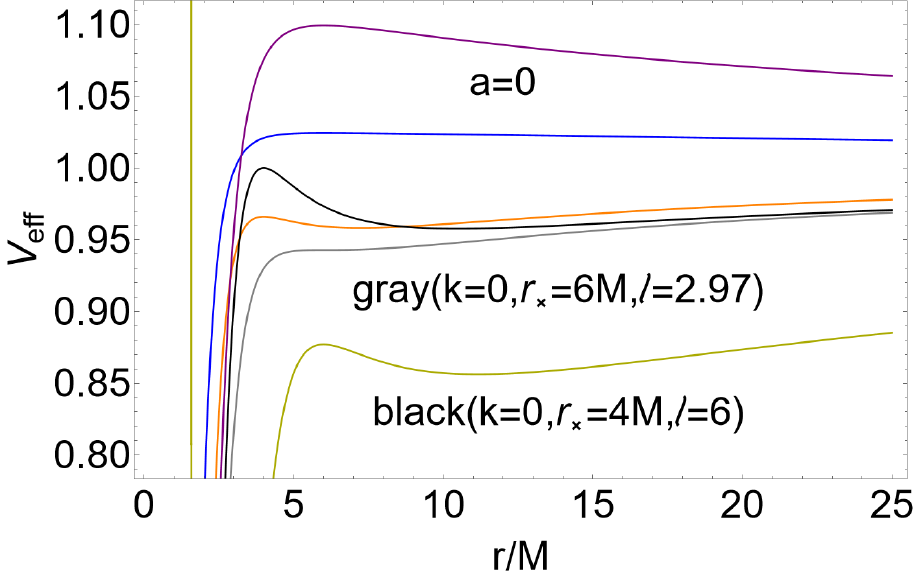}
\includegraphics[width=8cm]{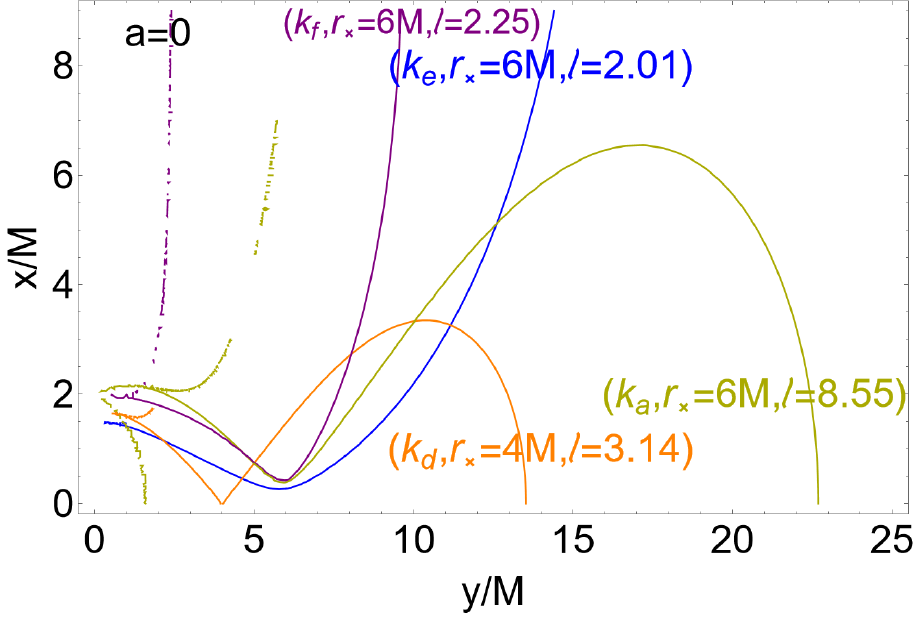}
           \caption{Case $a=0$. Tori orbiting BHs  in perfect fluid dark matter (PFDM) of  Eq.\il(\ref{Eq:metricDM}). Right panel shows the tori for selected values of the $k/M$ parameter, cusp location $r_\times$ and fluid specific angular momentum $\ell$, signed on the curves. Left panel shows the associated  tori effective potentials.    Values $\mathbf{k_6}$ of Eq.\il(\ref{Eq:stra-t-fut}) are considered. (For $k=0$
the line element describes the  Schwarzschild  geometry.)   There is $r=\sqrt{x^2+y^2}$ and $\sigma=y^2/(x^2+y^2)$, where $\sigma\equiv \sin^2\theta$.}\label{Fig:Plotselez}
\end{figure}
 \begin{figure}
\centering
       \includegraphics[width=8.5cm]{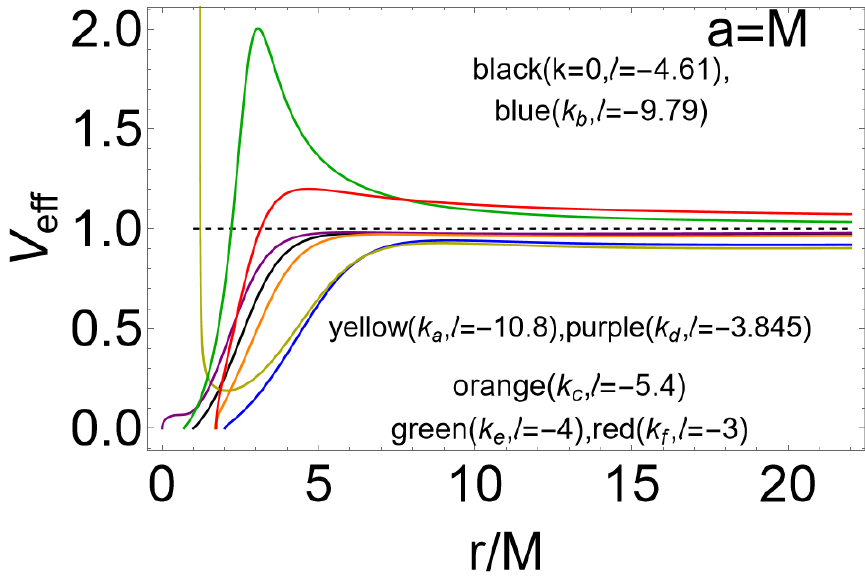}
         \includegraphics[width=8.5cm]{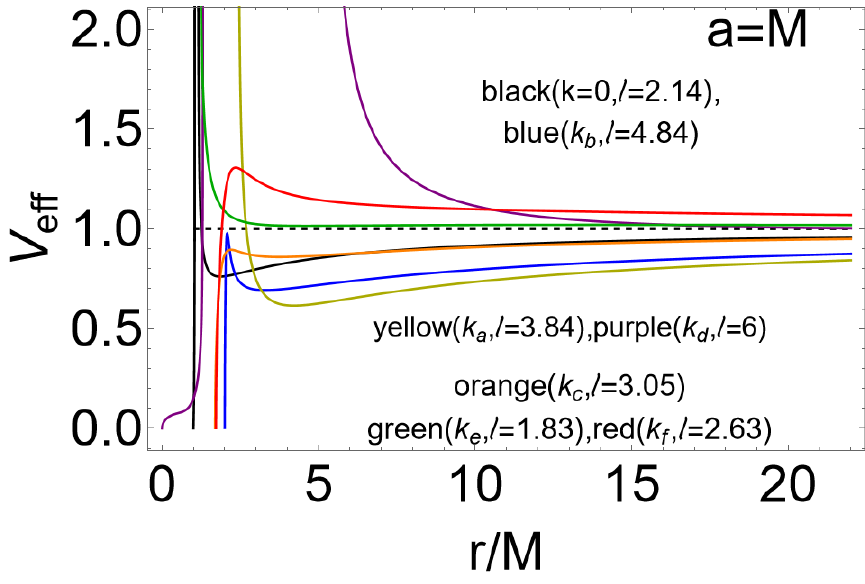}
           \caption{Case $a=M$. Effective potentials for the  tori orbiting BHs  in perfect fluid dark matter (PFDM) of  Eq.\il(\ref{Eq:metricDM}), for different values of the  parameter $k\in ]-7.18M, 2M[$,  describing the PFDM intensity and fluid angular momentum $\ell$, signed on the curves.   Values $\mathbf{k_6}$ of Eq.\il(\ref{Eq:stra-t-fut}) are considered. (For $k=0$
the line element describes the  extreme Kerr BH geometry).  Left (right) panel shows  the effective potentials for the counter-rotating (co-rotating) fluids.  There is $r=\sqrt{x^2+y^2}$ and $\sigma=y^2/(x^2+y^2)$, where $\sigma\equiv \sin^2\theta$. There are
black curves for $(k=0,\ell=-4.61,\ell=2.14)$, blue curves for $\left(k_b,\ell=-9.79,\ell=4.84\right)$, yellow curves for $\left(k_a,\ell=-10.8,\ell=3.84\right)$, purple curves for $\left(k_d,\ell=-3.845,\ell=6\right)$, orange curves for  $\left(k_c,\ell=-5.4,\ell=3.05\right)$, green curves for $\left(k_e,\ell=-4,\ell=1.83\right)$, red curves for $\left(k_f,\ell=-3,\ell=2.63\right)$.}\label{Fig:PlotDardieMVm}
\end{figure}
\begin{figure}
\centering
             \includegraphics[width=8cm]{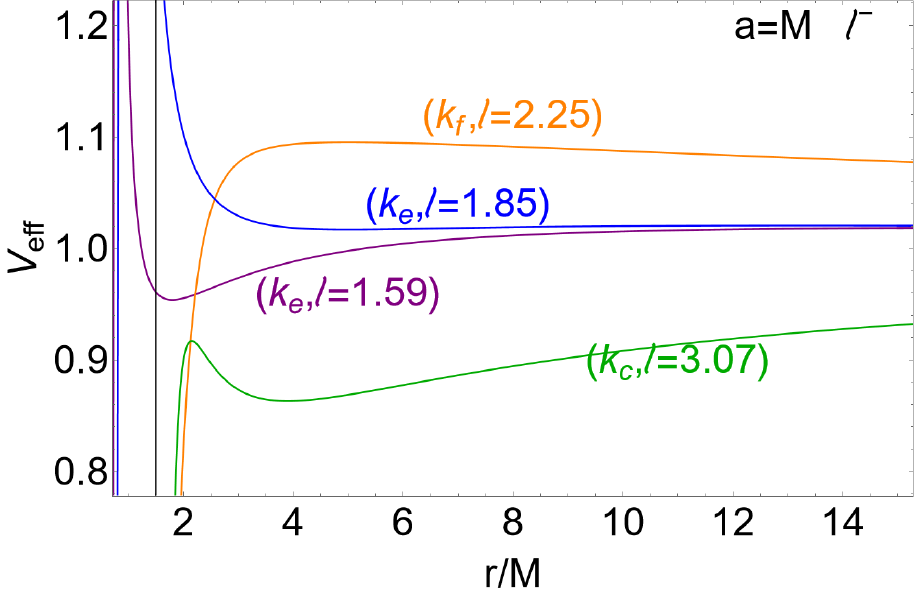}
           \includegraphics[width=8cm]{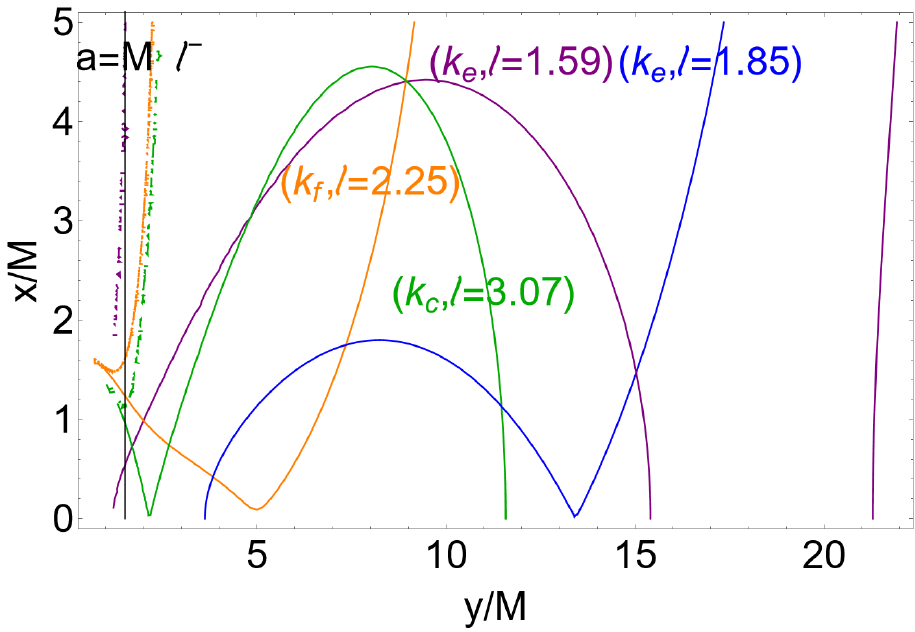}
           \caption{Case $a=M$.  Effective potentials and tori orbiting BHs  in perfect fluid dark matter (PFDM) of  Eq.\il(\ref{Eq:metricDM}), for different values of the  parameter $k\in ]-7.18M, 2M[$, describing the PFDM intensity  and fluid specific angular momentum $\ell$ signed on the curves.   Co-rotating $\ell^-=\ell>0$ cases are represented.  Values $\mathbf{k_6}$ of Eq.\il(\ref{Eq:stra-t-fut}) are considered. For $k=0$
the line element describes the  extreme  Kerr BH geometry.  There is $r=\sqrt{x^2+y^2}$ and $\sigma=y^2/(x^2+y^2)$, where $\sigma\equiv \sin^2\theta$. }\label{Fig:PlotDardieMVmw}
\end{figure}
\begin{figure}
\centering
                        \includegraphics[width=5.6cm]{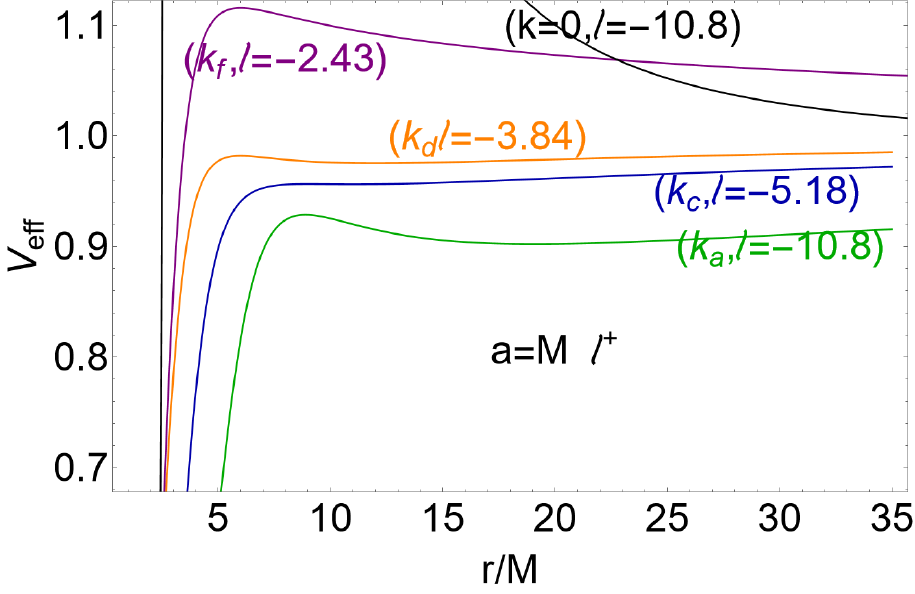}
      \includegraphics[width=5.6cm]{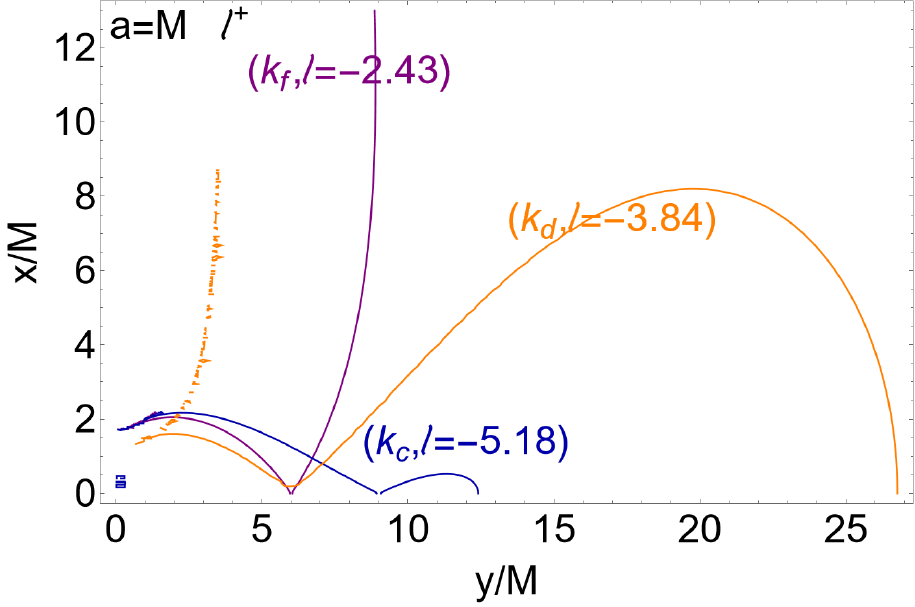}
      \includegraphics[width=5.6cm]{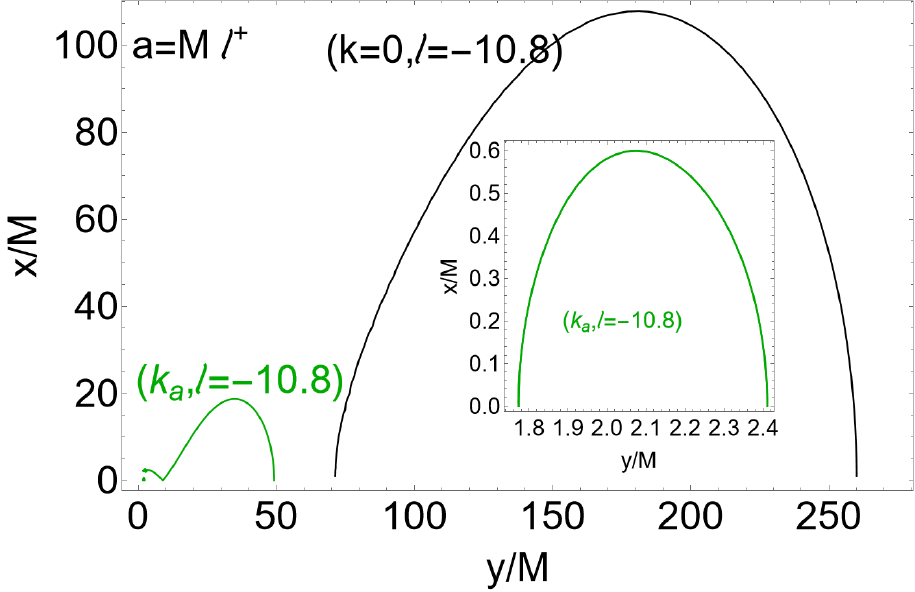}
           \caption{Case $a=M$.  Effective potentials and tori orbiting BHs  in perfect fluid dark matter (PFDM) of  Eq.\il(\ref{Eq:metricDM}), for different values of the  parameter $k\in ]-7.18M, 2M[$,   describing the PFDM intensity and fluid specific angular momentum $\ell$, signed on the curves.  Counter-rotating  $\ell^+=\ell<0$   cases are represented.    Values $\mathbf{k_6}$ of Eq.\il(\ref{Eq:stra-t-fut}) are considered. For $k=0$
the line element describes the  extreme  Kerr BH geometry.  Inside plot in the right panel is an enlarged view of the inner green torus. There is $r=\sqrt{x^2+y^2}$ and $\sigma=y^2/(x^2+y^2)$, where $\sigma\equiv \sin^2\theta$.}\label{Fig:PlotDardieMVmwmariu}
\end{figure}
 In general DM influence manifests also  with the existence of  extremely large cusped tori located considerably far from the central singularity as clear comparing  the geodesic structures in Figs\il(\ref{Fig:PlotconflKerr}) and Figs\il(\ref{Fig:PlotDardieM}).
In some cases, as clear from Figs\il(\ref{Fig:PlotDardieMVmw}), there are double configurations at equal $\ell$ (purple and blue curves of Figs\il(\ref{Fig:PlotDardieMVmw})  and green curve of Figs\il(\ref{Fig:PlotDardieMVmwmariu})),   with   considerable  larger  tori with respect to the case in  absence of dark matter.
Furthermore we note  the presence of outer cusps (blue curve in Figs\il(\ref{Fig:PlotDardieMVmw})) or possibly the emergence of double cusps   also in presence of BH solutions,  where the co-rotating marginally stable orbit shows some remarkable peculiarities at  $k\geq k_e$ for $a=M$ and $k\in[k_d,k_e]$ for $a=0$.

 The inter-disk cusp\footnote{Similarly to the   outer  cusps, the  inter--disks cusp is a tori cusp located between two configurations having  same   $(\ell,K)$ parameters, which eventually could be interpreted as an excretion cusp characterizing  some  cosmological models--see for example the double separated configurations (purple curves) or the blue curv in Figs\il(\ref{Fig:PlotDardieMVmw}).} (see also Figs\il(\ref{Fig:PlotDardieMVmw}))  can evolve, following  the  change in  one or  two of the tori parameters $(\ell,K)$ in an  inner cusp followed by an inner configuration (as the green curve in Figs\il(\ref{Fig:PlotDardieMVmwmariu})) or two separated configurations (as the purple curve in Figs\il(\ref{Fig:PlotDardieMVmw})).
For $k=k_e$, in the case $a=M$, there are no horizons consequently  the geometry, although is not over-spinning, can be considered a naked singularity, and  the  green curve in Figs\il(\ref{Fig:PlotDardieMVmwmariu})  can be seen as a typical double configuration characterizing certain NS geometries.
Also the presence of excretion cusps  could be a DM indicator.
Notably these  features are usually    read as tracers for the   possible NSs observations,  emerging as  consequences of the repulsive gravity effects characterizing NSs solutions,
and in this sense the  Kerr BH  immersed in  PFDM could  be a  "mimicker" of  super-spinars solutions.
\end{description}

\subsection{Cold   and scalar field dark matter}\label{Sec:SFDMCDM}
In this section we consider
 a spinning BH  in scalar  field dark matter (SFDM), addressed in Sec.\il(\ref{Sec:SFDM}) and in cold dark matter  (CDM), considered in Sec.\il(\ref{Sec:CDM})--see for example \cite{Enta,relative}.
\subsubsection{Scalar  field dark matter (SFDM)}\label{Sec:SFDM}
\begin{figure}
\centering
             \includegraphics[width=5.5cm]{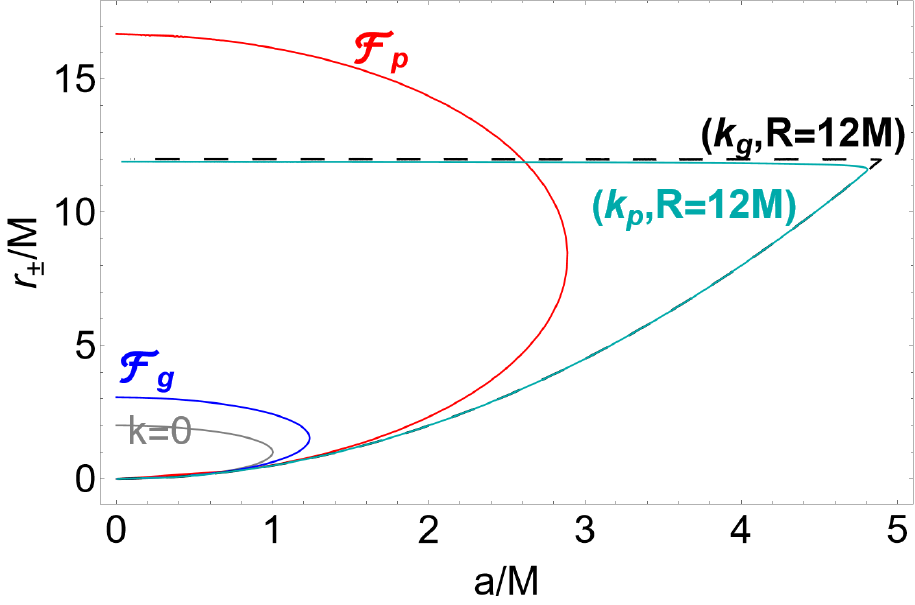}
       \includegraphics[width=5.5cm]{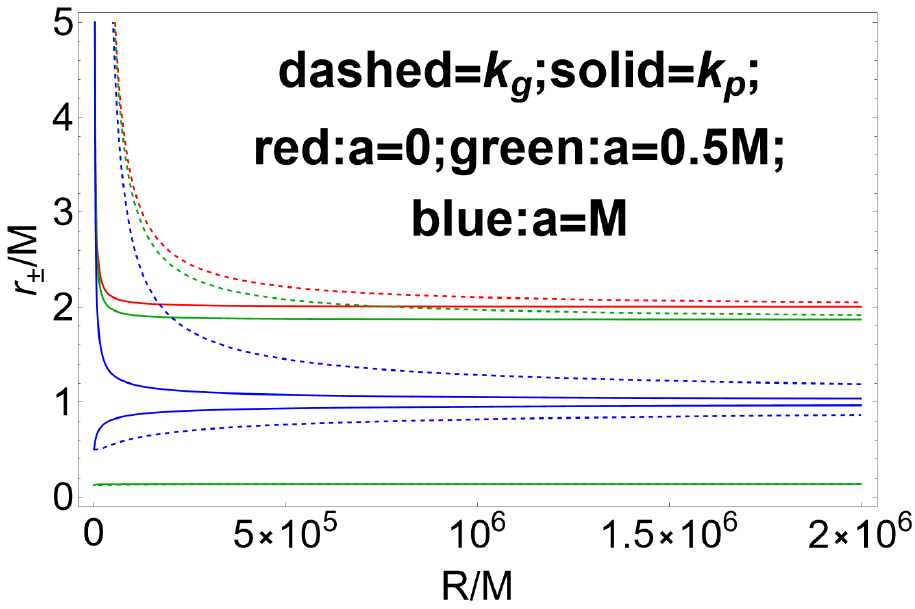}
              \includegraphics[width=5.5cm]{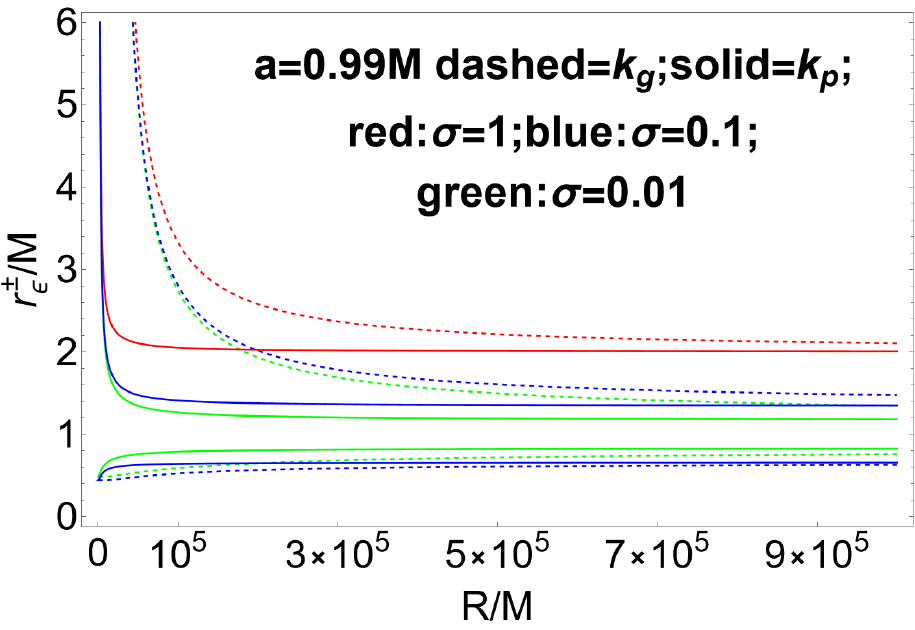}
                     \caption{Horizons and ergosurfaces of BHs geometries  in scalar field dark matter (SFDM) of  Eq.\il(\ref{Eq:SFDM-metric}).  Parameters $\mathcal{F}_p$ and $\mathcal{F}_g$ are defined in  Eqs\il(\ref{Eq:kpkg-definition}). The case $k=0$ corresponds to Kerr or Schwarzchild geometries. Left panel shows the BH horizons $r_\pm$ as functions of $a/M$ for different DM  parameters $(k,R)$. Gray curve is the Kerr BH  horizons. Center  panel: horizons radii $r_\pm$ as functions of the DM  parameter $R$, for different $k$ and spin $a$, red curve, for  $a=0$, is the static limiting case and $a=M$ is the blue curve, $a=0.5M$ is the green curve. Dashed curves correspond to $k_g$ and solid curves to $k_p$. Right  panel shows the ergosurfaces as functions of $R/M$ for different $k/M$, for spin $a=0.99M$ and different planes $\sigma\equiv\sin^2\theta$, $\sigma=1$ (red curve) is the equatorial plane, green curve is $\sigma =0.01$, blue curve is $\sigma =0.1$, dashed curves correspond to $k_g$ and solid curves to $k_p$  defined in  Eqs\il(\ref{Eq:kpkg-definition}).
}\label{Fig:Plotallows}
\end{figure}

There is
\bea\nonumber
&&  g_{tt}=-\left[1-\frac{{2  M r}+r^2(1-\xi_{SFDM})}{\Sigma}\right],\quad g_{t\phi}=-\frac{a \sigma  \left[{2  M r}+r^2(1-\xi_{SFDM})\right]}{\Sigma},\quad g_{\phi\phi}\equiv \frac{\sigma  \left[\left(a^2+r^2\right)^2-a^2 \sigma \Delta_{SFDM} \right]}{\Sigma}
\\
&&g_{rr}\equiv \frac{\Sigma}{\Delta_{SFDM}},\quad g_{\theta\theta}=\Sigma
\eea
where
\bea&&\label{Eq:SFDM-metric}
 \Delta_{SFDM}\equiv a^2-{2  M r}+r^2 \xi_{SFDM}
\quad\mbox{and}\quad  \xi_{SFDM}\equiv \exp \left[-\frac{8  \rho_c R^2 \sin \left(\frac{\pi  r}{R}\right)}{\frac{\pi  (\pi  r)}{R}}\right].
\eea
The metric components satisfy the asymptotically flatness condition, and  the fluid potential is well defined at infinity ($r\to +\infty$), where  $V_{eff}=1$.
     We introduce the quantity $k:\rho_c= {k}/{R^3}$. Here  $\rho_c$ is the central density and $R$ is the radius at which the pressure and density are zero\footnote{In the SFDM
 static solution, the Klein-Gordon equation and a quadratic
potential for the scalar field  have  been  considered \cite{Enta,relative}.} (where $\rho_c=0$ the metric reduces to the Kerr solution).
The Kerr limit occurs for  $k\to 0$ \emph{or} $R\to +\infty$.
The zeros of  $\Delta_{SFDM}$ distinguish the metric singularities  $r_\pm$  (SFDM deformed  horizons), while   the zeros of   $g_{tt}$  define a deformation of the Kerr ergosurfaces $r_\epsilon^\pm$ , which can be found by solving the equation $a=a_\pm$ and $a=a_\epsilon^\pm$  respectively\footnote{The accretion tori  considered here are geometrically thick and  characterized by  a pronounced verticality. The tori surfaces  can therefore approach the outer ergosurface  out of the equatorial plane (i.e. $\sigma<1$). The  ergosurface location (on planes $\sigma\neq1$) is  an  important factor  regulating  the Lense--Thirring effect on the disks and  on the  jets flows coming from the disks\cite{dragged}.}, where
\bea
a_\pm\equiv  \sqrt{r \left(2M-r  \xi_{sfdm}^{-1}\right)},\quad
a_{\epsilon}^\pm\equiv  \frac{a_\pm}{\sqrt{1-\sigma}}
\quad	\mbox{with}\quad \xi_{sfdm}\equiv e^{\frac{8 k \sin \left(\frac{\pi  r}{R}\right)}{\pi ^2 r}}.
\eea
%
In accordance with the study of the BHs horizons in Figs\il(\ref{Fig:Plotallows})  we  explore the following two cases:
\bea\label{Eq:kpkg-definition}
\mathcal{F}_p\equiv \{k_p\to 1000M,R_p\to 1200M\},\quad\mathcal{F}_g\equiv \{k_g\to 20000M,R_g\to 120000M\},
\eea
(see Figs\il(\ref{Fig:Plotallows})).
 \begin{figure}
\centering
    \includegraphics[width=5.5cm]{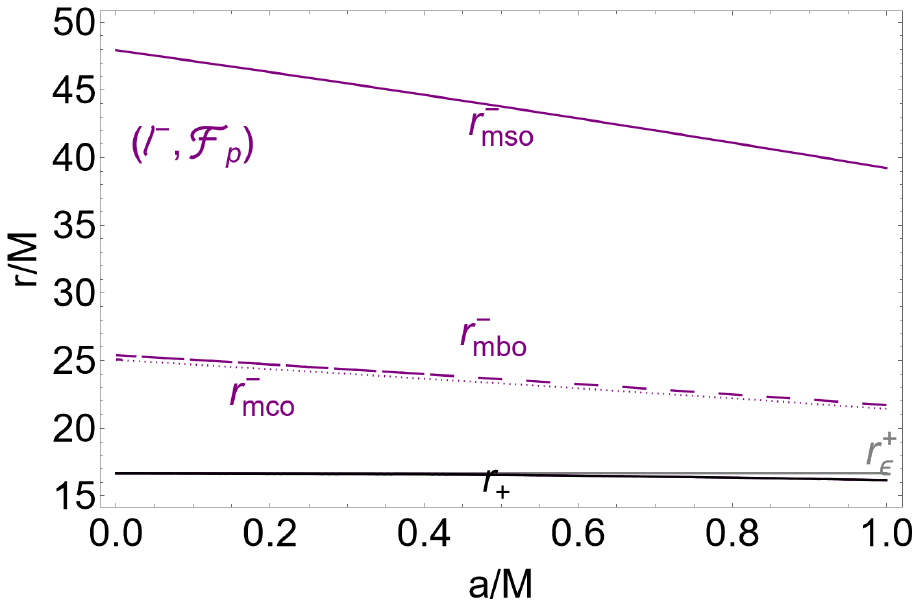}
                \includegraphics[width=5.5cm]{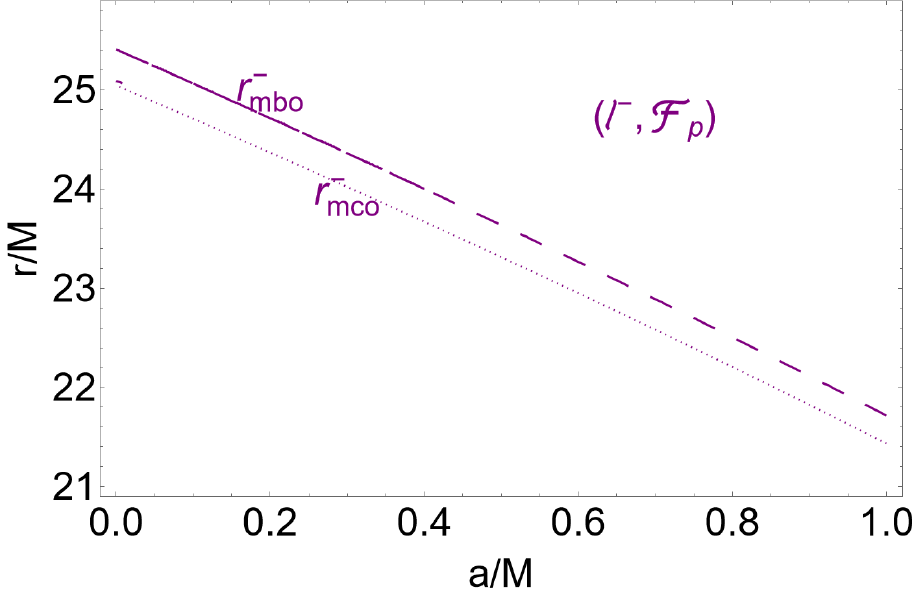}
                    \includegraphics[width=5.5cm]{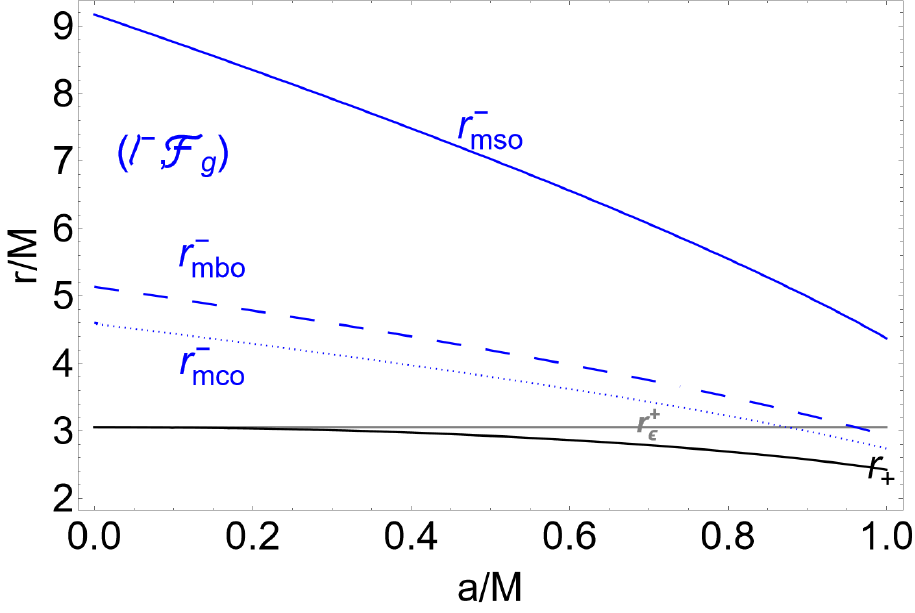}
                      \includegraphics[width=5.5cm]{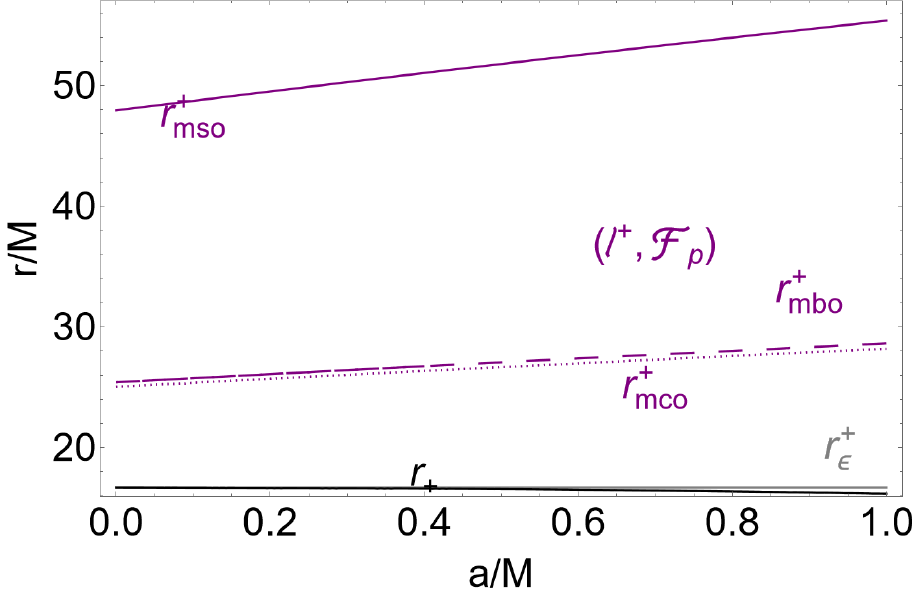}
       \includegraphics[width=5.5cm]{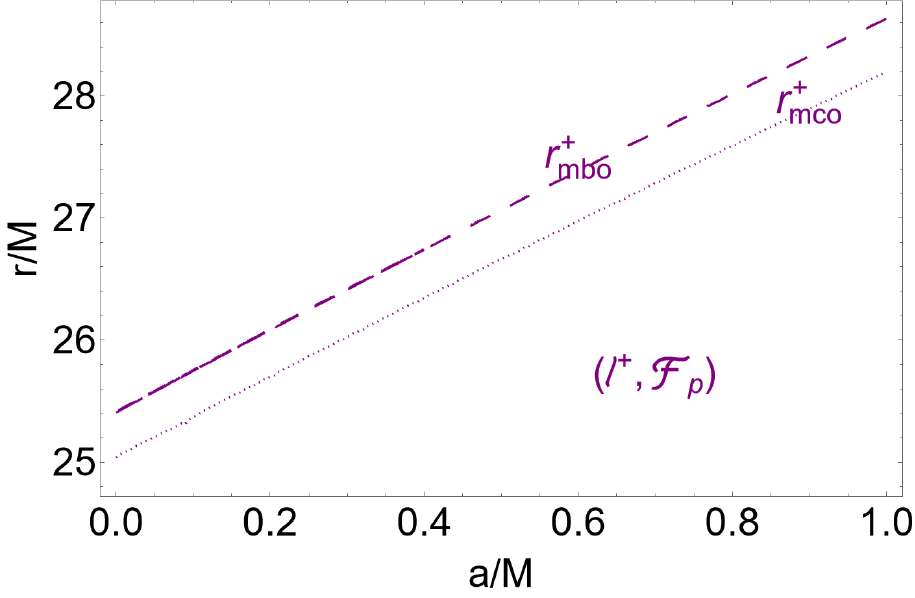}
         \includegraphics[width=5.5cm]{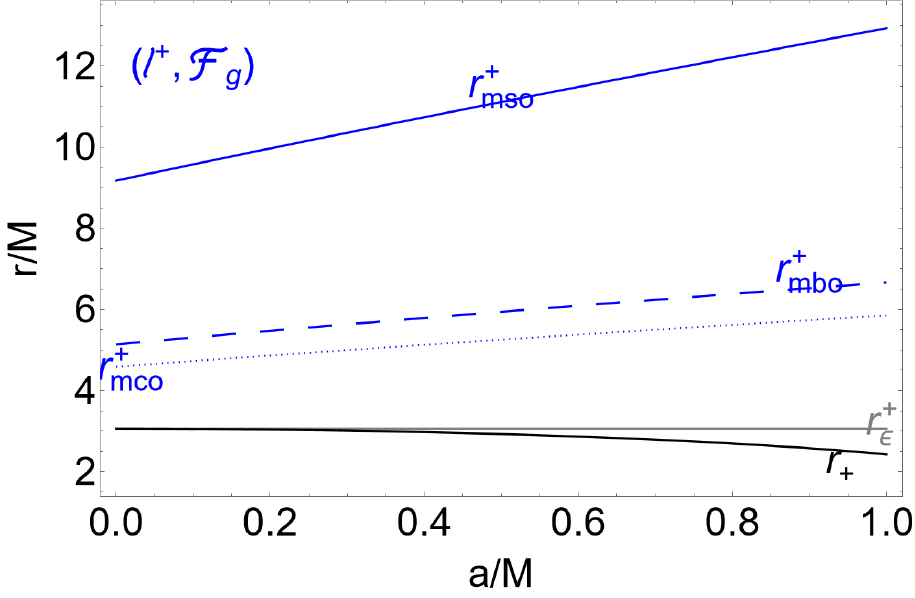}
           \caption{Geodesic equatorial  structure  of  scalar field dark matter (SFDM) geometry  of  Eq.\il(\ref{Eq:SFDM-metric}).
    Upper  (bottom) panels show  the situation for co-rotating (counter-rotating)  fluids  with $\ell^-$ ($\ell^+$)  for parameters $\mathcal{F}_p$ (purple curves) and $ \mathcal{F}_g$ (blue curves), defined in  Eqs\il(\ref{Eq:kpkg-definition}). Radius $r_+$ (black curve) is the outer horizon, $r_{\epsilon}^+$ (gray curve) is the outer ergosurface on the equatorial plane, $mso$ (solid purple and blue curves) is for marginally stable orbit, $mbo$  (dashed curves) is for the  marginally bounded orbits, $mco$  (dotted curves) is  for the marginally circular  orbits. The correspondent Kerr geodesic structure is  in Figs\il(\ref{Fig:PlotconflKerr}).
    Center panels are  close-up views  of the left panels.}\label{Fig:Plotallowsmall}
\end{figure}
\begin{figure}
\centering
  \includegraphics[width=8.5cm]{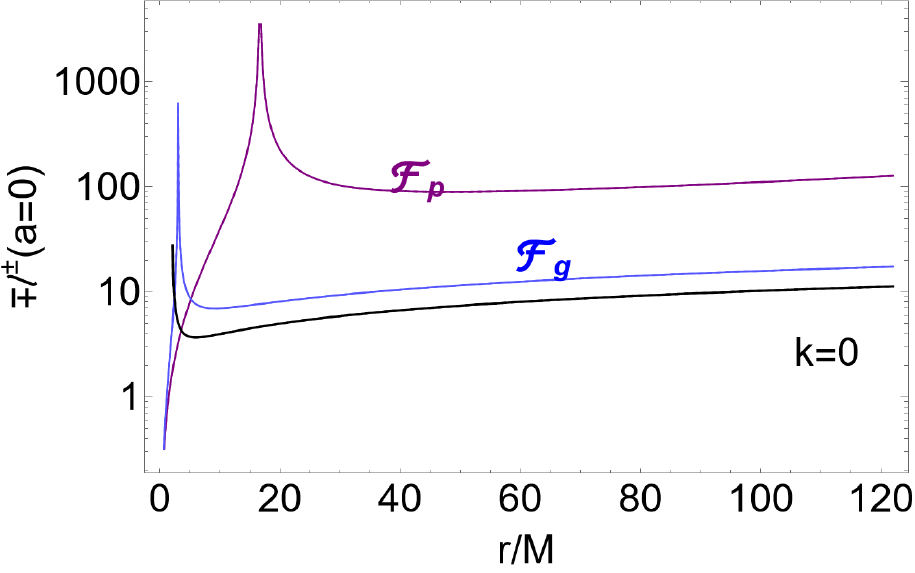} \includegraphics[width=8.5cm]{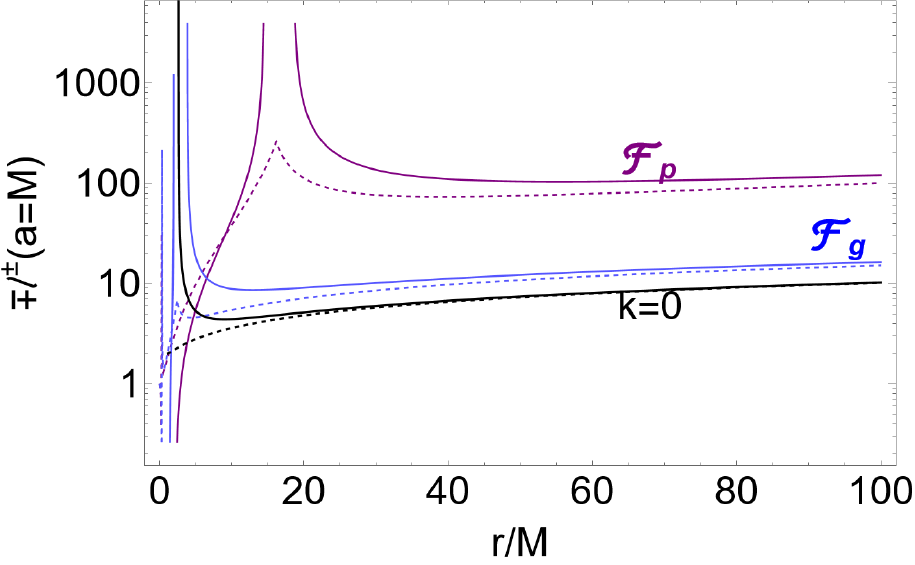}
\includegraphics[width=8.5cm]{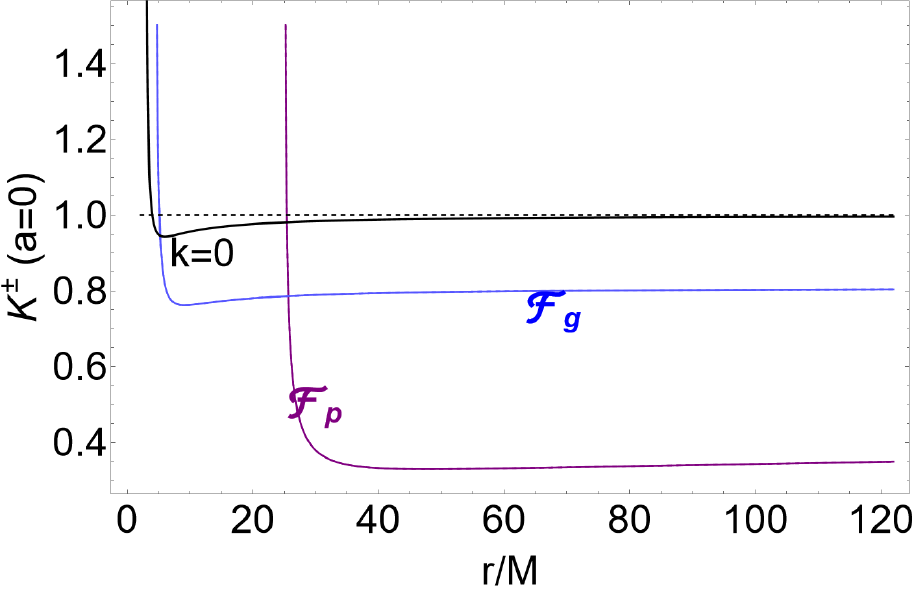}
\includegraphics[width=8.5cm]{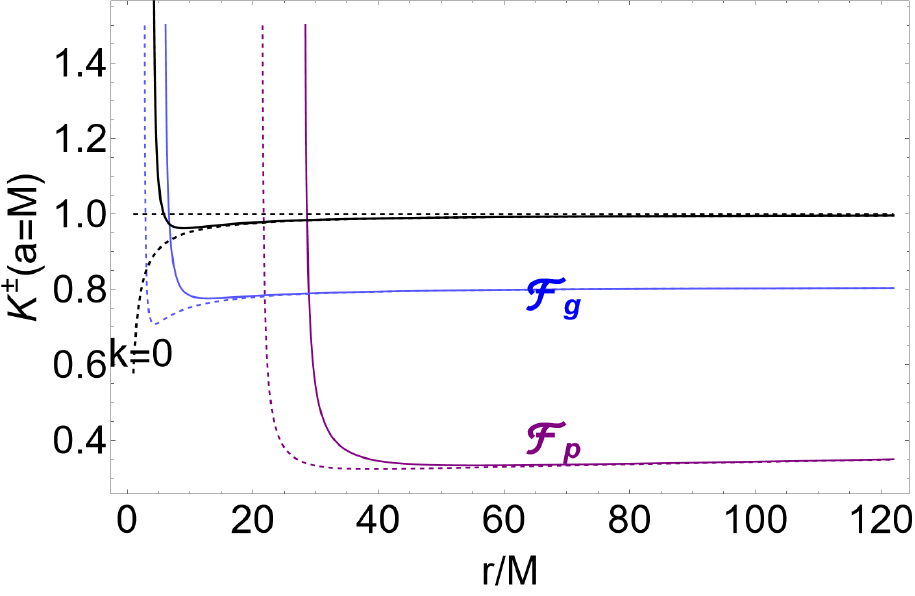}
  \includegraphics[width=8.5cm]{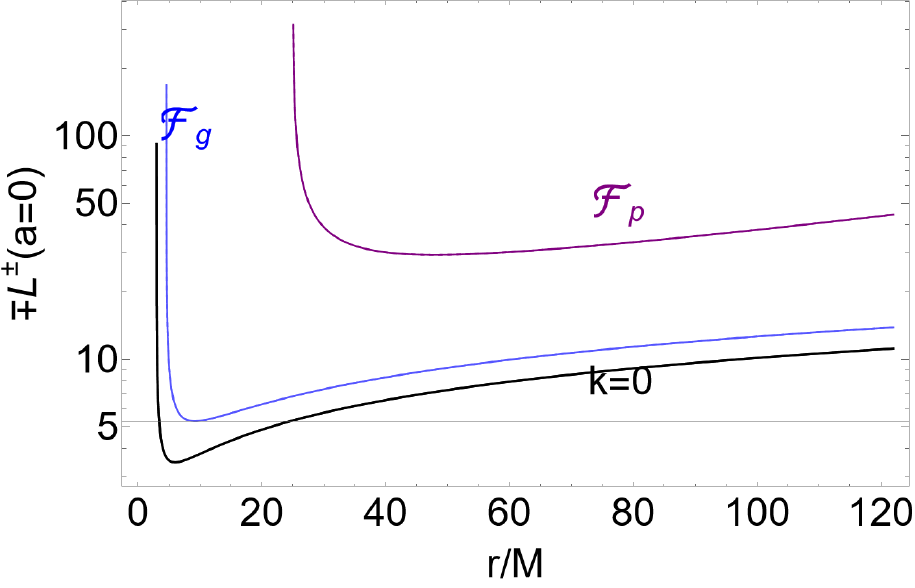}
  \includegraphics[width=8.5cm]{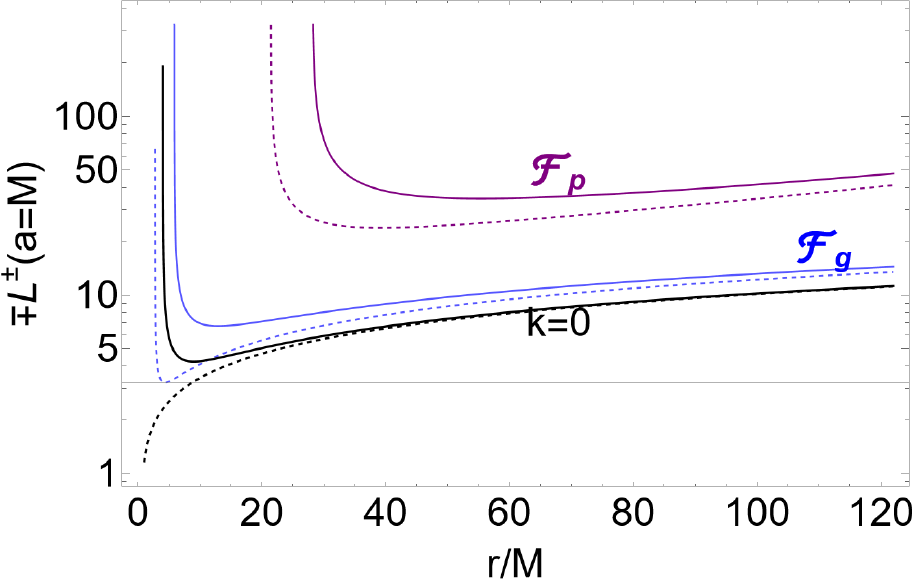}
                     \caption{Fluid specific angular momentum $\ell^\pm$, energy parameter $K^\pm$  and (test particles) Keplerian angular momentum $\mathcal{L}^\pm$ as function of $r/M$  for co-rotating $(-)$ and counter-rotating $(+)$ fluids and different  scalar field dark matter (SFDM) parameters $\mathcal{F}_p$ (purple) and $\mathcal{F}_g$ (blue), defined in  Eqs\il(\ref{Eq:kpkg-definition}). The value $k=0$ corresponds to the Schwarzschild ($a=0$) extreme  Kerr BH case ($a=M$). The columns are $a=0$ (left) and $a=M$ (right),
and rows are  for $\ell$ as function of $r$ (top), $K$ as function of $r$ (middle), $\La$  as function of $r$ (bottom).
}\label{Fig:PlotDardiesentisfdm0}
\end{figure}
 \begin{figure}
\centering
       \includegraphics[width=5.7cm]{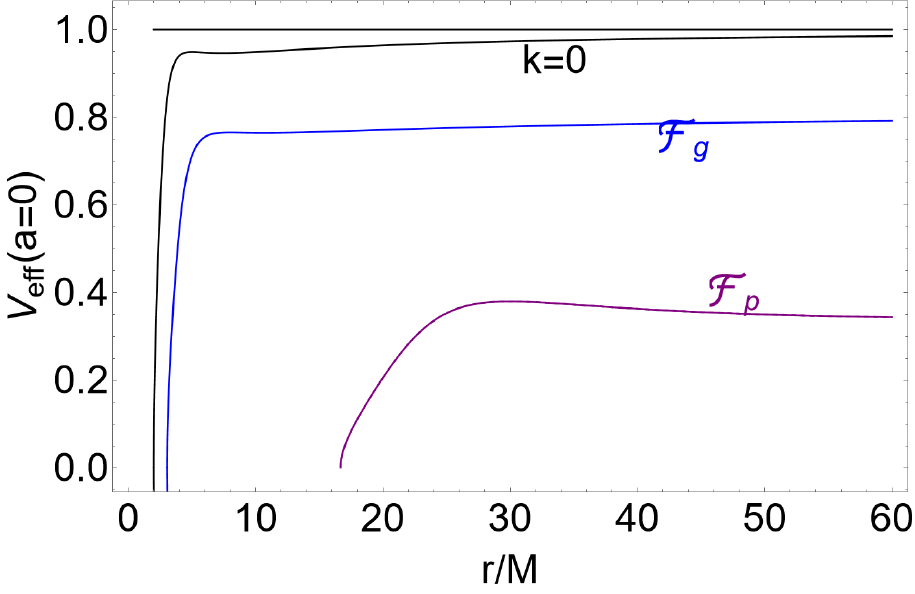}
         \includegraphics[width=5.7cm]{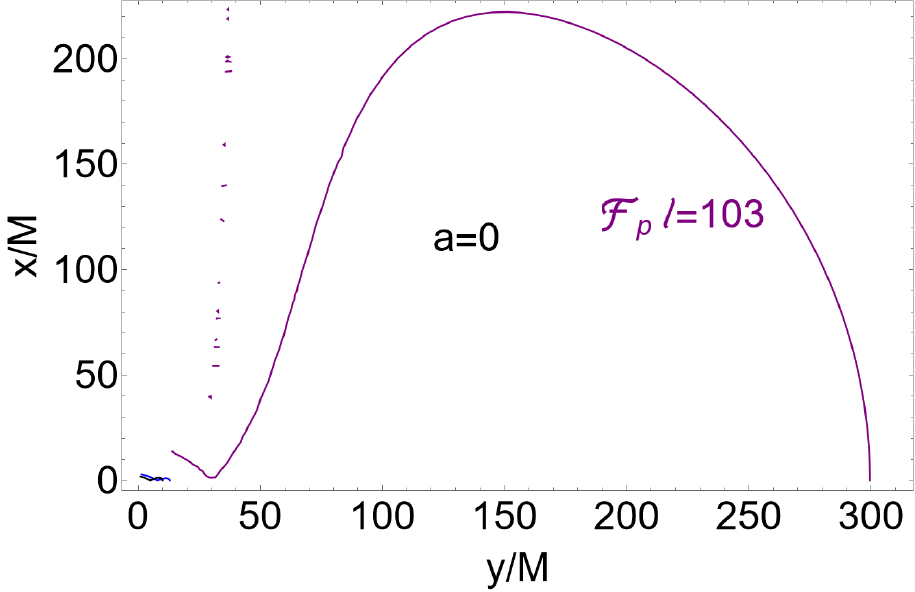}
               \includegraphics[width=5.7cm]{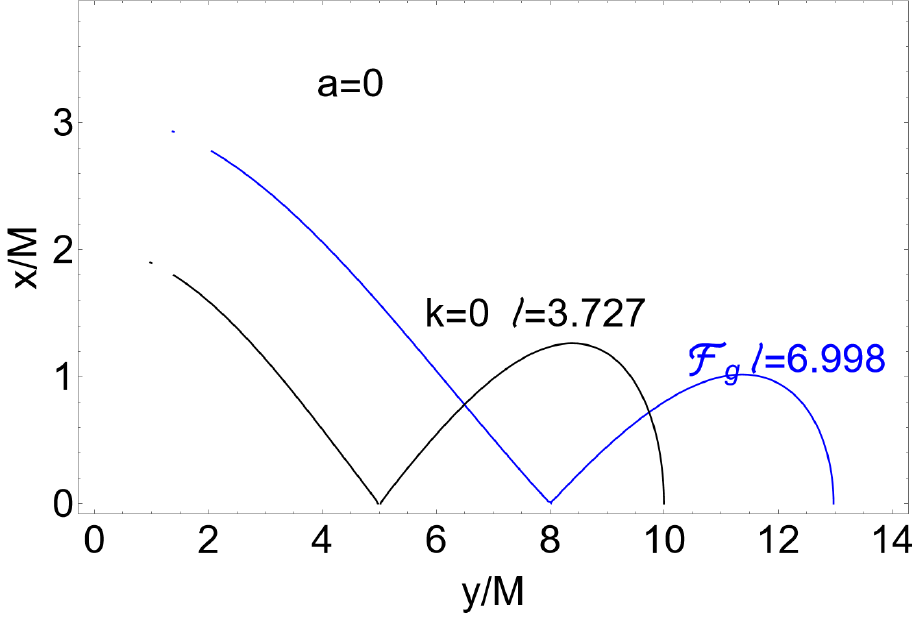}\\
                 \includegraphics[width=5.7cm]{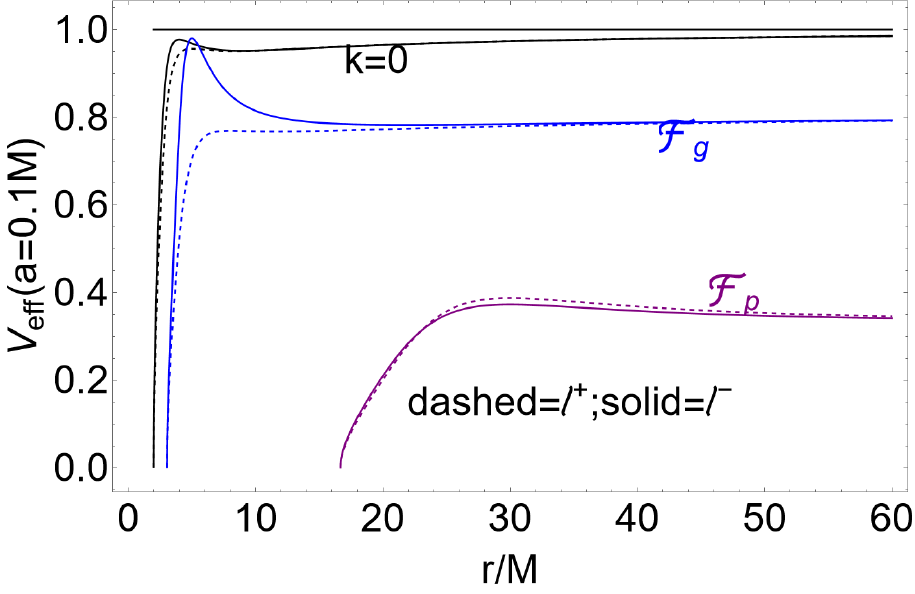}
         \includegraphics[width=5.7cm]{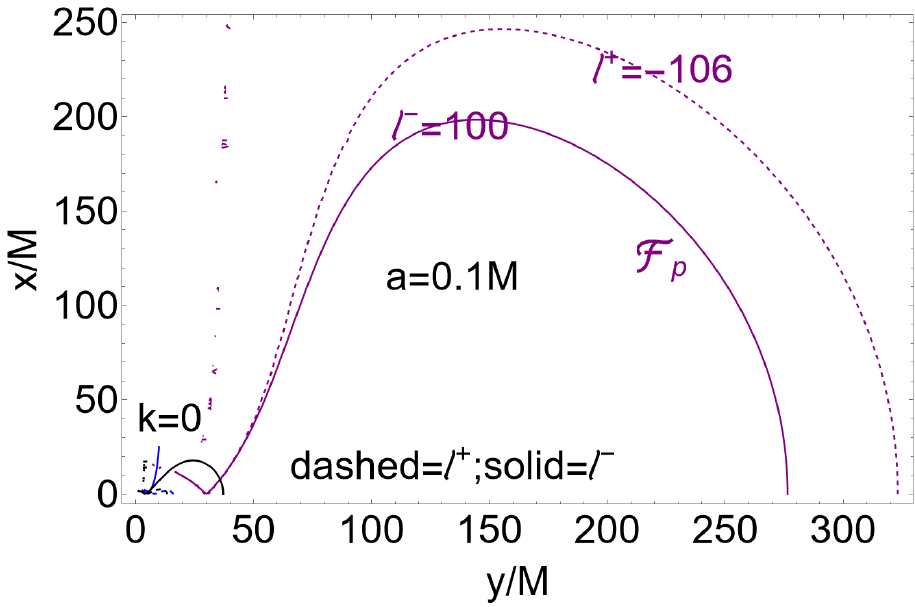}
               \includegraphics[width=5.7cm]{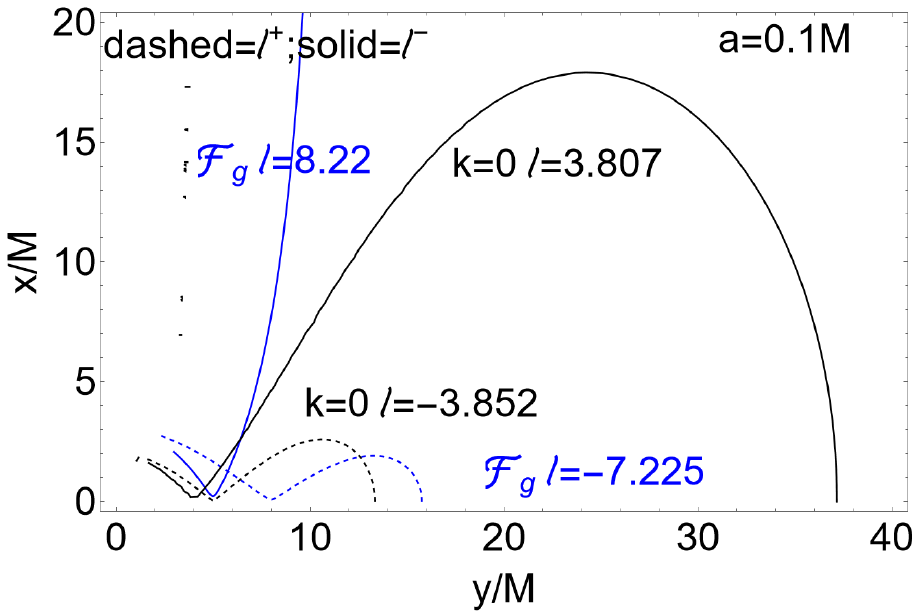}
         \includegraphics[width=5.7cm]{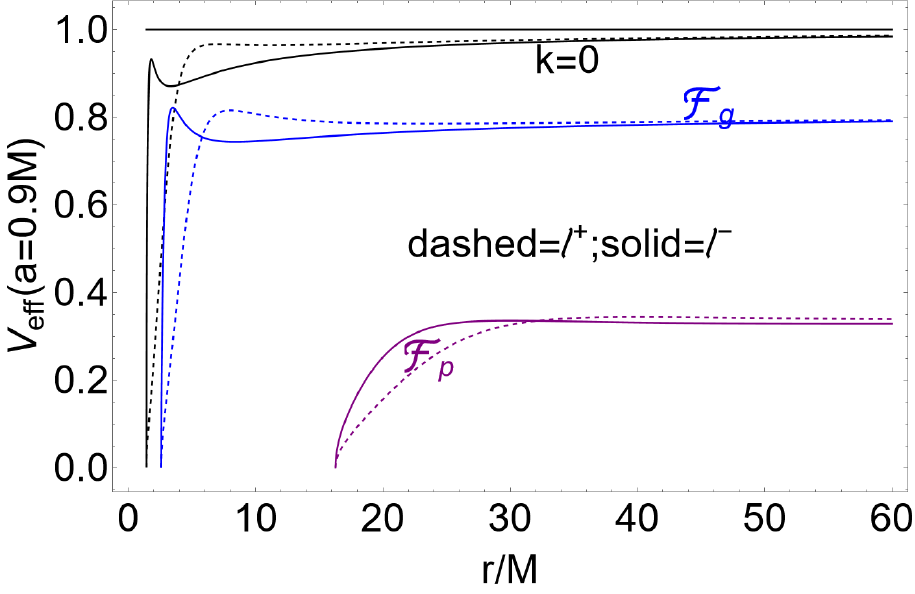}
               \includegraphics[width=5.7cm]{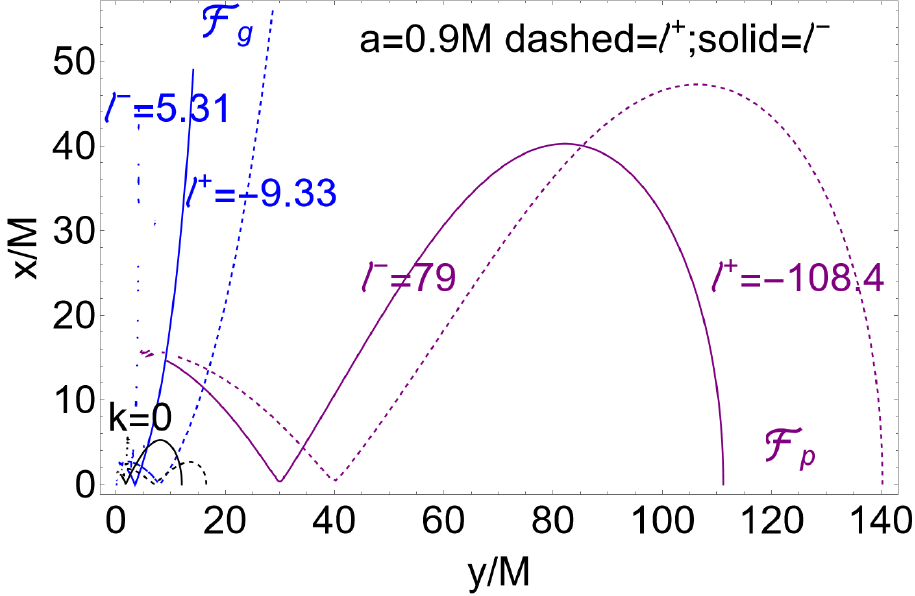}
         \includegraphics[width=5.7cm]{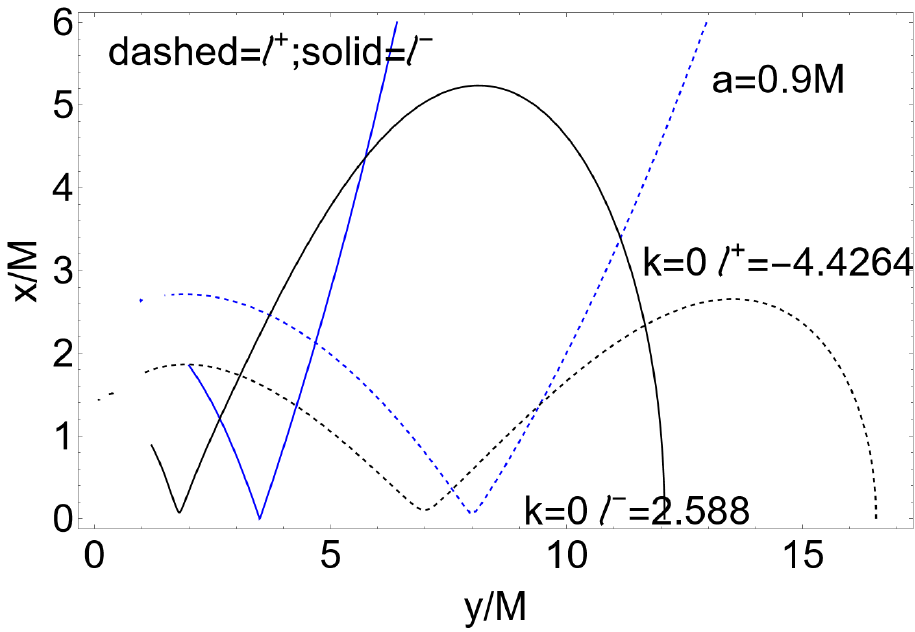}
           \caption{ Fluids effective potentials (left panels) and tori (center and right panels)  orbiting  in   scalar field dark matter (SFDM) geometry  of  Eq.\il(\ref{Eq:SFDM-metric}), with  parameters $\mathcal{F}_p$ (purple curves) and $\mathcal{F}_g$ (blue curves),  defined in  Eqs\il(\ref{Eq:kpkg-definition}).  The geodesic structure  is in  Figs\il(\ref{Fig:Plotallowsmall}).
            Rows are $a=0$ (top), $a=0.1M$ (middle), $a=0.9M$
(bottom).
  Black curves  for $k=0$ are the configurations for the case of Schwarzschild  and Kerr spacetimes in absence of DM. Tori are shown in the correspondent colors association relative to the effective potentials. Fluid specific angular momenta $\ell^+$ (dashed curves) and $\ell^-$ (solid curves) are signed close to each tori surface (center and right panels). Right panels are an enlarged view of the center panels in the region close to the central attractor. (The integration is in the entire orbital range). There is $r=\sqrt{x^2+y^2}$ and $\sigma=y^2/(x^2+y^2)$, where $\sigma\equiv \sin^2\theta$. }\label{Fig:Plotallowsmall50}
\end{figure}
However, as clear from  Figs\il(\ref{Fig:Plotallows}),  BH horizons exist also for
 $a>M$.

The geometry circular geodesic structure is  shown in   Figs\il(\ref{Fig:Plotallowsmall}) for the parameter $\mathcal{F}_g$  and $\mathcal{F}_p$, emphasizing  the differences for the cases $(k_p,R_p)$ and $(k_g,R_g)$ and the Kerr geometry circular structure of Figs\il(\ref{Fig:PlotconflKerr}). The orbital range locating the proto-jets  cusps, bounded by the radii $r_{mbo}^\pm$ and $r_{mco}^\pm$, for fluids specific angular momentum $\ell=\ell^\pm$ is very narrow, and in general the geodesic structure is shifted considerably outwardly with the respect to the Kerr geometry geodesic radii.
Therefore the range for the location of the accreting disks inner edges is considerably  larger  that the proto-jets cusp range, constituting possibly  a constraint on  the formation of proto-jets and tori with large angular momentum magnitude.
We note also that in the co-rotating case, for  $a\in]0,M]$,  and for $\mathcal{F}_p$ differently from the Kerr case in absence of DM, radii are located out of the ergoregion (and partially for  $\mathcal{F}_g$), this could imply a significant difference in the Lense-Thirring effects in presence of DM.
(There can be however over-spinning BHs (with $a>M$) where these effects could be present).
Differently from the   PFDM case, qualitatively the geodesic structure  is not differentiated with the respect to the case in absence of SFDM for  co-rotating  or counter-rotating  fluids, and for  static attractors  ($a=0$) and spinning attractors ($a\in]0,M]$) with SFDM.

We analyse below the case of the static attractor  i.e. $a=0$, having as limit in the vacuum the  Schwarzschild metric, we then studied the influence of the central attractor spin  combined with  dark matter effects, considering the case  $a=M$, corresponding to the DM deformation of  the vacuum  solution  (i.e. in absence of  DM) of the  extreme Kerr BH, and the case of the  slowly spinning attractor having $a=0.1M$.
More specifically:
\begin{description}
\item[\textbf{--The static attractor ($a=0$)}]
In Figs\il(\ref{Fig:PlotDardiesentisfdm0}) the fluid specific angular momentum $\ell^\pm$, tori  energy parameter $K^\pm$  and (test particles) Keplerian angular momentum $\mathcal{L}^\pm$ are shown as functions of $r/M$  for co-rotating and counter-rotating fluids, at different spins and  SFDM parameters, compared to the case $k=0$ corresponding  to the Schwarzschild  geometry. Tori and effective potentials are in Figs\il(\ref{Fig:Plotallowsmall50}).
\item[--The spinning attractors ($a=M$ and $a=0.1M$)]
We restrict our analysis to   $a=M$ and $a=0.1M$,   studying the cusped tori limiting the closed configurations, regulated by the effective potential function.
In Figs\il(\ref{Fig:PlotDardiesentisfdm0}),  there are the fluid specific angular momentum $\ell^\pm$, energy parameter $K^\pm$  and (test particles) Keplerian angular momentum $\mathcal{L}^\pm$ as function of $r/M$  for co-rotating and counter-rotating fluids, different spins and  SFDM parameters  of  Eq.\il(\ref{Eq:SFDM-metric}), with respect to the Kerr vacuum cases.
In Figs\il(\ref{Fig:Plotallowsmall50}) are  the fluids effective potentials and tori  compared to  the case of Kerr in absence of DM.
\end{description}
From  Figs\il(\ref{Fig:PlotDardiesentisfdm0}) we note that, within this parameter choice, differently from the case in absence of  DM, the fluids energy function can not converge to $1$ for large values of $r$.
In Figs\il(\ref{Fig:PlotDardiesentisfdm0})  the  fluid specific angular momentum distribution, compared to the distribution on the geometry in absence of DM, the  associated $K$ energy parameter and the Keplerian (test particle) angular momentum $\mathcal{L}^\pm\equiv \ell^\pm K^\pm$ are shown.
 From Figs\il(\ref{Fig:Plotallows}) it is clear how the  horizon curves in the plane $a-M$ are larger and shifted outwardly with respect to  the Kerr BH case, constituting  a discriminant for the  SFDM model and, for some values of the DM  parameters,  the BH  horizons   disappear giving rise to  a "DM--induced" NS.

Large tori orbiting SFDM spinning BHs are shown in
 Figs\il(\ref{Fig:Plotallowsmall50})  as, for example, the  purple surface for the case $a=0$ (from the tori effective potentials we can also note how  the tori  $K$ parameter for tori orbiting in SFDM are generally considerably lower than the $K$ parameter in absence of DM).

This feature of the DM model could also be an indication that such extremely huge   tori are actually not formed and similarly  the back-reaction on the metric is a predominant factor in these configurations, where self-gravity becomes a determinant  factor in the tori equilibrium.
\subsubsection{Cold Dark Matter (CDM)}\label{Sec:CDM}
The metric components  read
\bea\nonumber
&&g_{tt}=-\left[1-\frac{{2 M r}- \xi_{CDM}+r^2}{\Sigma}\right],
\quad g_{t\phi}=-\frac{a \sigma  \left[{2  M r}- \xi_{CDM}+r^2\right]}{\Sigma},\quad
g_{\phi\phi}\equiv\frac{\sigma  \left[\left(a^2+r^2\right)^2-a^2 \sigma\Delta_{CDM} \right]}{\Sigma};
\\\label{Eq:metric-CDM}
&&g_{rr}\equiv \frac{\Sigma}{\Delta_{CDM}},\quad g_{\theta\theta}=\Sigma
\eea
where
\bea
 \Delta_{CDM}\equiv a^2-{2  M r}+\xi_{CDM}\quad \mbox{and}\quad \xi_{CDM}\equiv r^2 \left(\frac{r}{R}+1\right)^{-\frac{8 \pi   \rho_c R^3}{r}}.
\eea
We adopt the parametrization
$\rho_c={k}/{R^3}$, where  $\rho_c$ is the density of the universe at
the moment when the  DM halo collapsed,  $R$ is a characteristic radius. The metric is asymptotically flat and we find the Kerr limit in  $k\to 0$ or $R \to +\infty$.
We first consider the metric singularities,  identifying  the  space of the parameters used in the tori analysis.
The  horizons $r_\pm$ can be written as  solutions of the equation $a=a_{\pm}(CDM)$, or
$R=R_{\pm}(CDM)$ or $k=k_{\pm}(CDM)$ where
\bea&&\nonumber
a_{\pm}(CDM)\equiv \sqrt{r \left[2M-r\wp^{-\frac{8 \pi  k}{r}}\right]},\quad R_{\pm}(CDM)\equiv  r \left[\frac{1}{1-\varsigma^{\frac{r}{8 \pi  k}}}-1\right],\quad
k_{\pm}(CDM)\equiv \frac{r \log\varsigma^{-1}}{8 \pi  \log \wp},
 \\\label{Eq:chan-kno}
 &&\mbox{with}\quad\wp\equiv \frac{r}{R}+1,\quad\mbox{and}\quad\varsigma\equiv \frac{2rM-a^2}{r^2}.
\eea
The metric horizons   are defined for
\bea&&\nonumber
a\in]0,M]: \quad r\in \left] r_{\circledast},{r_{-}}\right[;\quad a\in[0,M]: r>{r_+};\quad a>M: r> r_{\circledast},\quad \mbox{where}  \quad r_{\circledast}\equiv\frac{a^2}{2M},
\eea
 (assuming  $a>0,R>0,k>0$).

\begin{figure}
\centering
          \includegraphics[width=8cm]{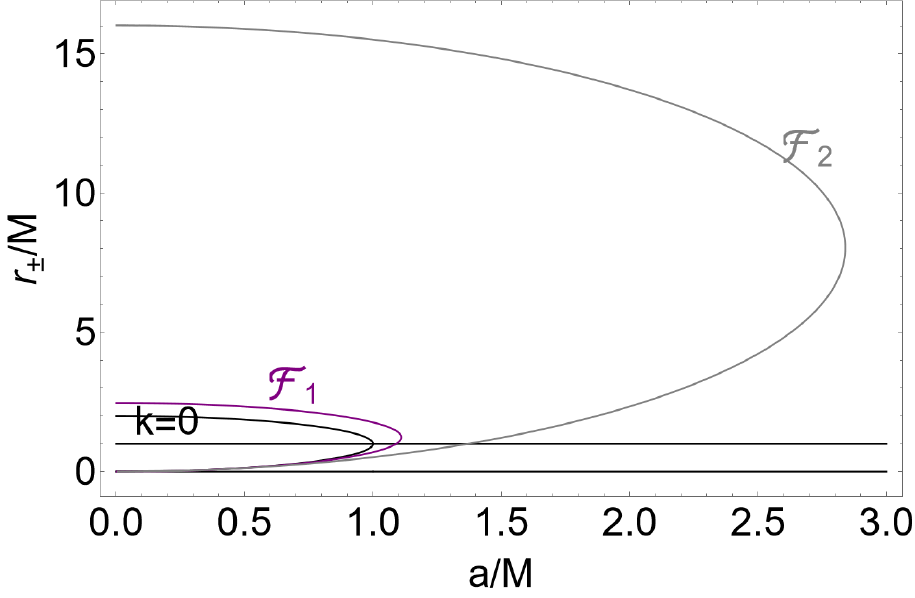}
                   \includegraphics[width=8cm]{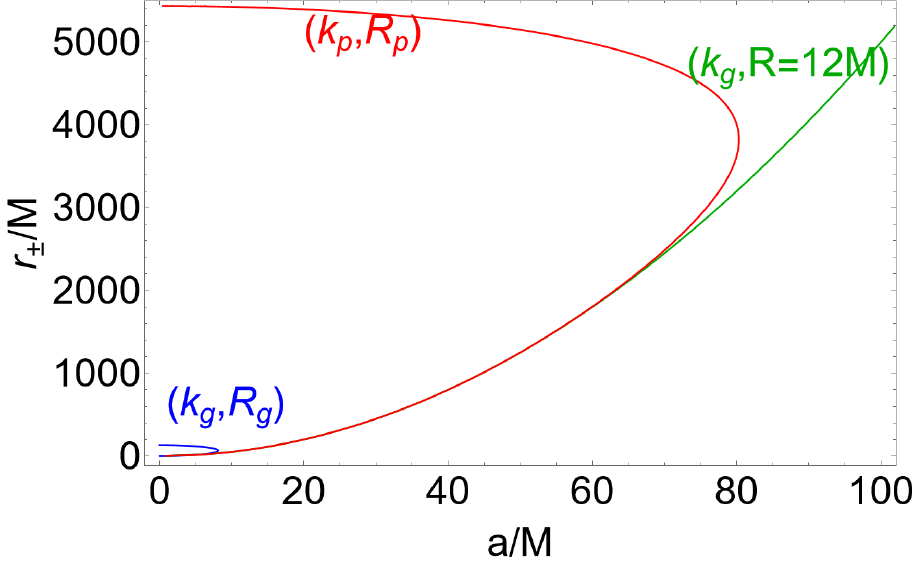}
 \includegraphics[width=8.cm]{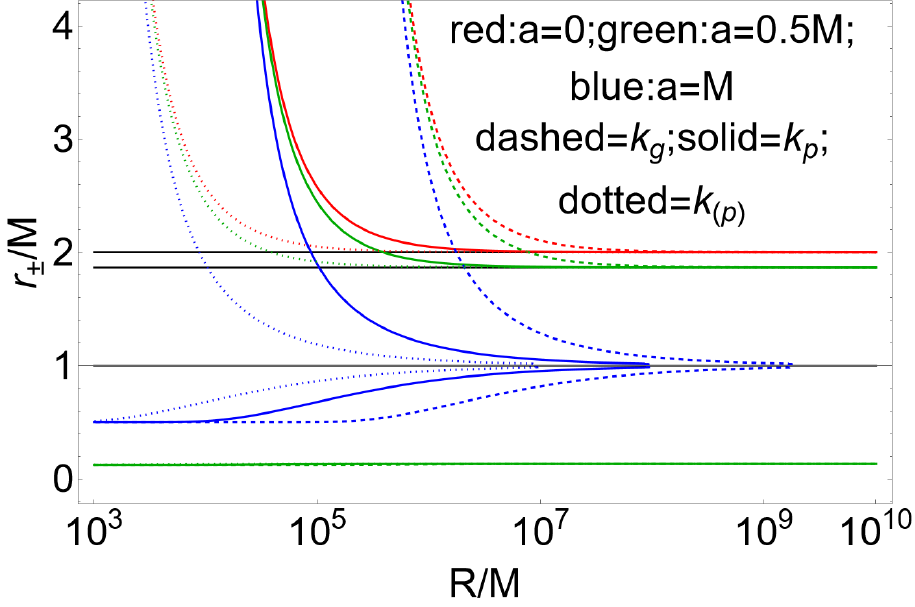}
                               \includegraphics[width=8cm]{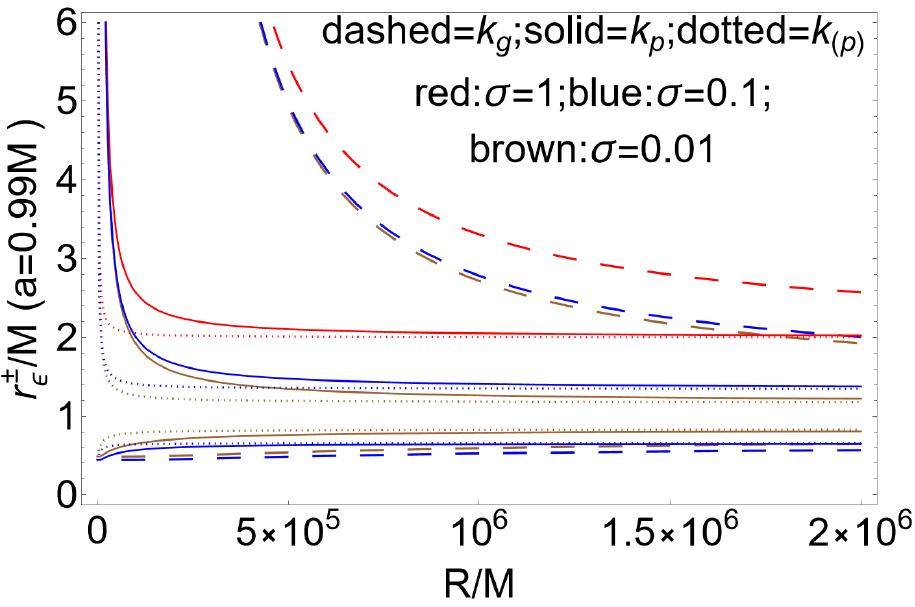}
           \caption{Horizons $r_\pm$ and ergosufaces $r_{\epsilon}^\pm$ of the cold dark matter CDM geometry  of Eqs\il(\ref{Eq:metric-CDM}).  Dark matter parameters $\mathcal{F}_1$, $\mathcal{F}_2$  and $k_{(p)}$ are in Eqs\il(\ref{Eq:F1-F2}), parameters $(k_p,k_g,R_g,R_p)$ are in Eqs\il(\ref{Eq:kpkg-definition}) . Black curves, $k=0$, correspond to the Kerr horizons. There is $\sigma\equiv\sin^2\theta$, where $\sigma=1$ is the equatorial plane. In the bottom left panel blue curves are $a=M$, red curves are $a=0$, green curves are $a=0.5M$, dotted curves are for  $k=k_{(p)}$, dashed curves are for $k=k_g$, solid curves are for $k=k_p$, black horizontal lines correspond to  $k=0$ for the Kerr spacetimes. Bottom right panel shows the ergosurfaces $r_\epsilon^\pm$ as function of  the dark matter parameter $R/M$ for spin $a=0.99M$ and for different planes $\sigma\in [0,1]$  and dark matter parameter $k$, where red curves correspond to  $\sigma =1$, blue curves correspond to $\sigma=0.1$, brown curves to $\sigma =0.01$,  dashed curves are for  $k=k_g$, solid curves for $k=k_p$ and dotted curves are for $k=k_{(p)}$.
}\label{Fig:Plotallowscold}
\end{figure}

The ergosurfaces  $r_\epsilon^\pm$ can be  found for $\sigma\neq 1$ as  solutions of the equation $a=a_\epsilon^\pm(CDM)\equiv {a_{\pm}(CDM)}/{\sqrt{1-\sigma}}$ and, on the equatorial plane  $(\sigma=1)$ as solutions of
$R=R_\epsilon^\pm$ or $k=k_\epsilon^\pm$ where\footnote{As clear from Figs\il(\ref{Fig:Plotallowscold}),
 these relations also represent some portions of  the ergosurfaces according to conditions on the DM parameters.}
 \bea&&
 k_{\epsilon}^\pm\equiv  \frac{r \log \frac{r}{2M}}{8 \pi  \log  \wp}\quad\mbox{and}\quad R_{\epsilon}^\pm\equiv  r \left[\frac{1}{1-\left(\frac{2M}{r}\right)^{\frac{r}{8 \pi  k}}}-1\right],
\eea
see Figs\il(\ref{Fig:Plotallowscold}).
On the equatorial plane, the outer ergosurface, independent on the spin $a$, corresponds to the metric singularity in the static ($a=0$) case--see Eqs\il(\ref{Eq:chan-kno}).

\begin{figure}
\centering
       \includegraphics[width=5.9cm]{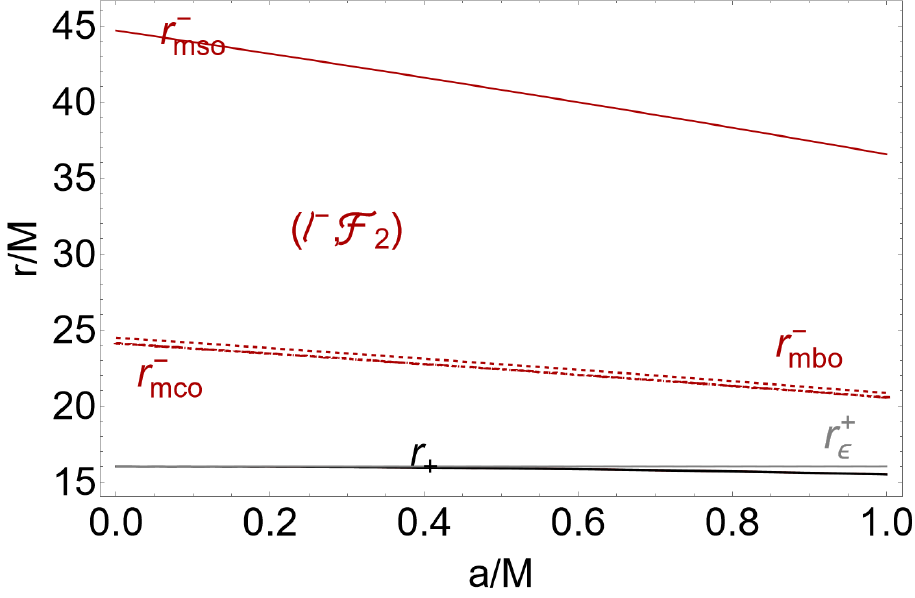}
              \includegraphics[width=5.9cm]{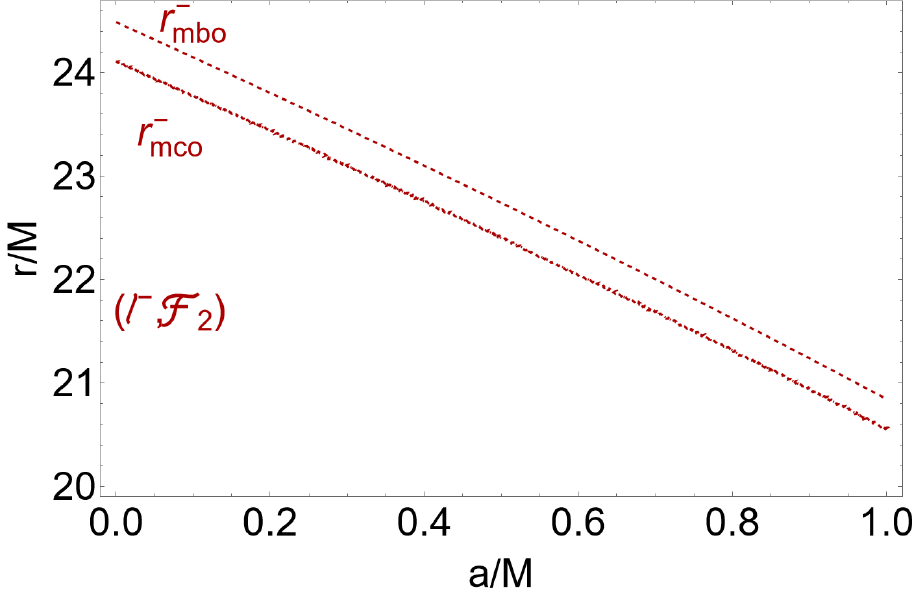}
                                                            \includegraphics[width=5.9cm]{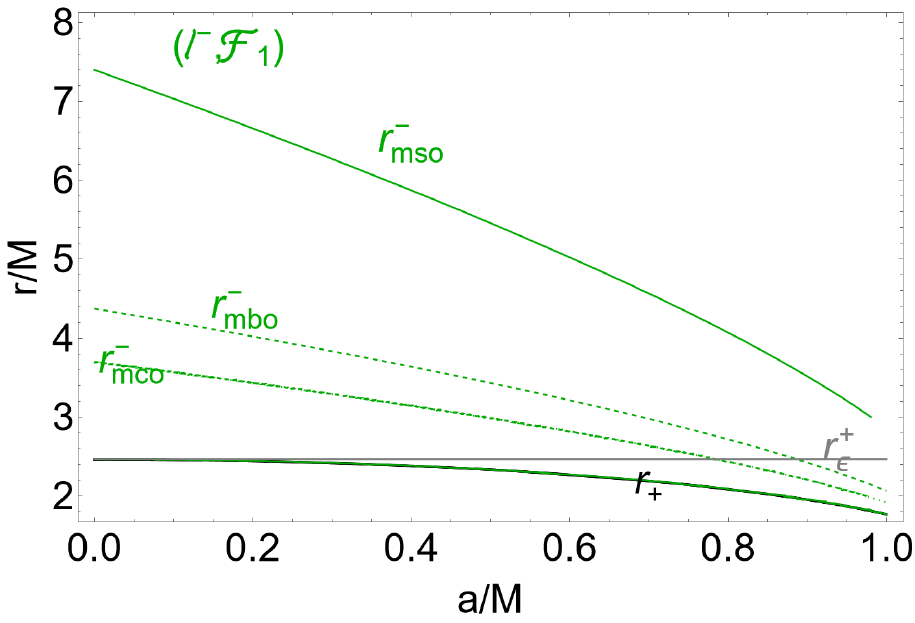}
                            \includegraphics[width=5.9cm]{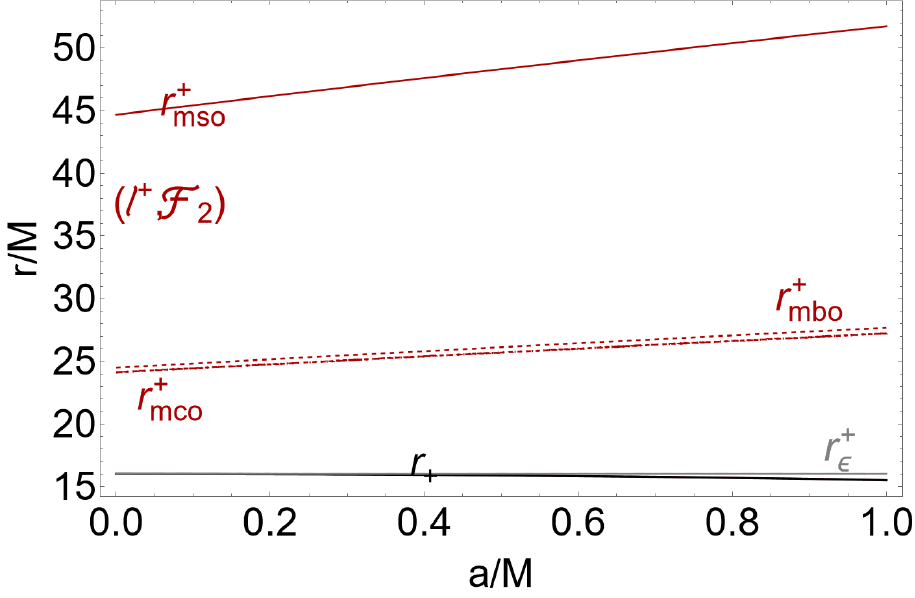}
              \includegraphics[width=5.9cm]{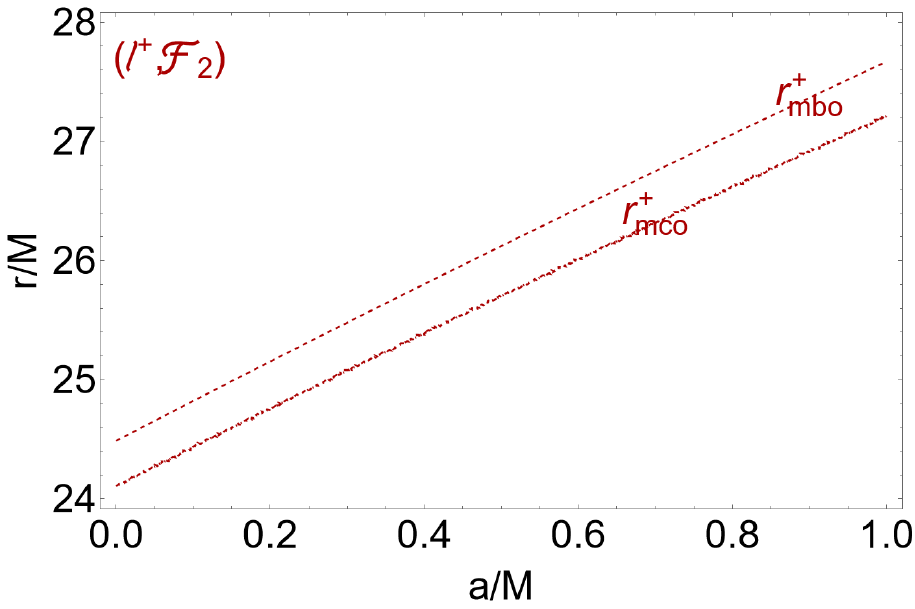}
                       \includegraphics[width=5.9cm]{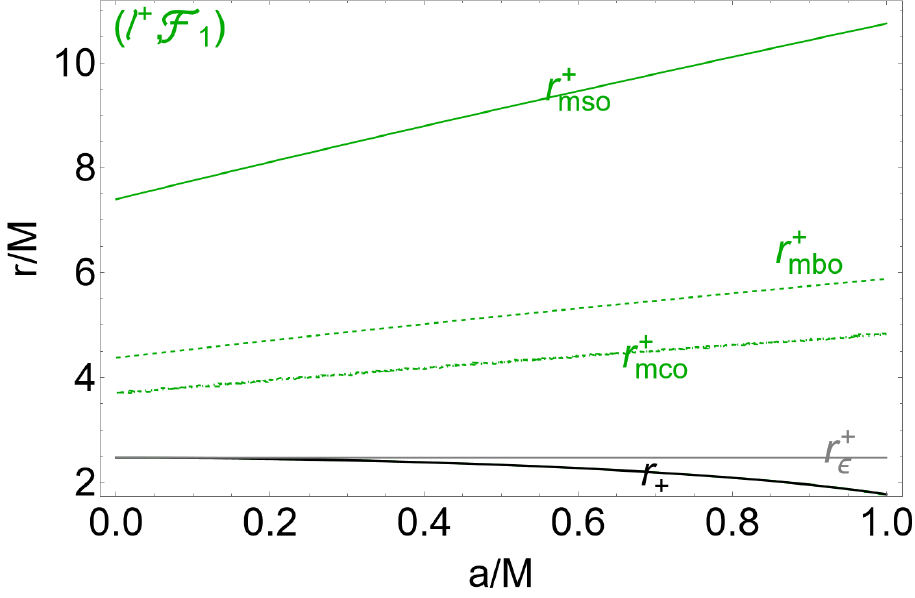}
                      \caption{Geodesic equatorial structure of  the cold dark matter CDM geometry of Eqs\il(\ref{Eq:metric-CDM}).  Dark matter parameters $\mathcal{F}_1$ (green curves) and $\mathcal{F}_2$ (red curves)  are in Eqs\il(\ref{Eq:F1-F2}). Upper (below) panels show the structure for  fluid specific angular momentum $\ell=\ell^->0$ ($\ell^+<0$) for co-rotating (counter-rotating) fluids. Center panels are an enlarged view of the right panels, showing the marginally stable  orbits ($mso$ solid curves)  marginally bounded orbits ($mbo$ dashed curves) and  marginally circular orbits $r_{mco}^\pm$ (dotted curves), where $r_+$ (black curves) is the outer horizon, $r_\epsilon^+$ (gray curves)  is the outer ergosurface on the equatorial plane. The correspondent Kerr geodesic structure is  in Figs\il(\ref{Fig:PlotconflKerr}). }\label{Fig:Plotchastrin}
\end{figure}

We consider the following  sets of parameters
\bea\label{Eq:F1-F2}
\mathcal{F}_1\equiv\{k\to 1000M,R\to 120000M\},\quad \mathcal{F}_2\equiv\{k\to 100M,R\to 1200M\},\quad \mbox{and}\quad k_{(p)}\equiv 100M
\eea
--Figs\il(\ref{Fig:Plotallowscold}).
The geodesic structure for this geometry is shown  in  Figs\il(\ref{Fig:Plotchastrin}).
Similarly to the SFDM model,  the equatorial geodesic structure shows that for the   $\mathcal{F}_2$ case, the range $[r_{mco}^\pm,r_{mbo}^\pm]$  of the   proto-jets  cusp  location is remarkably narrow. At $a\in]0,M]$,  for co-rotating fluids, the radii for $\mathcal{F}_1$  ($\mathcal{F}_2$) do not enter (are partially contained in) the outer ergoregion.
 \begin{figure}
\centering
  \includegraphics[width=8.5cm]{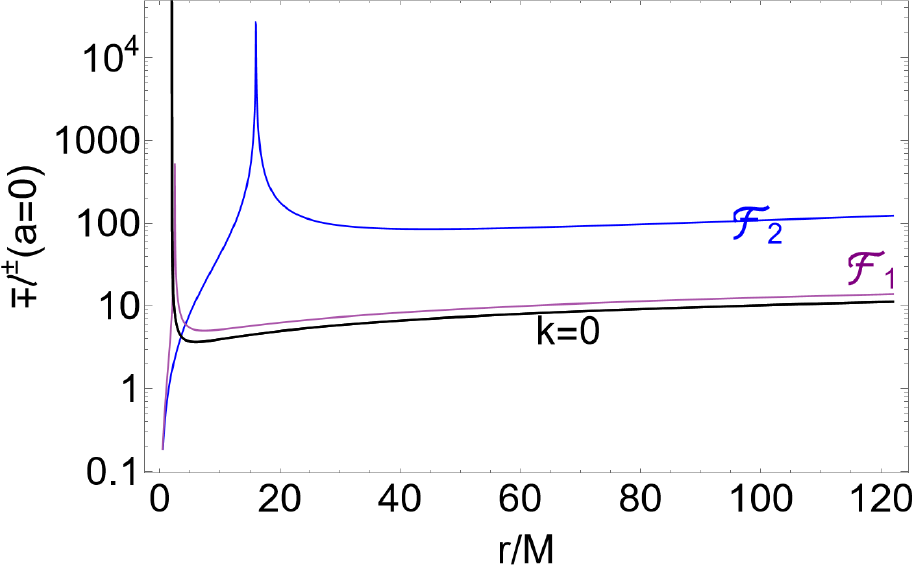} \includegraphics[width=8.5cm]{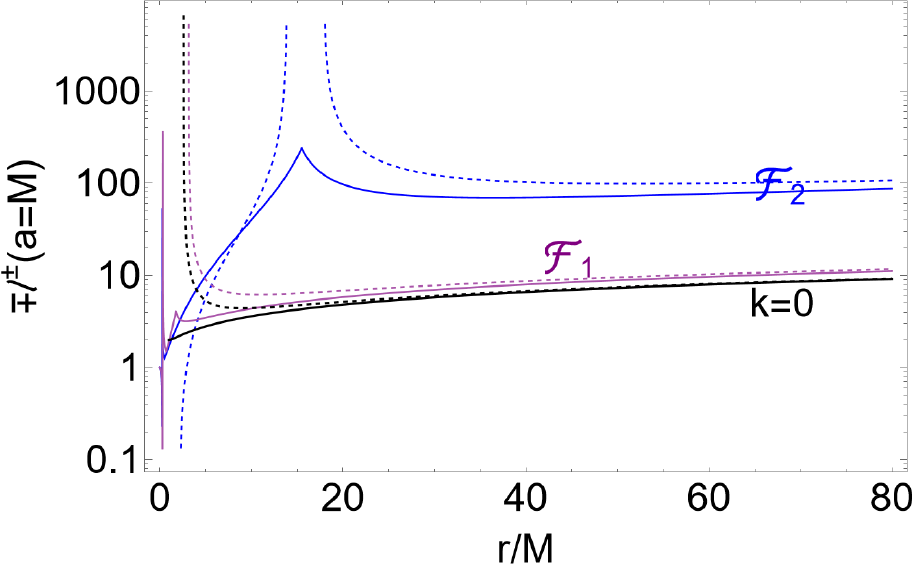}
 \includegraphics[width=8.5cm]{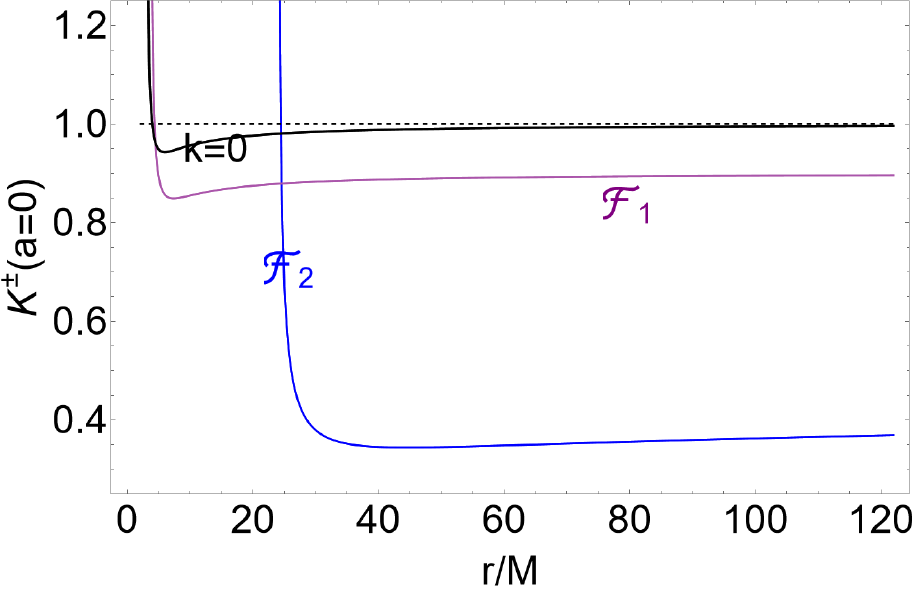}
\includegraphics[width=8.5cm]{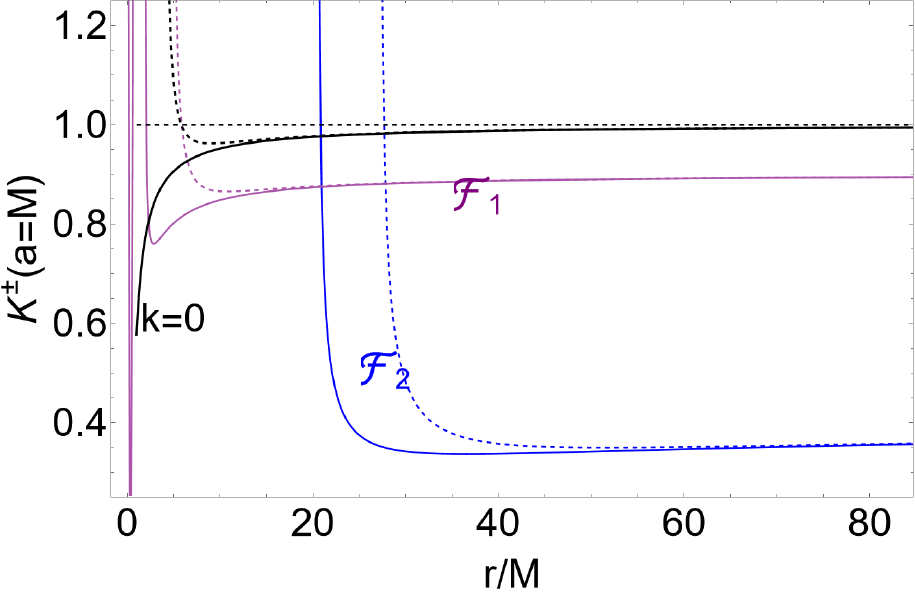}
\includegraphics[width=8.5cm]{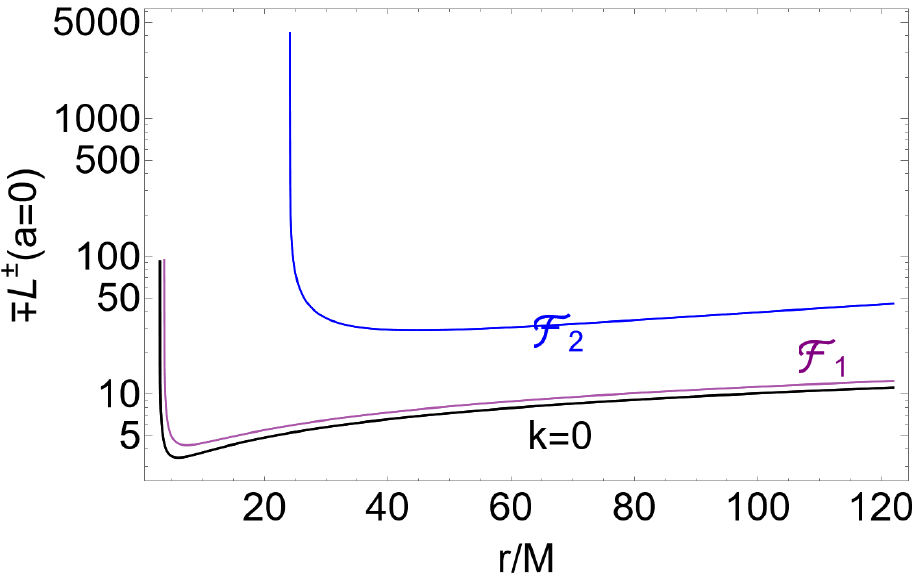}
\includegraphics[width=8.5cm]{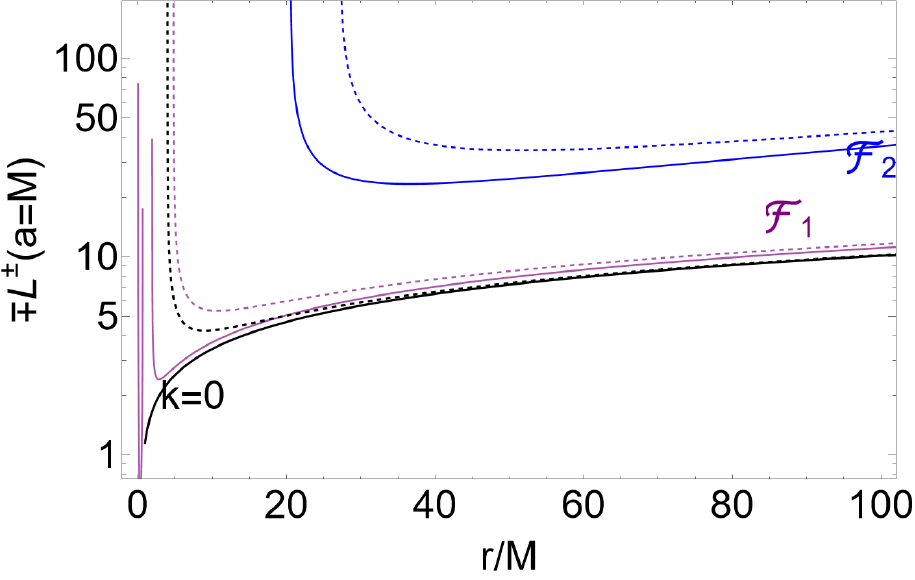}
           \caption{ Fluid specific angular momentum $\ell^\pm$ (upper row), energy parameter $K^\pm$ (middle row) and (test particles) Keplerian angular momentum $\mathcal{L}^\pm$ (bottom row) as function of $r/M$  for co-rotating and counter-rotating fluids, different  cold dark matter CDM parameters  of Eqs\il(\ref{Eq:metric-CDM}).  Dark matter parameters $\mathcal{F}_1$ (purple curves) and $\mathcal{F}_2$  (blue curves) are in Eqs\il(\ref{Eq:F1-F2}).   The columns are $a=0$ (left) and $a=M$ (right). Kerr and Schwarzschild  case are the black curves for $k=0$.
}\label{Fig:PlotDardiesenticfdm0}
\end{figure}
 \begin{figure}
\centering
         \includegraphics[width=7cm]{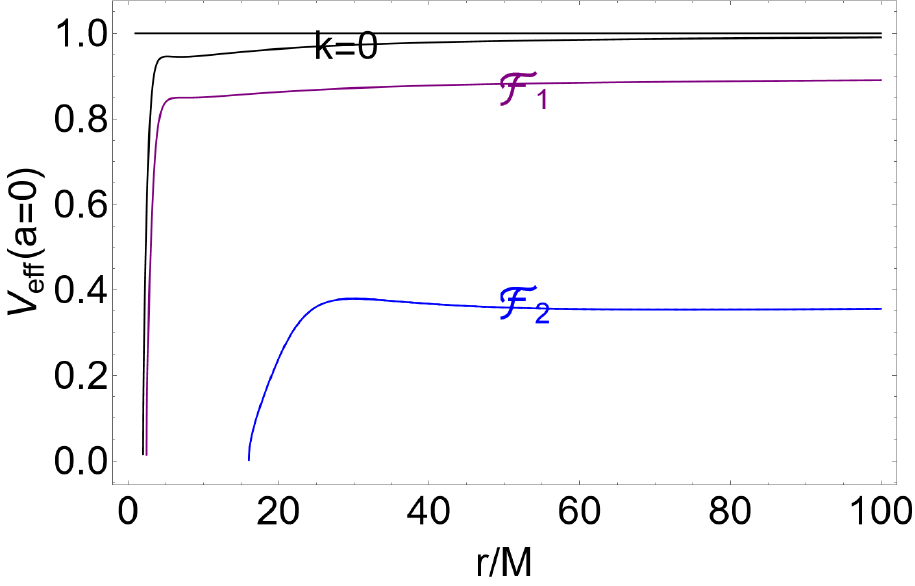}
    \includegraphics[width=7cm]{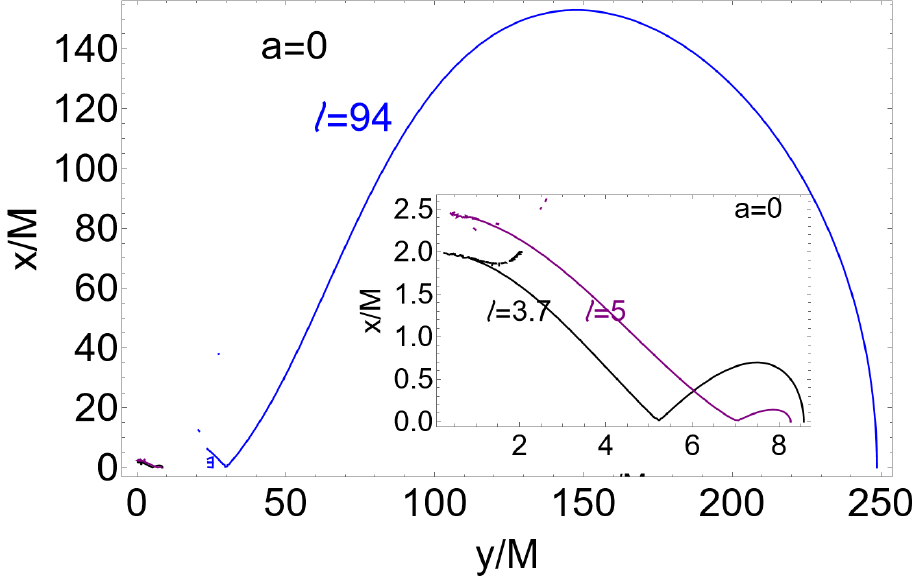}\\
             \includegraphics[width=7cm]{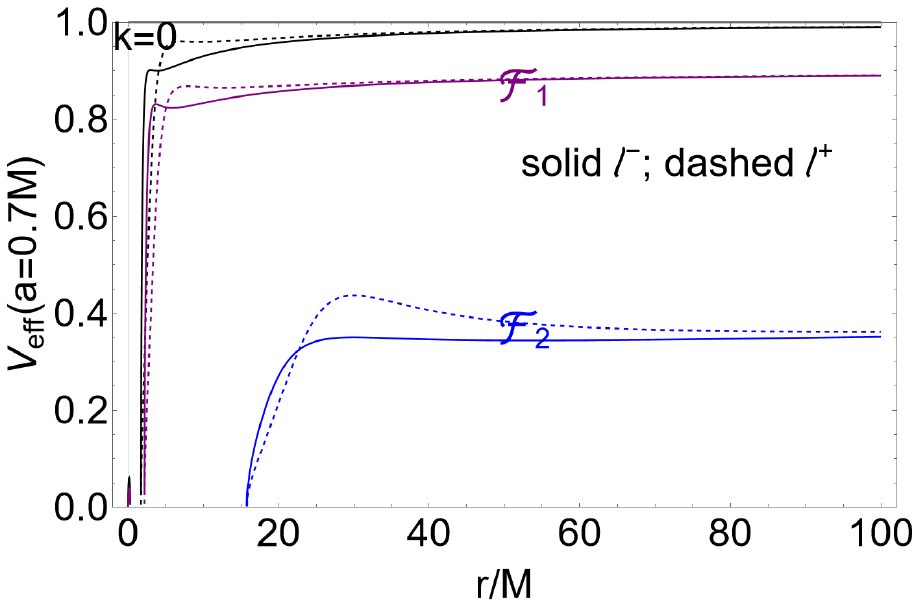}
       \includegraphics[width=7cm]{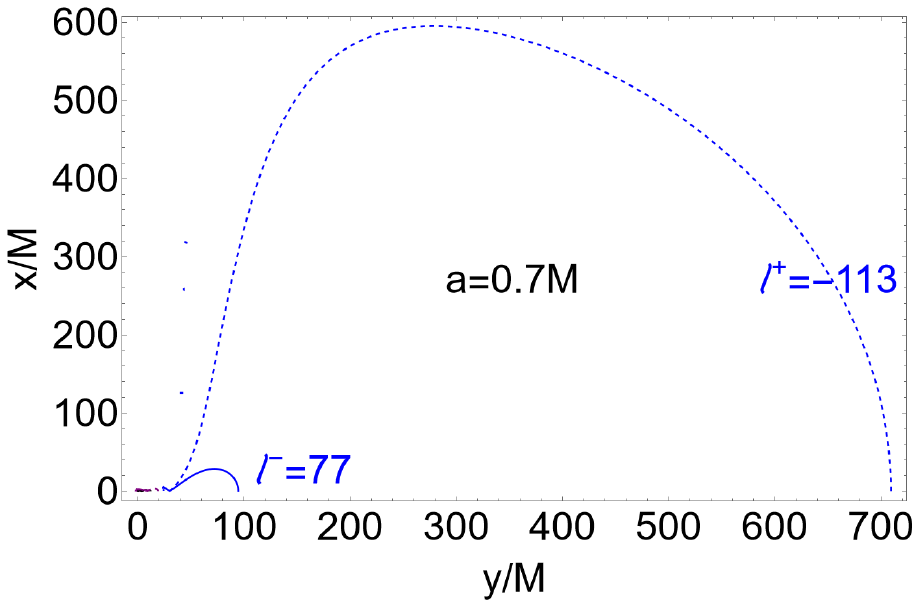}\\
       \includegraphics[width=7cm]{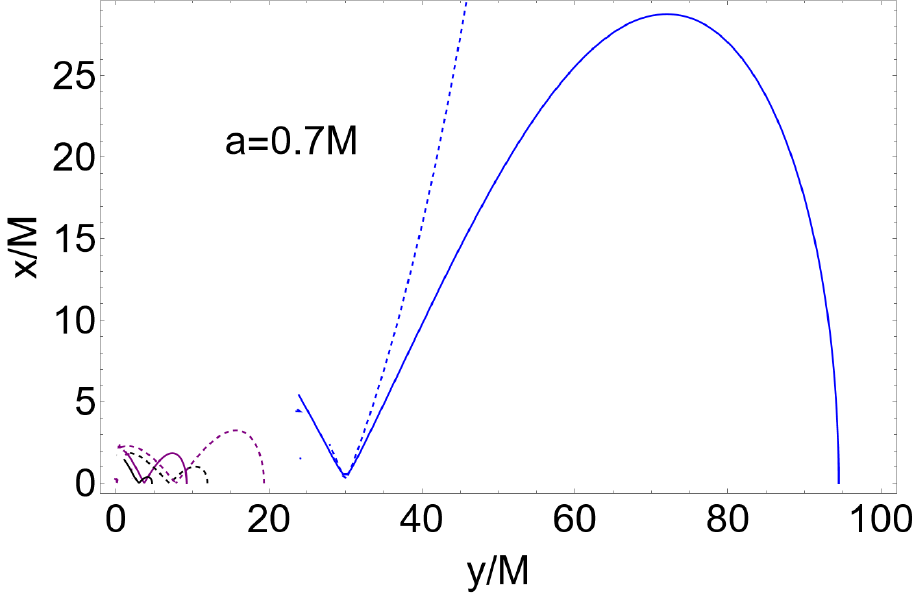}
       \includegraphics[width=7cm]{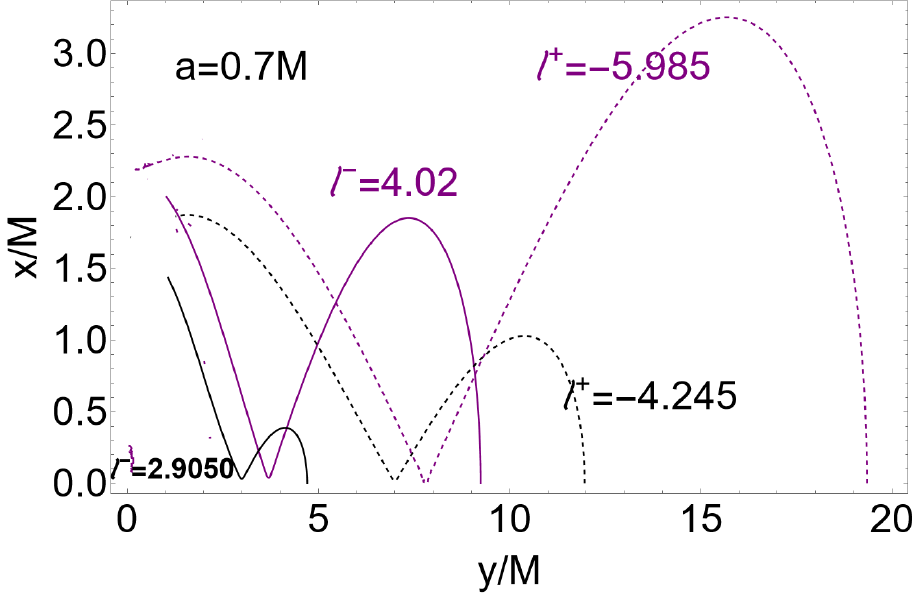}\\
          \includegraphics[width=7cm]{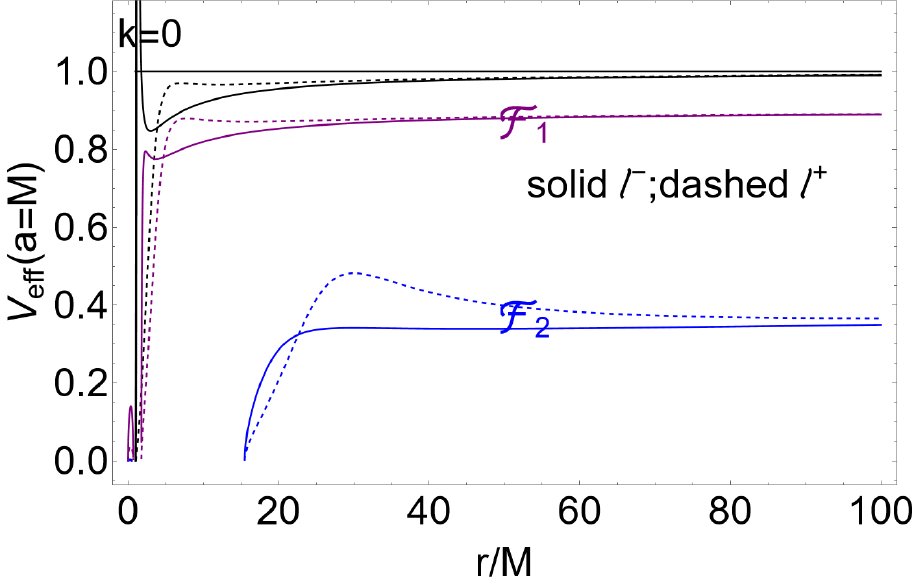}
       \includegraphics[width=7cm]{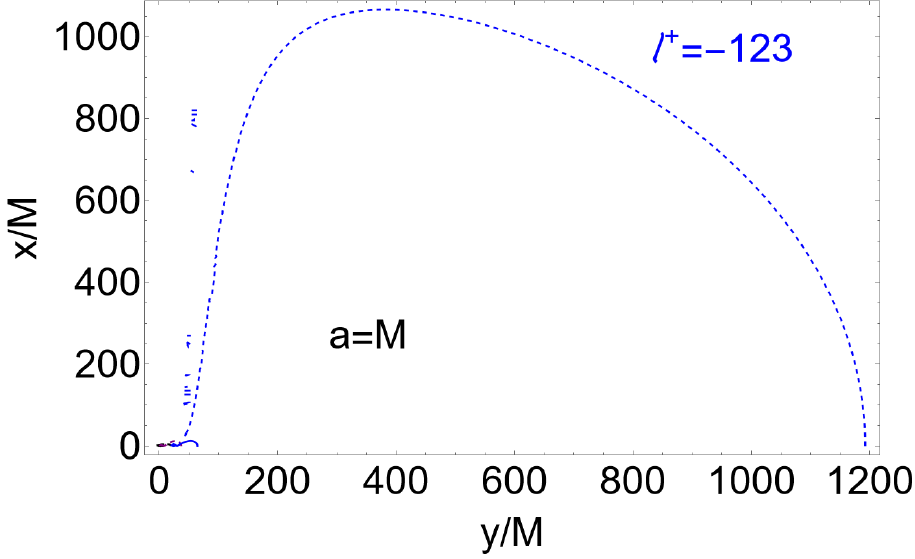}\\
       \includegraphics[width=7cm]{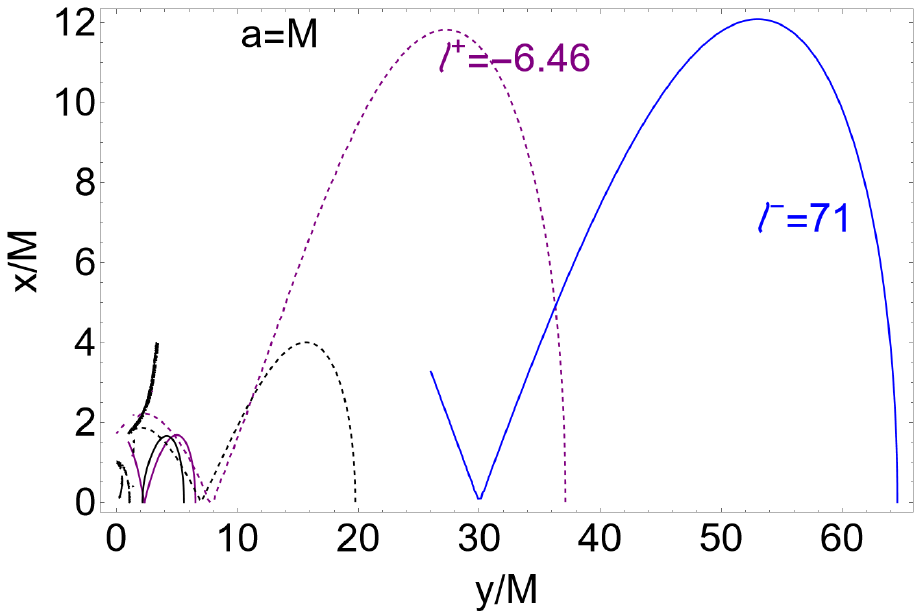}
       \includegraphics[width=7cm]{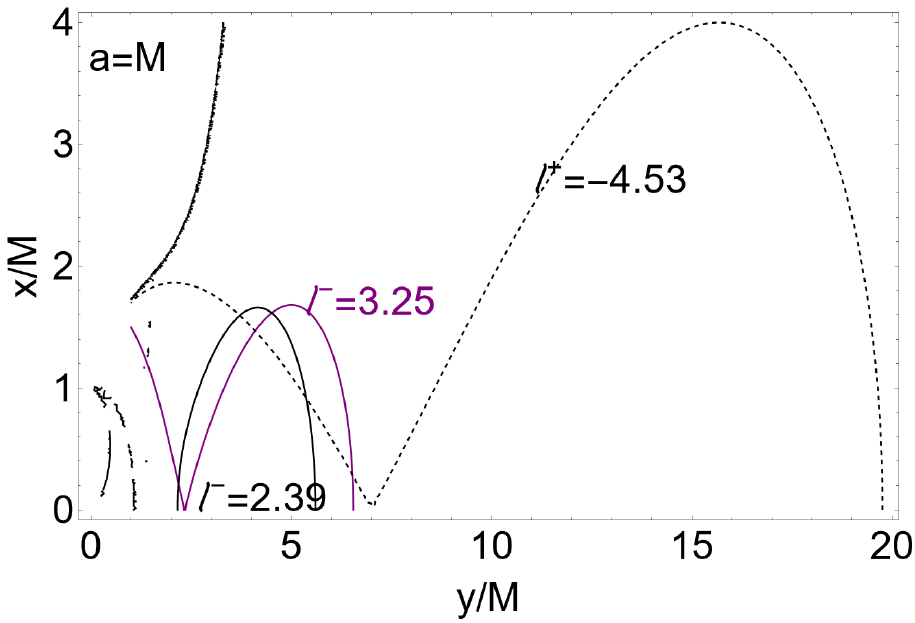}
 \caption{Effective potential and tori  of  the cold dark matter CDM geometry.  Dark matter parameters $\mathcal{F}_1$ (purple curves) and $\mathcal{F}_2$ (blue curves) are in Eqs\il(\ref{Eq:F1-F2}).   Rows are $a=0$ (top line), $a=0.7M$ (top second  and third lines), $a=M$
(top fourth and fifth lines).
  Black curves  for $k=0$ are the configurations for the case of Schwarzschild  and Kerr spacetimes in absence of DM. Tori are shown in the correspondent colors association relative to the effective potentials. Fluid specific angular momenta $\ell^+$ (dashed curves) and $\ell^-$ (solid curves) are signed close to each tori surface. Regions close to the central attractor are also shown in enlarged views panels. There is $r=\sqrt{x^2+y^2}$ and $\sigma=y^2/(x^2+y^2)$, where $\sigma\equiv \sin^2\theta$. }\label{Fig:Plotlavorc}
\end{figure}
We have analyzed the static and spinning attractors as follows:
\begin{description}
\item[--The static attractor ($a=0$)]
The fluid specific angular momentum $\ell^\pm$, energy parameter $K^\pm$  and (test particles) Keplerian angular momentum $\mathcal{L}^\pm$ as function of $r/M$,  and CDM parameters  $\mathcal{F}_1$ and $\mathcal{F}_2$  are in Figs\il(\ref{Fig:PlotDardiesenticfdm0}), compared to the  Schwarzschild case.
The effective potential and tori  are shown in   Figs\il(\ref{Fig:Plotlavorc})
 compared with the Schwarzschild   case.
\item[--The spinning attractor ($a\neq0$)]
Fluid specific angular momentum $\ell^\pm$, energy parameter $K^\pm$  and (test particles) Keplerian angular momentum $\mathcal{L}^\pm$ as function of $r/M$  are represented in Figs\il(\ref{Fig:PlotDardiesenticfdm0})  for co-rotating and counter-rotating fluids, different spins and CDM parameters $\mathcal{F}_1$  in comparison with  the  extreme Kerr BH case in absence of DM.
Effective potential and tori  for  CDM parameters $\mathcal{F}_1$ and $\mathcal{F}_2$  are in Figs\il(\ref{Fig:Plotlavorc}), for $a=M$ and $a=0.7M$ in Figs\il(\ref{Fig:Plotlavorc}), for  fluid specific angular momentum $\ell=\ell^->0$ ($\ell^+<0$) for co-rotating (counter-rotating) fluids, compared with  the case in the vacuum Kerr geometry
\end{description}
 It can be proved that, in all the cases considered, the limit  $K\to 1$ for large $r$ holds, where $\mathcal{L}^\pm\equiv \ell^\pm K^\pm$.
 Large tori orbiting CDM spinning BHs are shown in
 Figs\il(\ref{Fig:Plotlavorc})   as, for example, the  blue surface for the case $a=0$, dashed -blue curve for $a=0.7M$  and $a=M$ (similarly to the SFDM case,  the tori  $K$ parameter for tori orbiting in CDM are generally considerably lower than the $K$ parameter in absence of DM).

From Figs\il(\ref{Fig:Plotlavorc}) we note, as for  SFDM, the presence of  larger cusped  tori located far from the central spinning attractor, distinguishing  the DM deformed geometry from     the Kerr case.
\section{Discussion and Final Remarks}\label{Sec:discussion}
 In   DM  models considered here there are NS  solutions, solutions with one horizon  and two horizons,  according to the DM parameters. There are also BH spacetime solutions with horizons at $a>M$ or NSs for $a<M$.
The geodesic structure regulating the accretion physics  and  the tori location around the central spinning  attractor  can be  shifted considerably outwardly with the respect to the Kerr geometry. DM effects mimic Kerr attactors with
altered spin to mass ratio $a/M$. For example, in all the models presented, the DM
affects the disk inner edge which is a tracer of the $a/M$ in the Kerr geometry. The presence of an excretion cusp, double cusps, or double tori, which are also
typical of Kerr NSs solutions, could indicate the presence of DM. Consequently DM  affects BH horizon physics, considering   DM (models) as NSs  mimickers, or vice versa DM (models) as   BHs  mimickers   for super-spinars  (cosmological)  solutions.
DM could also affect the jet emission.
The orbital range locating the proto-jets  cusps  can be also very small, as discussed in Sec.\il(\ref{Sec:SFDM})
  for SFDM and  in Sec.\il(\ref{Sec:CDM}) for CDM. The open cusped solutions (constraining also the jet emission) are very different from their  counterparts in the Kerr spacetime in absence of DM.
In general DM  manifests also  with the existence of  extremely large cusped tori orbiting   very far from the central singularity. From Figs\il(\ref{Fig:Plotallowsmall50}) we see  the large dimensions of the cusped tori orbiting SFDM spinning BHs. The equilibrium of these  tori  may be hugely affected by the their self-gravity.

 In all  DM  models considered here,  however, DM affects the  geometric and causality properties, while there is no   coupling with ordinary matter (nor an hypothetical accretion disk consisting of dark matter in orbit), considering  gravity modified by the effects of  the dark matter--see also \cite{LH}.
We addressed three models,  drawing qualitative and comparative considerations,  ruling out some solutions  and  tracing  some common patterns. We have taken as a selection criterion in the space of  the metric parameters the observation that
there are two expected regimes, where there  is  a fully modified geometry,  qualitatively divergent with respect to the general relativistic onset, as a strongly  different horizons structures compared to the reference Kerr solution,   and the second scenarios   consisting in an appreciable quantitative  deformation of the orbiting structures,  but not a qualitatively significant change of the background geometry defined by the spinning BH.
The current methods of measuring and identifying BHs are  based also  on the physics of
   accretion, being related to the  accretion disk inner edge, which we prove to be  distorted by the DM
  treated, in the metric models considered here, as a background deformation.

Spherically symmetric black hole solutions in PFDM
have been  considered to be adapted to the  observed asymptotically flat rotation velocity in spiral galaxies and a possible  interaction between the DM  {halo and  central  BH has been  differently theorized. However it has been supposed that  SMBHs could enhance
the DM density significantly\footnote{Producing a so-called spike" phenomenon \cite{spike}.}.}
The  results of our analysis  prove  accretion to  be  a good indicator  of the divergences  induced by the  DM presence and the study of the accretion disks in DM models  to represent a valid  DM models discriminant.
The tori dimensions provide an indication  of the possible effects of DM for the energetics  associated with the physics of accretion around BHs.  In the  P-D models for example, the  thickness of the accretion throat (opening of the cusp for tori with specific fluid angular momentum in  \textbf{L1} with $K\in]K_\times,1[$) determines (in the assumptions of  vanishing pressure at the inner edge), many characteristics of  tori  energetics such as mass accretion rates and cusp luminosity,  the
rate of the thermal-energy    carried at the  cusp,    the  mass flow rate through the cusp (i.e., mass loss accretion rate),  the  fraction of energy produced inside the flow and not radiated through the surface but swallowed by central BH,
mass-flux, the  enthalpy-flux (related to the  temperature parameter),
  depending also on  the  EoS, as the  polytropic index and constant\footnote{Configurations considered here  have been often adopted as the initial conditions in the set up for simulations of the general relativistic magnetohydrodynamic (GRMHD) accretion structures \cite{Igumenshchev,Shafee,Fragile:2007dk,DeVilliers}.
 The geometrically thick axial symmetric  hydrodynamical models are widely adopted  in many contexts showing a remarkably good fitting with  the more complex  dynamical models  as  discussed  for example in \cite{Lei:2008ui}. In the current analysis of dynamical  systems of   both  general relativistic hydrodynamic (GRHD) and  GRMHD set-up, these tori are  commonly adopted as initial configurations for the numerical  analysis--\cite{Fragile:2007dk,DeVilliers,Porth:2016rfi} constituting also a  comparative model in  many  numerical analysis of  complex situations  sharing  the same symmetries.
Indeed the   general relativistic thick  tori {morphological}  features, related to  the equilibrium (quiescent) and accretion phases as the cusp emergence,  are   predominantly determined    by the centrifugal and gravitational components of the force balance in the disks rather then the dissipative ones.}. It has been shown  in \cite{submitted,retro-inversion} that the maximum  of flow thickness and the maximum amount of matter swallowed by the central BH is determined by the  attractor spin--mass ratio only, being defined by the location of the marginally stable circular orbit, and therefore  DM influences on  the tori dimension and the marginally stable orbits in these models can be searched  in a variation of the central BH energetics\footnote{
A  further relevant issue  in the  analysis  of the DM effects on  BHs accretion, is if   the
features  shown  as track  of the DM presence may  be used to distinguish  the DM models. An answer to this question is  immediate from the comparison  of  the horizons structures for the  PFDM model in Figs\il(\ref{Fig:PlotDardie}),
SFDM model in Figs\il(\ref{Fig:Plotallows}),
CDM model in Figs\il(\ref{Fig:Plotallowscold}),
and  of the geodesic structure (constraining  tori morphology and formation) for the PFDM  model
in Figs\il(\ref{Fig:PlotDardieM}),
SFDM model
in Figs\il(\ref{Fig:Plotallowsmall}),
CDM model
 in Figs\il(\ref{Fig:Plotchastrin}).
Thus, it is immediate  to determine that  in general   the main differences are  between PFDM model on the one side and CDM and SFDM models on the other, also for  small  values of  DM parameters.
In these  DM  models we have pointed out "DM-induced" NSs (slower spinning attractors  without BH horizons) and  in all  cases also "DM-induced" BHs,
(over spinning solutions with one or two horizons).
 Here,  we take into account  DM models  differentiation by means of the tori  characteristics   determining   possiply the DM presence, as the  inter-disk  cusps, double accretion tori, presence of
extremely large and far  tori, limited proto-jets ranges, DM differentiation according  to  fluid rotation orientation,  Lense--Thirring effects in DM presence.
While the  SFDM and CDM models show qualitatively similar characteristics, not
distinguishing substantially  DM effects for fluid rotation orientation  and  slowly spinning  from faster spinning attractors,
the situation for PFDM is clearly  different.
PFDM  model shows remarkable differences  also for a small variation of the DM $k$ parameter,
 distinguishing   DM effects on  co-rotating and counter-rotating fluids  and  between slowing spinning attractors and faster spinning  attractors. The
  CDM and SFDM models
  show, for the considered parameters  ranges ($a\in[0,M]$)  a  qualitatively  similar  geodesics  structure compared with  the  BH case in absence of DM.
 There are very large tori  located far from the attractor, and  proto-jets cusps constrained in a  narrow  orbital range around the attractor,  constraining  proto-jets emission and  the formation of tori
 with large angular  momentum in magnitude. We have also noticed  indications of a possible alteration of the   Lense--Thirring effects   on the disks and flows with respect to the Kerr case without DM.
In the considered parameters ranges,  major differences of  the PFDM  models with respect to the case in absence of DM appear in the formation of the inter-cusps, double configurations and possible excretion tori.
It must be stressed however  that,   while we have drawn here a DM models comparative analysis,  an in-depth exploration of   more extensive DM parameters ranges  in all models,  would  further narrow the DM parameters with the DM effects on the  BHs accretion.} \cite{Japan,letter,retro-inversion}.

It should be emphasized then that since DM-BHs can exhibit features associated with Kerr NSs,
it has implications on cosmic censorship, in the fact that observing a compact object with such
tracers (excretion cusp, double cusps,  double tori) would not require the
breaking of cosmic censorship (viewing a Kerr NSs), but instead could mean
one is observing a BH surrounded by  DM.
Finally in this work we developed a comparative analysis of accretion disks in different dark matter models while we reserve    the in-depth explorations of  different DM  parametric values for   future analysis.

\end{document}